\documentclass[a4paper,english,longauth]{aa}
\usepackage{mathptmx}
\usepackage[T1]{fontenc}
\usepackage[latin9]{inputenc}
\usepackage{subfigure}
\usepackage{color}
\usepackage{graphicx}
\usepackage{amssymb}
\usepackage[authoryear]{natbib}

\makeatletter

\newcommand{\noun}[1]{\textsc{#1}}
\providecommand{\tabularnewline}{\\}
\newcommand{\lyxdot}{.}

\titlerunning{The evolution of the mass-metallicity relation in VVDS up to $z\sim0.9$}

\makeatother

\usepackage{babel}

\begin{document}
\abstract{}{We want to derive the mass-metallicity relation of star-forming
galaxies up to $z\sim0.9$, using data from the VIMOS VLT Deep Survey.
The mass-metallicity relation is commonly understood as the relation
between the stellar mass and the gas-phase oxygen abundance.\\
}{Automatic measurement of emission-line fluxes and equivalent widths
have been performed on the full spectroscopic sample of the VIMOS
VLT Deep Survey. This sample is divided into two sub-samples depending
on the apparent magnitude selection: wide ($I_{\mathrm{AB}}<22.5$)
and deep ($I_{\mathrm{AB}}<24$). These two samples span two different
ranges of stellar masses. Emission-line galaxies have been separated
into star-forming galaxies and active galactic nuclei using emission
line ratios. For the star-forming galaxies the emission line ratios
have also been used to estimate gas-phase oxygen abundance, using
empirical calibrations renormalized in order to give consistent results
at low and high redshifts. The stellar masses have been estimated
by fitting the whole spectral energy distributions with a set of stellar
population synthesis models.\\
}{We assume at first order that the shape of the mass-metallicity
relation remains constant with redshift. Then we find a stronger metallicity
evolution in the wide sample as compared to the deep sample. We thus
conclude that the mass-metallicity relation is flatter at higher redshift.
At $z\sim0.77$, galaxies at $10^{9.4}$ solar masses have $-0.18$
dex lower metallicities than galaxies of similar masses in the local
universe, while galaxies at $10^{10.2}$ solar masses have $-0.28$
dex lower metallicities. By comparing the mass-metallicity and luminosity-metallicity
relations, we also find an evolution in mass-to-light ratios: galaxies
at higher redshifts being more active. The observed flattening of
the mass-metallicity relation at high redshift is analyzed as an evidence
in favor of the open-closed model.\\
}{}


\title{Physical properties of galaxies and their evolution\\
in the VIMOS VLT Deep Survey%
\thanks{based on data obtained with the European Southern Observatory Very
Large Telescope, Paranal, Chile, program 070.A-9007(A), and on data
obtained at the Canada-France-Hawaii Telescope, operated by the CNRS
in France, CNRC in Canada and the University of Hawaii.%
}}

\subtitle{I. The evolution of the mass-metallicity relation up to $z\sim0.9$}

\author{F. Lamareille \inst{1,3}\and J. Brinchmann \inst{19} \and
T. Contini \inst{1} \and C.J. Walcher \inst{7} \and S. Charlot
\inst{10,8} \and E. Pérez-Montero \inst{1} \and G. Zamorani
\inst{3} \and L. Pozzetti \inst{3} \and M. Bolzonella \inst{3}
\and B. Garilli \inst{2} \and S. Paltani \inst{15,16} \and
A. Bongiorno \inst{22} \and O. Le Fèvre \inst{7} \and D. Bottini
\inst{2} \and V. Le Brun \inst{7} \and D. Maccagni \inst{2}
\and R. Scaramella \inst{4,13} \and M. Scodeggio \inst{2}
\and L. Tresse \inst{7} \and G. Vettolani \inst{4} \and A.
Zanichelli \inst{4} \and C. Adami \inst{} \and S. Arnouts
\inst{23,7} \and S. Bardelli \inst{3} \and A. Cappi \inst{3}
\and P. Ciliegi \inst{3} \and S. Foucaud \inst{21} \and P.
Franzetti \inst{2} \and I. Gavignaud \inst{12} \and L. Guzzo
\inst{9} \and O. Ilbert \inst{20} \and A. Iovino \inst{9}
\and H.J. McCracken \inst{10,11} \and B. Marano \inst{6} \and
C. Marinoni \inst{18} \and A. Mazure \inst{7} \and B. Meneux
\inst{22,24} \and R. Merighi \inst{3} \and R. Pellò \inst{1}
\and A. Pollo \inst{7,17} \and M. Radovich \inst{5} \and
D. Vergani \inst{2} \and E. Zucca \inst{3} \and A. Romano
\inst{5} \and A. Grado \inst{5} \and L. Limatola \inst{5}}

\institute{Laboratoire d'Astrophysique de Toulouse-Tarbes, Université de Toulouse,
CNRS, 14 av. E. Belin, F-31400 France \and IASF-INAF, Via Bassini
15, I-20133, Milano, Italy \and INAF-Osservatorio Astronomico di
Bologna, Via Ranzani 1, I-40127, Bologna, Italy \and IRA-INAF, Via
Gobetti 101, I-40129, Bologna, Italy \and INAF-Osservatorio Astronomico
di Capodimonte, Via Moiariello 16, I-80131, Napoli, Italy \and Università
di Bologna, Dipartimento di Astronomia, Via Ranzani 1, I-40127, Bologna,
Italy \and Laboratoire d'Astrophysique de Marseille, UMR 6110 CNRS-Université
de Provence, BP8, F-13376 Marseille Cedex 12, France \and Max Planck
Institut für Astrophysik, D-85741, Garching, Germany \and INAF-Osservatorio
Astronomico di Brera, Via Brera 28, I-20021, Milan, Italy \and Institut
d'Astrophysique de Paris, UMR 7095, 98 bis Bvd Arago, F-75014, Paris,
France \and Observatoire de Paris, LERMA, 61 Avenue de l'Observatoire,
F-75014, Paris, France \and Astrophysical Institute Potsdam, An der
Sternwarte 16, D-14482, Potsdam, Germany \and INAF-Osservatorio Astronomico
di Roma, Via di Frascati 33, I-00040, Monte Porzio Catone, Italy \and
Universitá di Milano-Bicocca, Dipartimento di Fisica, Piazza delle
Scienze 3, I-20126, Milano, Italy \and Integral Science Data Centre,
ch. d'Écogia 16, CH-1290, Versoix, Switzerland \and Geneva Observatory,
ch. des Maillettes 51, CH-1290, Sauverny, Switzerland \and Astronomical
Observatory of the Jagiellonian University, ul Orla 171, PL-30-244,
Krak{ó}w, Poland \and Centre de Physique Théorique, UMR 6207 CNRS-Université
de Provence, F-13288, Marseille, France \and Centro de Astrof{í}sica
da Universidade do Porto, Rua das Estrelas, P-4150-762, Porto, Portugal
\and Institute for Astronomy, 2680 Woodlawn Dr., University of Hawaii,
Honolulu, Hawaii, 96822, USA \and School of Physics \& Astronomy,
University of Nottingham, University Park, Nottingham, NG72RD, UK
\and Max Planck Institut für Extraterrestrische Physik (MPE), Giessenbachstrasse
1, D-85748 Garching bei München,Germany \and Canada France Hawaii
Telescope corporation, Mamalahoa Hwy, Kamuela, HI-96743, USA \and
Universitätssternwarte München, Scheinerstrasse 1, D-81679 München,
Germany }

\maketitle

\section{Introduction}

The stellar mass and the gas-phase metallicity of a galaxy are two
of the main parameters involved in the study of galaxy formation and
evolution. As cosmological time progresses, theory predicts that both
the mean metallicity and stellar mass of galaxies increase with age
as galaxies undergo chemical enrichment and grow through merging processes.
At any given epoch, the accumulated history of star formation, gas
inflows and outflows, affects a galaxy mass and its metallicity. Hence
one expects theses quantities to show some correlation and this will
provide crucial information about the physical processes that govern
galaxy formation.

First discovered for irregular galaxies \citep{Lequeux:1979A&A....80..155L},
the mass-metallicity relation has been intensively studied \citep[among others]{Skillman:1989ApJ...347..875S,Brodie:1991ApJ...379..157B,Zaritsky:1994ApJ...420...87Z,Richer:1995ApJ...445..642R,Garnett:1997ApJ...489...63G,Pilyugin:2000A&A...358...72P}
and is now well established in the local universe by the work of \citet{Tremonti:2004astro.ph..5537T}
with SDSS data and \citet{Lamareille:2004MNRAS.350..396L} with 2dFGRS
data, the latter done on the luminosity-metallicity relation which
is easier to derive when small number of photometric bands are available.
These two studies have shown in two different ways that the mass-metallicity
relation is mainly driven by the decrease of metal loss when stellar
mass increases. \citet{Tremonti:2004astro.ph..5537T} have indeed
observed an increase of the effective yield with stellar mass, while
\citet{Lamareille:2004MNRAS.350..396L} have shown the increase of
the slope of the luminosity-metallicity relation. This last trend
has also been observed down to much lower galaxy masses by \citet{Lee:2006ApJ...647..970L}.
We nevertheless note that they also observe a large scatter in the
effective yield, which they find difficult to analyse in the context
of more efficient mass loss among low mass galaxies.

Hierarchical galaxy formation models, that take into account the chemical
evolution and feedback processes, are able to reproduce the observed
mass-metallicity relation in the local universe \citep[e.g.][]{DeLucia:2004MNRAS.349.1101D,deRossi:2007MNRAS.374..323D,Finlator:2008MNRAS.385.2181F}.
However these models rely on free parameters, such as feedback efficiency,
which are not yet well constrained by observations. Alternative scenarios
have been proposed to explain the mass-metallicity relation including
low star formation efficiency in low-mass galaxies caused by supernova
feedback \citep{brooks:2006astro.ph..9620B} and a variable stellar
initial mass function being more top-heavy in galaxies with higher
star formation rates, thereby producing higher metal yields \citep{Koeppen:2006astro.ph.11723K}.

The evolution of the mass-metallicity relation on cosmological timescales
is now predicted by semi-analytic models of galaxy formation, that
include chemical hydrodynamic simulations within the standard $\Lambda$-CDM
framework \citep{DeLucia:2004MNRAS.349.1101D,Dave:2007MNRAS.374..427D}.
Reliable observational estimates of the mass-metallicity relation
of galaxies at different epochs (and hence different redshifts) may
thus provide important constraints on galaxy evolution scenarios.
Estimates of the mass-metallicity - or luminosity-metallicity - relation
of galaxies up to $z\sim1$ are limited so far to small samples \citep[among others]{Hammer:2005A&A...430..115H,Liang:2004A&A...423..867L,Maier:2004A&A...418..475M,Kobulnicky:2003ApJ...599.1006K,Maier:2005astro.ph..8239M}.
Recent studies have been performed on larger samples ($>100$ galaxies)
but with contradictory results. \citet{Savaglio:2005ApJ...635..260S}
concluded with a steeper slope in the distant universe, interpreting
these results in the framework of the closed-box model. On the contrary,
\citet{Lamareille:2006A&A...448..907L} did not find any significant
evolution of the slope of the luminosity-metallicity relation, while
the average metallicity at $z\approx0.9$ is lowered by $0.55$ dex
at a given luminosity, and by $0.28$ dex after correction for luminosity
evolution. \citet{Shapley:2005ApJ...635.1006S} and \citet{Liu:2008ApJ...678..758L}
have also found $0.2-0.3$ dex lower metallicities at $z=1$in the
DEEP2 sample. At higher redshifts \citet{erb:2006ApJ...644..813E}
derived a mass-metallicity relation at $z\sim2$ lowered by $0.3$
dex in metallicity compared with the local estimate, a trend which
could extend up to $z\sim3.3$ \citep{Maiolino:2007arXiv0712.2880M}.
Please note that the above numbers are given after the metallicities
have been renormalized to the same reference calibration in order
to be comparable \citep{Kewley:2008arXiv0801.1849K}.

In this work, we present the first attempt to derive the mass-metallicity
relation at different epochs up to $z\sim1$ using a unique, large
(almost 20~000 galaxies) and homogeneous sample of galaxies selected
in the VIMOS VLT Deep Survey (VVDS). The paper is organized as follows:
the sample selection is described in Sect.~\ref{sec:Description-of-the},
the estimation of stellar masses and magnitudes from SED fitting are
described in Sect.~\ref{sec:masses}, the estimation of metallicities
from line ratios is described in Sec.~\ref{sec:metals}, and we finally
study the luminosity-metallicity (Sect.~\ref{sec:The-luminosity-metallicity-relation}),
and mass-metallicity (Sect.~\ref{sec:The-mass-metallicity-relation})
relations. Throughout this paper we normalize the derived stellar
masses, and absolute magnitudes, with the standard $\Lambda$-CDM
cosmology, i.e. $h=0.7$, $\Omega_{\mathrm{m}}=0.3$ and $\Omega_{\Lambda}=0.7$
\citep{Spergel:2003ApJS..148..175S}.

\section{Description of the sample\label{sec:Description-of-the}}

\begin{figure}
\begin{centering}
\includegraphics[width=1\linewidth]{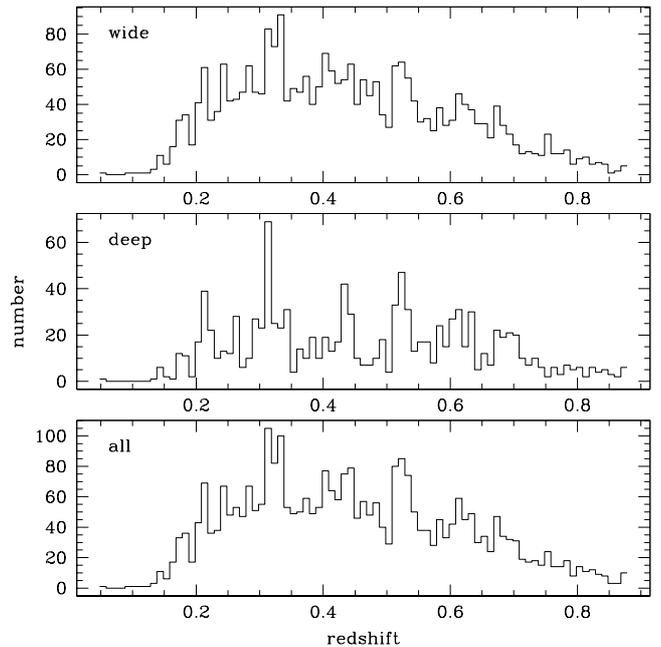} 
\par\end{centering}

\caption{Redshift distributions of our wide and deep samples of star-forming
galaxies, and of their union (see text for details).}

\label{fighistoz}
\end{figure}

The VIMOS VLT Deep Survey \citep[VVDS,][]{LeFevre:2003SPIE.4834..173L}
is one of the widest and deepest spectrophotometric surveys of distant
galaxies with a mean redshift of $z\approx0.7$. The optical spectroscopic
data, obtained with the VIsible Multi-Object Sectrograph \citep[VIMOS,][]{LeFevre:2003SPIE.4841.1670L}
installed at ESO/VLT (UT3), offers a great opportunity to study the
evolution of the mass-metallicity relation on a statistically significant
sample up to $z\approx0.9$. This limitation is imposed, in the current
study, by the bluest emission lines needed to compute a metallicity
(i.e. {[}O\noun{ii}]$\lambda$3727, H$\beta$ and {[}O\noun{iii}]$\lambda$5007)
being redshifted out of the wavelength range of the survey (approximately
$5500\mbox{\AA}<\lambda<9500\mbox{\AA}$). Metallicities measurements
up to $z\approx1.24$ using a different set of lines will be provided
in a subsequent paper (Perez-Montero et al. 2008). The spectra have
been taken in one arcsecond width slits, under a spectral resolution\textbf{
$R_{\mathrm{s}}\approx230$.}

The first epoch VVDS spectroscopic sample is purely apparent magnitude-selected
and is divided into two deep fields ($17.5\le I_{\mathrm{AB}}\le24$):
VVDS-02h \citep[herefater F02, ][]{LeFevre:2005A&A...439..845L} and
CDFS \citep{LeFevre:2004A&A...428.1043L}; and three wide fields ($17.5\le I_{\mathrm{AB}}\le22.5$):
VVDS-22h, VVDS-10h and VVDS-14h \citep[hereafter F22, F10, and F14 respectively, ][]{Garilli:2008arXiv0804.4568G}.
Spectroscopic data have been reduced using the \emph{VIPGI} pipeline
\citep{Scodeggio:2005PASP..117.1284S}, which performs automatic 1D
spectra extraction, correction for telluric absorption lines, and
flux calibration. We only kept objects with a redshift known at a
confidence level greater than 75\% (i.e. VVDS redshift flags 9, 2,
3 and 4). Duplicated observations were not used (some objects have
been observed twice or more, either by chance or intentionally), we
always kept the main observation. We also removed, for the purpose
of our study (i.e. star-forming galaxies), all stars (zero redshift)
and the sample of broad-line Active Galactic Nuclei \citep[hereafter AGNs; ][]{Gavignaud:2006A&A...457...79G}.

The photometric coverage of the VVDS spectroscopic sample is as follows:
CFH12k observations in $BVRI$ bands for F02, F22, F10 and F14 fields
\citep{McCracken:2003A&A...410...17M,LeFevre:2004A&A...417..839L},
completed by $U$ band observations in the F02 field \citep{Radovich:2004A&A...417...51R},
CFHTLS (Canada-France-Hawaï Telescope Legacy Survey) $u*g'r'i'z'$
bands for F02 and F22 fields, and $JKs$ photometry available in F02
field \citep{Iovino:2005A&A...442..423I,Temporin:2008A&A...482...81T}.
For the CDFS field, we use CFH12k $UBVRI$ \citep{Arnouts:2001A&A...379..740A}
and HST $bviz$ \citep{Giavalisco:2004ApJ...600L..93G} observations.
Finally, objects in F02 and F22 fields have been cross-matched with
the UKIDSS public catalog \citep{Warren:2007astro.ph..3037W}, providing
additional observations in $JK$ bands.

Throughout this paper, we will use a \emph{deep} and a \emph{wide}
samples. The deep sample is made of F02 and CDFS fields. Up to $z<1.4$
(the limit for the {[}O\noun{ii}]$\lambda$3727 emission line to be
in the observed wavelength range), the deep sample contains 7404 galaxies
(non-broad-line AGN) with a redshift measured at a confidence level
greater than 75\%. The wide sample is made of F22, F10, F14, F02 and
CDFS fields (the last two being limited out to $I_{\mathrm{AB}}\le22.5$).
Up to $z<1.4$ the wide sample contains 13978 galaxies (non-broad-line
AGN) with a redshift measured at a confidence level greater than 75\%.
We draw the reader's attention on the fact that the two samples overlap
for galaxies observed in F02 or CDFS fields at $I_{\mathrm{AB}}\le22.5$,
giving a total number of 18648 galaxies. Fig.~\ref{fighistoz} shows
the redshift distribution of the star-forming galaxies (see Sect.~\ref{sub:starforming})
in the wide and deep samples and in their union.

\subsection{Automatic spectral measurements}

The emission lines fluxes and equivalent widths, in all VVDS galaxy
(non-broad-line AGN) spectra, have been measured with the \emph{platefit\_vimos}
pipeline. Originally developed for the high spectral resolution SDSS
spectra \citep{Tremonti:2004astro.ph..5537T,Brinchmann:2004MNRAS.351.1151B},
the \emph{platefit} software has first been adapted to fit accurately
all emission lines after removing the stellar continuum and absorption
lines from lower resolution and lower signal-to-noise spectra \citep{Lamareille:2006A&A...448..893L}.
Finally, other improvements have been made to the new \emph{platefit\_vimos}
pipeline thanks to tests performed on the VVDS and zCOSMOS \citep{Lilly:2006astro.ph.12291L}
spectroscopic samples. A full discussion of \emph{platefit\_vimos}
pipeline will be presented in Lamareille et al. (in preparation) but,
for the benefit of the reader, we outline some of the main features
in this section.

The stellar component of the spectra is fitted as a non-negative linear
combination of $30$ single stellar population templates with different
ages ($0.005$, $0.025$, $0.10$, $0.29$, $0.64$, $0.90$, $1.4$,
$2.5$, $5$ and $11$Gyr) and metallicities ($0.2$, $1$ and $2.5$
$Z_{\odot}$). These templates have been derived using the \citet{Bruzual:2003MNRAS.344.1000B}
library and resampled to the velocity dispersion of VVDS spectra.
The dust attenuation in the stellar population model is left as a
free parameter. Foreground dust attenuation from the Milky Way has
been corrected using \citet{Schlegel:1998ApJ...500..525S} maps.

After removal of the stellar component, the emission lines are fitted
together as a single \emph{nebular spectrum} made of a sum of gaussians
at specified wavelengths. All emission lines are tied to have the
same width, with exception of the {[}O\noun{ii}]$\lambda$3727 line
which is actually a doublet of two lines at 3726 and 3729 \AA{} and
appear broadened compared to the other single lines. The spectral
resolution is also too low to clearly separate {[}N\noun{ii}]$\lambda$6584
and H$\alpha$ emission lines. It has been shown however by \citet{Lamareille:2006A&A...448..893L}
that the {[}N\noun{ii}]$\lambda$6584/H$\alpha$ emission-line ratio,
which is used as a metallicity calibrator, can be reliably measured
above a sufficient signal-to-noise ratio even at the resolution of
VVDS.

Note that because of the limited observed wavelength coverage of the
spectra, we do not observe all well-known optical lines at all redshifts.
Thanks to the stellar-part subtraction, no correction for underlying
absorption has to be applied to the Balmer emission lines.

The error spectrum which is needed for both fits of the stellar or
nebular components is calculated as follows: a first guess is obtained
from photons statistics and sky subtraction and is calculated directly
by the \emph{VIPGI} pipeline. A fit of the stellar and nebular components
is performed with \emph{platefit\_vimos} using this first error spectrum.
The residuals of this first fit are then smoothed and added to the
error spectrum quadratically, and a new fit is performed with \emph{platefit\_vimos}.

\begin{table}
\caption{Definition of the H$\delta_{\mathrm{W}}$ index}

\label{tab:hdw}

\begin{centering}
\begin{tabular}{cc}
\hline 
\hline
bandpass & $\lambda$ (\AA)\tabularnewline
\hline
central & $\left[4060.5;4145\right]$\tabularnewline
blue & $\left[4014;4054\right]$\tabularnewline
red & $\left[4151;4191\right]$\tabularnewline
\hline
\end{tabular}
\par\end{centering}
\end{table}

\begin{figure}
\begin{centering}
\includegraphics[width=1\linewidth,keepaspectratio]{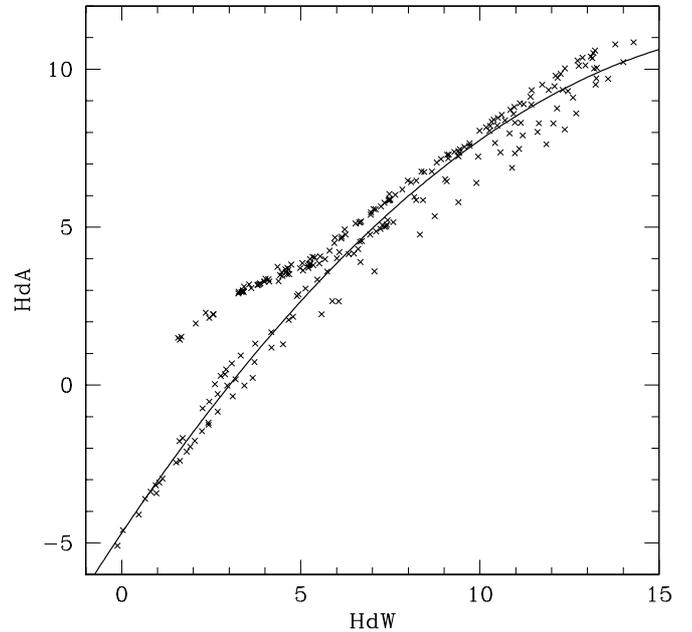} 
\par\end{centering}

\caption{Comparison between the standard Lick H$\delta_{\mathrm{A}}$ index
and the new broad index defined for low resolution VVDS spectra, namely
H$\delta_{\mathrm{W}}$ (see Tab.~\ref{tab:hdw}), for a set of model
templates taken from \citet{Bruzual:2003MNRAS.344.1000B} library
and covering a wide range of ages and metallicities. The solid curve
is the least square fit, which follows Eq.~\ref{eq:hdahdw}. Values
are given in \AA.}

\label{fighdahdw}
\end{figure}

Finally, detected emission lines may also be removed from the original
spectrum in order to get the observed \emph{stellar spectrum} and
measure indices from it, as well as emission-line equivalent widths.
The underlying continuum is indeed obtained by smoothing the stellar
spectrum. Then, equivalent widths are measured as direct integration
over a $5$ Gaussian-sigma bandpass of the emission-line Gaussian
model divided by the underlying continuum.

For the absorption lines, one has to be aware that the resolution
of VVDS spectra is too low in order to measure accurately Lick indices:
the bandpasses of these indices are narrower than the width of the
absorption lines at this resolution. In order to measure the absorption
component in the H$\delta$ line which is used to derive stellar masses
(see below), we have defined a new broadened index for this specific
line called H$\delta_{\mathrm{W}}$. Table~\ref{tab:hdw} gives the
three bandpasses of this index. Figure~\ref{fighdahdw} shows the
comparison between this index and the standard Lick H$\delta_{\mathrm{A}}$
index measured on various model templates taken from \citet{Bruzual:2003MNRAS.344.1000B}
library and covering a wide range of ages and metallicities. For future
comparison purpose between VVDS measurements and other studies, we
derive the following relation:\begin{equation}
\mathrm{H\delta_{\mathrm{A}}}=-4.69+1.691\times\mathrm{H\delta_{\mathrm{W}}}-0.044711\times\mathrm{H}\delta_{\mathrm{W}}^{2}\label{eq:hdahdw}\end{equation}

Note that the majority of points lying outside the fitted relation
in Fig.~\ref{fighdahdw} (for $\mathrm{H\delta_{\mathrm{W}}}<5\mbox{\AA}$)
are models with a very young stellar population ($<10$ Myr), hence
this relation is only valid for stellar populations older than $10$
Myr.

\subsection{Selection of star-forming galaxies\label{sub:starforming}}

\begin{table}
\caption{Sets of emission lines which are required to have $S/N>4$ in various
redshift ranges associated to the various diagnostics used in this
study. }

\begin{centering}
\begin{tabular}{lrrrrrr}
\hline 
\hline
Redshift & {[}O\noun{ii}] & H$\beta$ & {[}O\noun{iii}] & H$\alpha$ & {[}N\noun{ii}] & {[}S\noun{ii}]\tabularnewline
\hline
$0.0<z<0.2$ &  &  &  & $\checkmark$ & $\checkmark$ & $\checkmark$\tabularnewline
$0.2<z<0.4$ &  & $\checkmark$ & $\checkmark$ & $\checkmark$ & $\checkmark^{\star}$ & $\checkmark^{\star}$\tabularnewline
$0.4<z<0.5$ &  & $\checkmark$ & $\checkmark$ &  &  & \tabularnewline
$0.5<z<0.9$ & $\checkmark$ & $\checkmark$ & $\checkmark$ &  &  & \tabularnewline
\end{tabular}
\par\end{centering}

\medskip{}

{\footnotesize $^{\star}$ For the red diagnostic the {[}N}\noun{\footnotesize ii}{\footnotesize ]$\lambda$6584
and {[}S}\noun{\footnotesize ii}{\footnotesize ]$\lambda\lambda$6717+6731
emission lines may not be used at the same time.}{\footnotesize \par}

\label{tab:redranges}
\end{table}

\begin{table*}
\caption{Statistics of star-forming galaxies and narrow-line AGNs among emission-line
galaxies for various diagnostics depending on which emission lines
are observed (see text for details). The results are presented for
the two wide and deep samples used in this paper, and for their union
(some objects are in common). We also mention for each sample the
total number of objects which includes emission-line, faint and early-type
galaxies.}

\begin{centering}
\begin{tabular}{lrrrrrrrrrr}
\hline 
\hline
Sample/Diagnostic & red & \emph{\%} & blue & \emph{\%} & H$\alpha$ & \emph{\%} & H$\beta$ & \emph{\%} & all & \emph{\%}\tabularnewline
\hline
\textbf{wide} (13978) &  &  &  &  &  &  &  &  &  & \tabularnewline
\ emission-line (total) & 455 & \emph{100} & 924 & \emph{100} & 47 & \emph{100} & 1402 & \emph{100} & 2828 & \emph{100}\tabularnewline
\ star-forming & 415 & \emph{91} & 768 & \emph{83} & 30 & \emph{64} & 978 & \emph{70} & 2191 & \emph{77}\tabularnewline
\ candidate s.-f. & 0 & \emph{0} & 103 & \emph{11} & 0 & \emph{0} & 306 & \emph{22} & 409 & \emph{14}\tabularnewline
\ candidate AGN & 0 & \emph{0} & 35 & \emph{4} & 0 & \emph{0} & 74 & \emph{5} & 109 & \emph{4}\tabularnewline
\ Seyfert 2 & 18 & \emph{4} & 18 & \emph{2} & 17 & \emph{36} & 44 & \emph{3} & 97 & \emph{3}\tabularnewline
\ LINER & 22 & \emph{5} & 0 & \emph{0} & 0 & \emph{0} & 0 & \emph{0} & 22 & \emph{1}\tabularnewline
 &  &  &  &  &  &  &  &  &  & \tabularnewline
\textbf{deep} (7404) &  &  &  &  &  &  &  &  &  & \tabularnewline
\ emission-line (total) & 151 & \emph{100} & 568 & \emph{100} & 11 & \emph{100} & 574 & \emph{100} & 1304 & \emph{100}\tabularnewline
\ star-forming & 136 & \emph{90} & 412 & \emph{73} & 10 & \emph{91} & 312 & \emph{54} & 870 & \emph{67}\tabularnewline
\ candidate s.-f. & 0 & \emph{0} & 92 & \emph{16} & 0 & \emph{0} & 166 & \emph{29} & 258 & \emph{20}\tabularnewline
\ candidate AGN & 0 & \emph{0} & 47 & \emph{8} & 0 & \emph{0} & 47 & \emph{8} & 94 & \emph{7}\tabularnewline
\ Seyfert 2 & 5 & \emph{3} & 17 & \emph{3} & 1 & \emph{9} & 49 & \emph{9} & 72 & \emph{6}\tabularnewline
\ LINER & 10 & \emph{7} & 0 & \emph{0} & 0 & \emph{0} & 0 & \emph{0} & 10 & \emph{1}\tabularnewline
 &  &  &  &  &  &  &  &  &  & \tabularnewline
\textbf{union of wide and deep }(18648) &  &  &  &  &  &  &  &  &  & \tabularnewline
\ emission-line (total) & 469 & \emph{100} & 1213 & \emph{100} & 47 & \emph{100} & 1671 & \emph{100} & 3400 & \emph{100}\tabularnewline
\ star-forming & 427 & \emph{91} & 945 & \emph{78} & 30 & \emph{64} & 1074 & \emph{64} & 2476 & \emph{73}\tabularnewline
\ candidate s.-f. & 0 & \emph{0} & 166 & \emph{14} & 0 & \emph{0} & 404 & \emph{24} & 570 & \emph{17}\tabularnewline
\ candidate AGN & 0 & \emph{0} & 70 & \emph{6} & 0 & \emph{0} & 111 & \emph{7} & 181 & \emph{5}\tabularnewline
\ Seyfert 2 & 20 & \emph{4} & 32 & \emph{3} & 17 & \emph{36} & 82 & \emph{5} & 151 & \emph{4}\tabularnewline
\ LINER & 22 & \emph{5} & 0 & \emph{0} & 0 & \emph{0} & 0 & \emph{0} & 22 & \emph{1}\tabularnewline
\end{tabular}
\par\end{centering}

\label{tab:classif}
\end{table*}

\begin{figure*}
\begin{centering}
\subfigure[]{\includegraphics[width=0.45\linewidth,keepaspectratio]{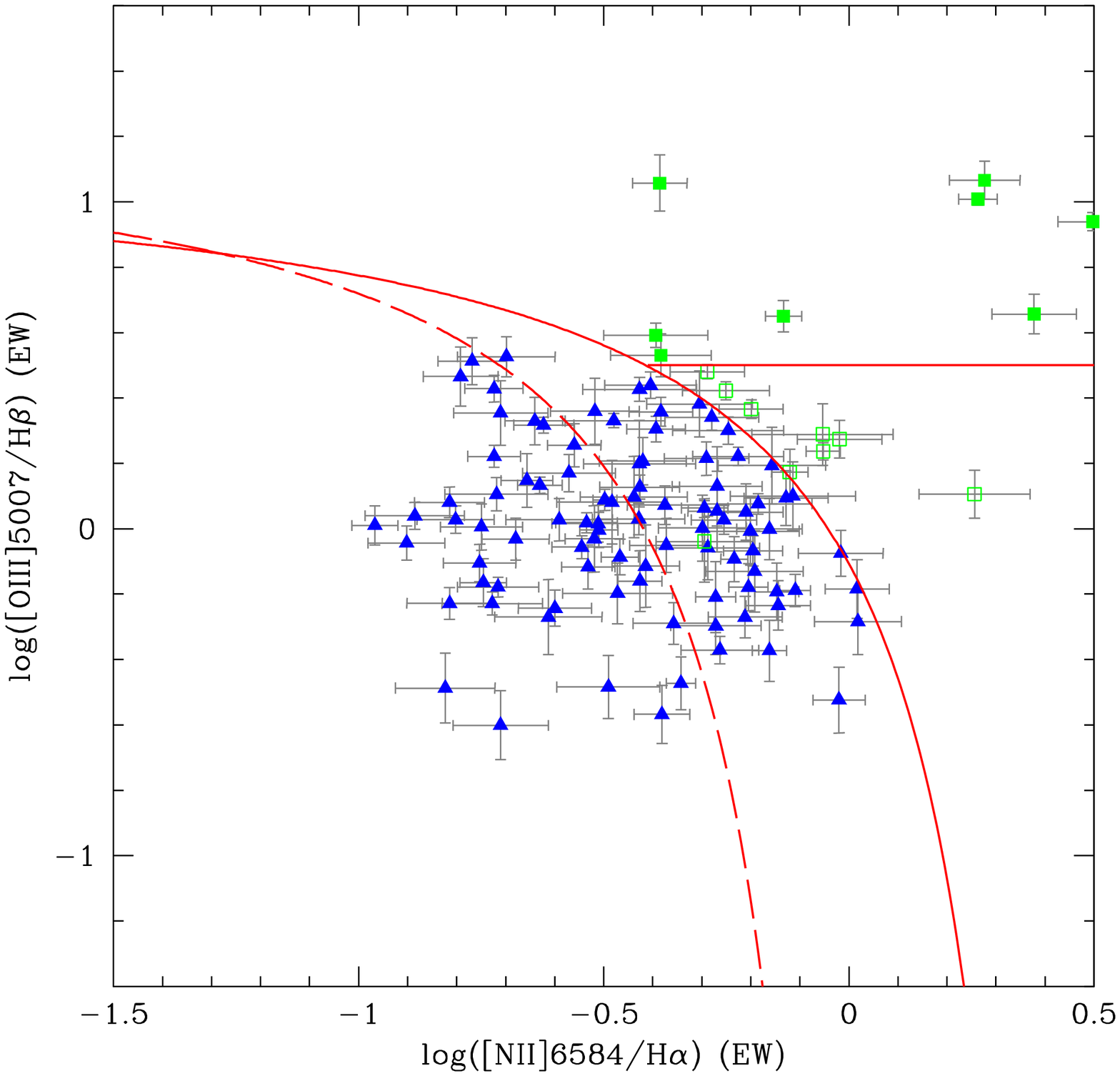}}
\subfigure[]{\includegraphics[width=0.45\linewidth,keepaspectratio]{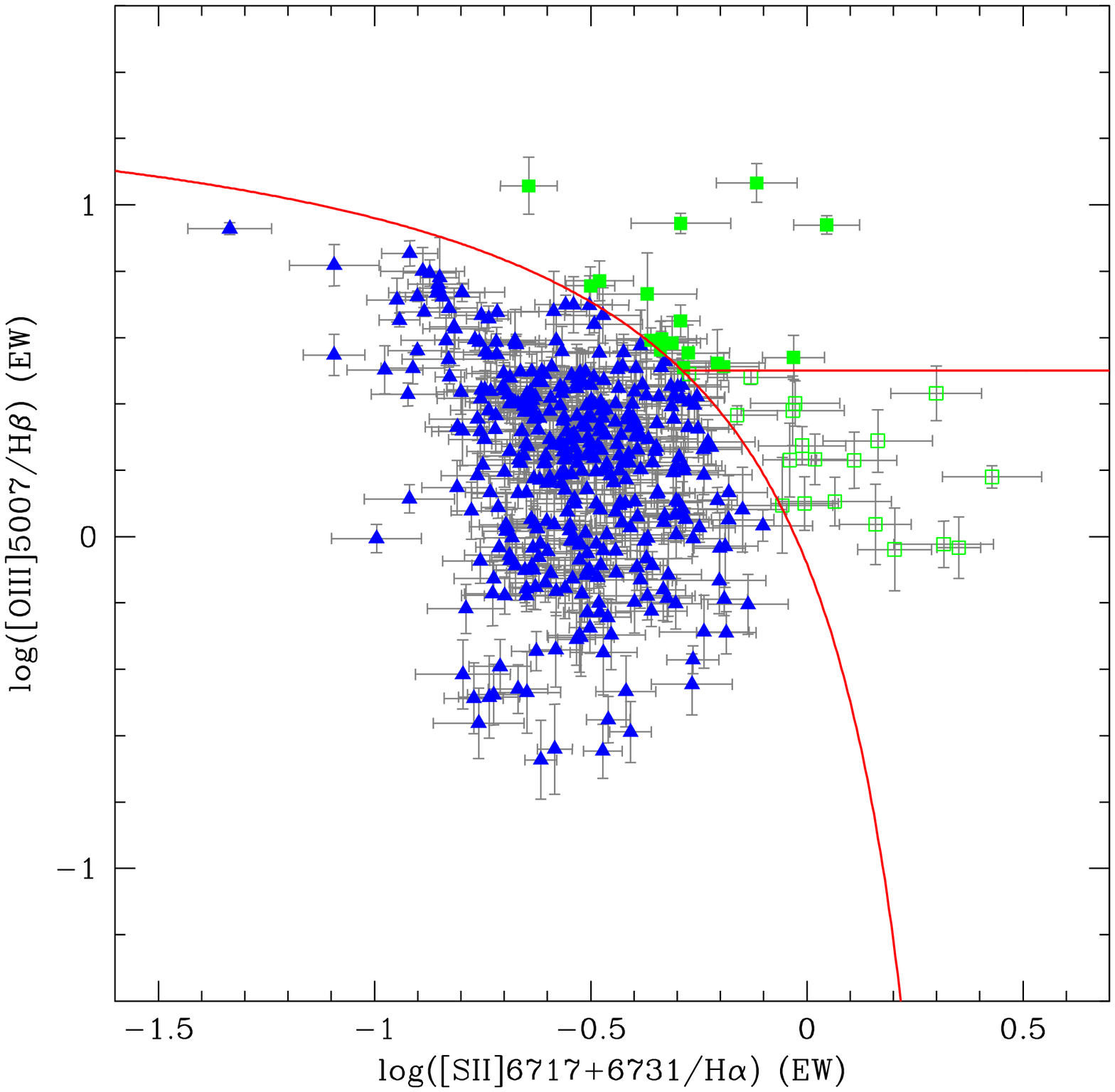}}
\par\end{centering}

\caption{Red spectral classification of $412$ narrow emission-line galaxies
in the redshift bin $0.2<z<0.4$. The emission-line ratios -- i.e.
{[}O\noun{iii}]$\lambda$5007/H$\beta$ and {[}N\noun{ii}]$\lambda$6584/H$\alpha$
(a) or {[}O\noun{iii}]$\lambda$5007/H$\beta$ and {[}S\noun{ii}]$\lambda\lambda$6717+6731/H$\alpha$
(b) -- are calculated using equivalent widths. The red solid curves
are the semi-empirical separations defined by \citet{Kewley:2001ApJS..132...37K}.
The dashed line is the separation proposed by \citet{Kauffmann:2003MNRAS.346.1055K}.
The star-forming galaxies are plotted as blue triangles, the Seyfert
2 galaxies as green solid squares, the LINERs as green open squares.
The error bars are shown in grey.}

\label{figdiagred}
\end{figure*}

\begin{figure}
\begin{centering}
\includegraphics[width=0.95\linewidth,keepaspectratio]{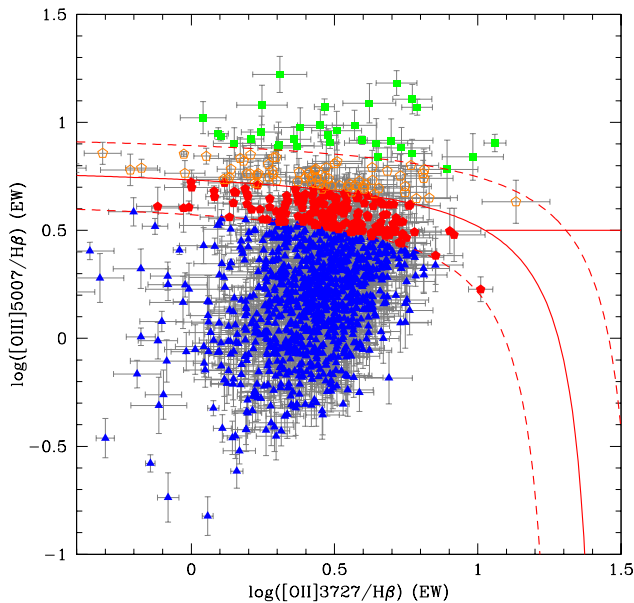}
\par\end{centering}

\caption{Blue spectral classification of $1060$ narrow emission-line galaxies
in the redshift bin $0.5<z<0.9$. The emission-line ratios -- i.e.
{[}O\noun{iii}]$\lambda$5007/H$\beta$ and {[}O\noun{ii}]$\lambda$3727/H$\beta$
-- are calculated using equivalent widths. The red solid curve is
the empirical separation defined by \citet{Lamareille:2004MNRAS.350..396L}.
The dashed curves delimits the error domain where both star-forming
and AGNs galaxies show similar blue line ratios. The star-forming
galaxies are plotted as blue triangles, the Seyfert 2 galaxies as
green solid squares, the candidate star-forming galaxies as red solid
pentagons, and the candidate AGNs as orange open pentagons. The error
bars are shown in grey.}

\label{figdiagblue}
\end{figure}

\begin{figure*}
\begin{centering}
\subfigure[]{\includegraphics[width=0.45\linewidth,keepaspectratio]{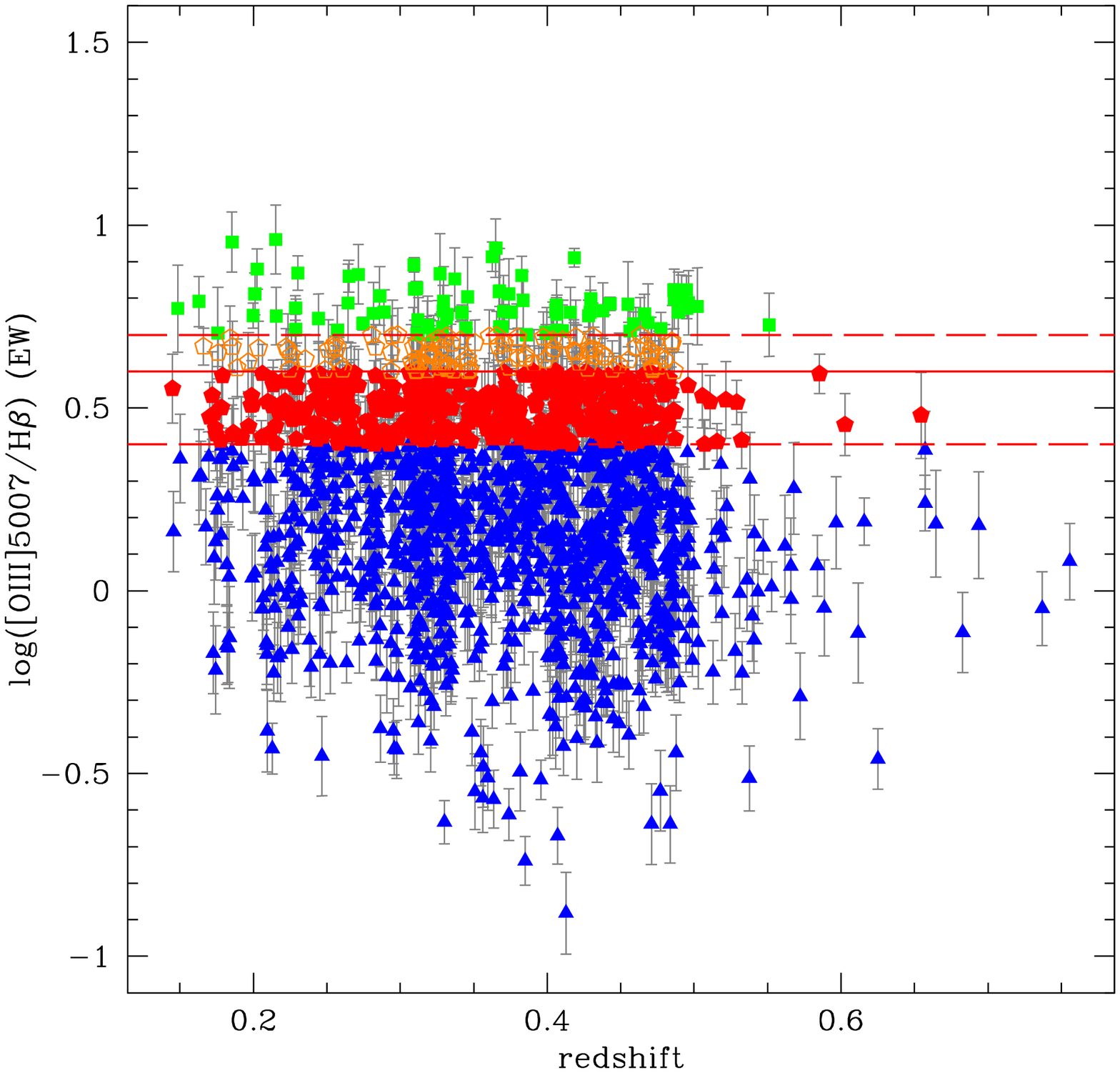}}
\subfigure[]{\includegraphics[width=0.45\linewidth,keepaspectratio]{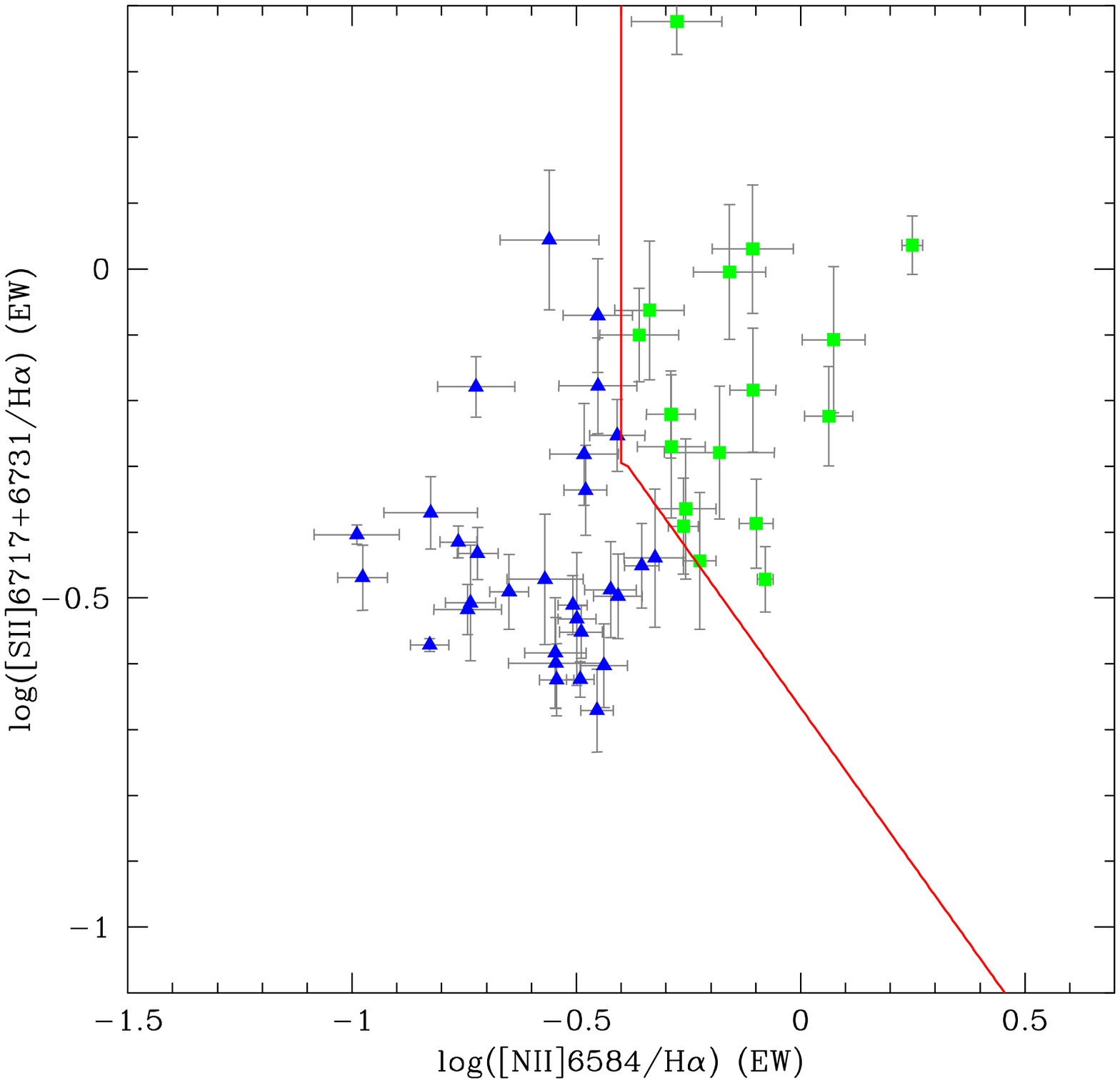}}
\par\end{centering}

\caption{Intermediate spectral classifications of $1479$ (a) and $43$ (b)
narrow emission-line galaxies. The two plots show the intermediate
diagnostic diagrams, used in the redshift bins $0.4<z<0.5$ (a) and
$0.0<z<0.2$ (b), or in other redshifts bins for galaxies with insufficient
number of detected emission lines (therefore not classified with the
red or blue diagnostic). The emission-line ratios -- i.e. {[}O\noun{iii}]$\lambda$5007/H$\beta$,
{[}S\noun{ii}]$\lambda\lambda$6717+6731/H$\alpha$, and {[}N\noun{ii}]$\lambda$6584/H$\alpha$
-- are calculated using equivalent widths. The red solid lines are
the empirical separations defined by \citet{Lamareille:2006PhDT}.
In the left panel, the dashed lines delimits the error domain where
both star-forming and AGNs galaxies show similar blue line ratios.
The star-forming galaxies are plotted as blue triangles, the Seyfert
2 galaxies as green solid squares, the candidate star-forming galaxies
as red solid pentagons, and the candidate AGNs as orange open pentagons.
The error bars are shown in grey.}

\label{figdiagmin}
\end{figure*}

\subsubsection{Description of the various diagnostics}

To ensure accurate abundance determinations, we restrict our attention
to galaxies with emission lines detected at $S/N>4$, where the set
of lines considered for this S/N cut varies with redshift. Table~\ref{tab:redranges}
provides a summary of the emission lines required to have $S/N>4$
in the various redshift ranges associated to the various diagnostics
defined below.

Our work intends to understand the star-formation process in galaxies.
Thus we have to remove from the sample of emission-line galaxies those
for which the source of ionized gas, responsible for these lines,
is not hot young stars. Emission-line galaxies can be classified in
various spectral types, which depend on the nature of their nebular
spectrum. The two main categories are the star-forming and AGN galaxies.
Their source of ionizing photons are respectively hot young stars,
or the accretion disk around a massive black-hole. As stated before,
broad-line AGNs (also called Seyfert 1 galaxies) are already taken
out of our sample. The narrow-line AGNs are divided in two sub-categories:
Seyfert 2 galaxies , and LINERs .

Both Seyfert 2 galaxies and LINERs can be distinguished from star-forming
galaxies using standard diagnostic diagrams \citep{Baldwin:1981PASP...93....5B,Veilleux:1987ApJS...63..295V},
which are based on emission-line ratios. The most commonly used classification
is provided by the {[}O\noun{iii}]$\lambda$5007/H$\beta$ vs. {[}N\noun{ii}]$\lambda$6584/H$\alpha$
diagram (Fig.~\ref{figdiagred}a), for which a semi-empirical limit
between star-forming, Seyfert 2 galaxies, and LINERs has been derived
by \citet{Kewley:2001ApJS..132...37K} from photo-ionization models.
Due to our low spectral resolution, the {[}N\noun{ii}]$\lambda$6584
line is not always detected and its detection is less reliable than
other lines, because of possible deblending problems with the bright
H$\alpha$ line. Consequently, most of the galaxies in the same redshift
range are actually classified with the {[}O\noun{iii}]$\lambda$5007/H$\beta$
vs. {[}S\noun{ii}]$\lambda\lambda$6717+6731/H$\alpha$ diagram (see
Fig.~\ref{figdiagred}b).

The two diagrams shown in Fig.~\ref{figdiagred}, which we also refer
to as the red diagnostics, can only be used with our data in the $0.2<z<0.4$
redshift range, as the desired emission lines are not visible outside
this domain. At higher redshifts, one can use the alternative blue
diagnostic which was empirically calibrated by \citet{Lamareille:2004MNRAS.350..396L}
from 2dFGRS data, in order to give similar results as the red diagnostics.
The blue diagnostic is based on the {[}O\noun{iii}]$\lambda$5007/H$\beta$
vs. {[}O\noun{ii}]$\lambda$3727/H$\beta$ diagram (line ratios calculated
on rest-frame equivalent widths), and may be applied in the $0.5<z<0.9$
redshift range (see Fig.~\ref{figdiagblue}). 

For the other redshift ranges, where neither the red nor the blue
diagnostic applies, we use a minimal classification based on H$\alpha$,
{[}N\noun{ii}]$\lambda$6584 and {[}S\noun{ii}]$\lambda\lambda$6717+6731
for $0.0<z<0.2$, and on H$\beta$ and {[}O\noun{iii}]$\lambda$5007
for $0.4<z<0.5$. We propose the following selection (called H$\beta$
diagnostic) for star-forming galaxies in the $0.4<z<0.5$ redshift
range: $\log(\mathrm{[O}\mathsc{iii}\mathrm{]}\lambda5007/\mathrm{H}\beta)<0.6$
(rest-frame equivalent widths, see Fig.~\ref{figdiagmin}a). We note
that real star-forming galaxies which are lost with the H$\beta$
diagnostic are mainly low metallicity ones. In any case a quick check
on SDSS data has told us that no more than $40$\% of star-forming
galaxies with $12+\log(\mathrm{O/H})<8.1$ actually fall in the AGN
region of the H$\beta$ diagnostic, this proportion being negligible
at higher metallicities.

Figure~\ref{figdiagmin}b shows the minimum classification for the
$0.0<z<0.2$ redshift range (called H$\alpha$ diagnostic). The proposed
separation has been derived based on the 2dFGRS data \citep{Lamareille:2006PhDT}
and follows the equation (rest-frame equivalent widths):\begin{equation}
\begin{array}{l}
\log(\mathrm{[N}\mathsc{ii}\mathrm{]}\lambda6584/\mathrm{H}\alpha)=\\
\left\{ \begin{array}{l}
-0.4\mathrm{\quad if\quad}\log(\mathrm{[S}\mathsc{ii}\mathrm{]}\lambda\lambda6717+6731/\mathrm{H}\alpha)\ge-0.3\\
-0.7-1.05\log(\mathrm{[S}\mathsc{ii}\mathrm{]}\lambda\lambda6717+6731/\mathrm{H}\alpha)\mathrm{\quad otw.}\end{array}\right.\end{array}\label{eq:diagmin}\end{equation}

This equation have been derived to efficiently reduce both the contamination
by real AGNs in the star-forming galaxies region ($<1\%$), and the
fraction of real star-forming galaxies which are lost ($<4\%$).

The results of all diagnostics, in the wide, deep, and global samples,
are shown in Table.~\ref{tab:classif}.

\subsubsection{Discussion of possible biases}

It is very important to evaluate the possible sources of biases, which
may affect the spectral classification, coming from the use of various
diagnostics at different redshift ranges.

As the {[}O\noun{ii}]$\lambda$3727/H$\beta$ line ratio is less accurate
than {[}N\noun{ii}]$\lambda$6584/H$\alpha$, or {[}S\noun{ii}]$\lambda\lambda$6717+6731/H$\alpha$,
to distinguish between star-forming and Seyfert 2 galaxies, the blue
diagnostic is defined with an error domain (see the dashed curves
in Fig.~\ref{figdiagblue}), inside which individual galaxies cannot
be safely classified. We thus introduce two new categories of galaxies:
candidate star-forming galaxies and candidate AGNs, which fall respectively
in the lower-half or the upper-half part of the error domain.

We know however that AGNs are in minority in the universe. Thus we
emphasize that \emph{i)} still no reliable classification can be performed
for any individual galaxy falling inside the error domain of the blue
diagnostic, \emph{ii)} the search for AGNs is highly contaminated
in the candidate AGN region, \emph{but} \emph{iii)} any study involving
statistically significant samples of star-forming galaxies should
not be biased when candidate star-forming galaxies and candidate AGNs
are included. As it will be shown in Sect.~\ref{sec:The-luminosity-metallicity-relation}
and~\ref{sec:The-mass-metallicity-relation}, this ends up to a negligible
bias on the derived metallicities.

As for the blue diagnostic, we have defined an error domain for the
H$\beta$ diagnostic in the following range: $0.4<\log(\mathrm{[O}\mathsc{iii}\mathrm{]}\lambda5007/\mathrm{H}\beta)<0.6$.
Star-forming galaxies or AGNs falling inside this domain are classified
as candidates.

One can see in Table.~\ref{tab:classif} that the red diagnostic
is the only one able to find LINERs. Indeed, LINERs fall in the star-forming
galaxies region with the blue or H$\beta$ diagnostic, while they
fall in the Seyfert 2 region with the H$\alpha$ diagnostic. We know
however that the contamination of star-forming galaxies by LINERs,
in the blue or H$\beta$ diagnostic, is less than 1\%.

Another bias could come from the population of composites galaxies,
i.e. for which the ionized gas is produced by both an AGN and some
star-forming regions. When looking for AGNs in the SDSS data, \citet{Kauffmann:2003MNRAS.346.1055K}
have defined a new, less restrictive, empirical separation between
star-forming galaxies and AGNs (see the dashed curve in Fig.~\ref{figdiagred}).
They have then classified as composites all galaxies between this
new limit and the old one by \citet{Kewley:2001ApJS..132...37K}.
This result is confirmed by theoretical modeling: \citet{Stasinska:2006MNRAS.371..972S}
have found that composite galaxies are indeed falling in the region
between the curves of \citet{Kauffmann:2003MNRAS.346.1055K} and \citet{Kewley:2001ApJS..132...37K}
in the {[}O\noun{iii}]$\lambda$5007/H$\beta$ vs. {[}N\noun{ii}]$\lambda$6584/H$\alpha$
diagram. Moreover, they have found that composite galaxies fall in
the star-forming galaxy region in the other diagrams ({[}O\noun{iii}]$\lambda$5007/H$\beta$
vs. {[}S\noun{ii}]$\lambda\lambda$6717+6731/H$\alpha$ or {[}O\noun{ii}]$\lambda$3727/H$\beta$).

It is difficult to evaluate the actual bias due to contamination by
composite galaxies, in the red or blue diagnostics that we use in
this study. This difficulty comes mainly from the fact that, if composite
galaxies actually fall in the composite region as defined by \citet{Kauffmann:2003MNRAS.346.1055K},
not all galaxies inside this region are necessarily composites. A
large majority of them might be normal star-forming galaxies. One
way to evaluate the contamination by composite galaxies is to look
for star-forming galaxies with an X-ray detection. Such work has been
performed with zCOSMOS data with similar selection criteria than the
wide sample: Bongiorno et al.\textbf{ }(in preparation) have found
that the contamination of star-forming galaxies by composites is approximately
10\%.

\section{Estimation of the stellar masses and absolute magnitudes\label{sec:masses}}

\subsection{The Bayesian approach}

The stellar masses are estimated by comparing the observed Spectral
Energy Distribution (hereafter SED), and two spectral features (H$\delta_{\mathrm{W}}$
absorption line and $D_{n}(4000)$ break), to a library of stellar
population models. The use of two spectral features reduces the well-known
age-dust-metallicity degeneracy in determining the mass-to-light ratio
of a galaxy. Compared to pure photometry, Balmer absorption lines
are indeed less sensitive to dust, while the $D_{n}(4000)$ break
is less sensitive to metallicity, both being sensitive to the age.

One observation is defined by a set $F_{i}$ of observed fluxes in
all photometric bands, a set $\sigma F_{i}$ of associated errors,
a set $I_{i'}$ of observed indices, and a set $\sigma I_{i'}$of
associated errors. The $\chi^{2}$ of each model, described by a set
$F_{i}^{0}$ of theoretical photometric points, and a set $I_{i'}^{0}$
of theoretical indices, is calculated as:\begin{equation}
\chi^{2}=\sum_{i}\frac{\left(F_{i}-A\cdot F_{i}^{0}\right)^{2}}{\sigma F_{i}^{2}}+\sum_{i'}\frac{\left(I_{i'}-I_{i'}^{0}\right)}{\sigma I_{i'}^{2}}\label{eq:chi2}\end{equation}
where $A$ is the normalization constant that minimizes the $\chi^{2}$.
Note that the normalization constant has not to be applied to the
spectral indices as they are already absolute.

Still for one observation, the set of $\chi_{j}^{2}$ calculated on
all models are summarized in a PDF (Probability Distribution Function)
which gives the probability of each stellar mass, given the underlying
library of models (the prior). Each stellar mass $M_{\star}$ is assigned
a probability described by the following normalized sum: \begin{equation}
P(M_{\star}|\{F_{i}\})=\frac{\sum_{j}\delta(M_{\star}-A\cdot M_{\star}^{j})\cdot\exp\left(-\chi_{j}^{2}/2\right)}{\sum_{j}\exp\left(-\chi_{j}^{2}/2\right)}\label{eq:pdf}\end{equation}
where $M_{\star}^{j}$ is the stellar mass of the model associated
to $\chi_{j}^{2}$. In practice, the PDF is discretized into bins
of stellar masses. Our stellar mass estimate is given by the median
of the PDF .

This method, based on the Bayesian approach, has been introduced by
the SDSS collaboration \citep{Kauffmann:2003MNRAS.341...33K,Tremonti:2004astro.ph..5537T,Brinchmann:2004MNRAS.351.1151B}
in order to carry SED fitting estimates of the physical properties
of galaxies, and starts now to be widely used on this subject. Its
main advantage is that it allows us, for each parameter, to get a
reliable estimate, \emph{independently} for each parameter, which
takes all possible solutions into account, not only the best-fit.
Thus, this method takes into account degeneracies between observed
properties in a self-consistent way. It also provides an error estimate
of the derived parameter from the half-width of the PDF .

\subsection{Description of the models\label{sub:Description-of-the}}

We use a library of theoretical spectra based on \citep[hereafter BC03]{Bruzual:2003MNRAS.344.1000B}
stellar population synthesis models, calculated for various star formation
histories. Unless many other studies which only use a standard declining
exponential star formation history, we have used an improved grid
including also secondary bursts \citep[stochastic library,][]{Salim:2005ApJ...619L..39S,Gallazzi:2005MNRAS.362...41G}.
The result of the secondary bursts, compared to previous methods,
is mainly to get higher masses as we are now able to better reproduce
the colors of galaxies containing both old and recent stellar populations.
When one uses a prior with a smooth star formation history, such objects
are better fitted by young models which correctly reproduce the recent
star formation history and the colors in the bluer bands. But such
models fail to reproduce the underlying old stellar population which
affects the colors of the redder bands, therefore they lead to an
underestimate of the final stellar masses by a factor $\approx1.4$
\citep[see][for a discussion of the different methods]{Pozzetti:2007A&A...474..443P}.

Masses derived with this method are also relative to the chosen IMF
\citep{Chabrier:2003PASP..115..763C} . The models include self-consistent
two-component dust corrections \citep{Charlot:2000ApJ...539..718C}.

\begin{figure}
\begin{centering}
\includegraphics[width=1\columnwidth]{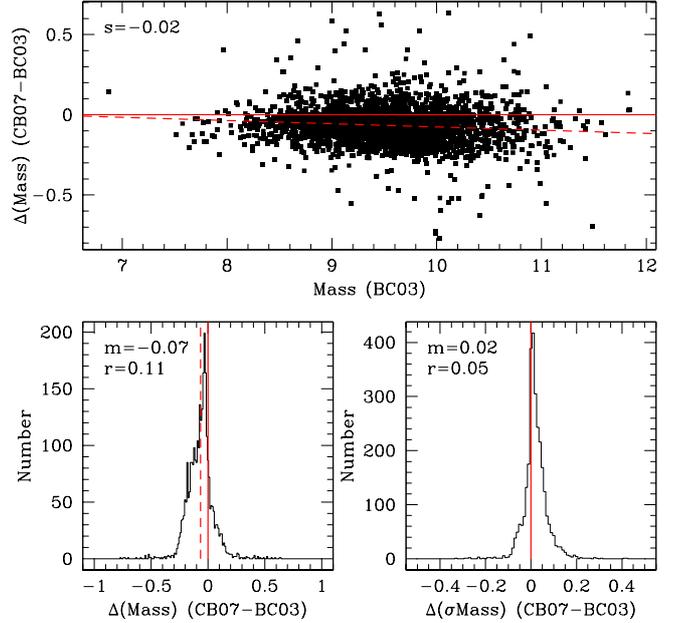}
\par\end{centering}

\caption{Comparison of the stellar masses of star-forming galaxies in our sample,
obtained with BC03 and CB07 models, using both photometry and spectroscopy.
The latter provides a better treatment of the TP-AGB stars in the
near-infrared. The top panel shows the difference between the two
masses, the bottom panels show the associated histogram (left) and
the histogram of the differences between the two error estimates (right).
The slope of the difference (s) is shown in the plot. The mean value
(m) and the rms (r) of the two histograms are given in the plots.}

\label{fig:compmass1}
\end{figure}

In addition to the secondary bursts, the models used in this study
also include a new treatment of the TP-AGB stars \citep{Marigo:2007A&A...469..239M}
(Charlot \& Bruzual, in preparation, hereafter CB07), in order to
provide a better fit of the near-infrared photometric bands \citep[see also][]{Maraston:2005MNRAS.362..799M,Maraston:2006ApJ...652...85M}.
We have derived the stellar masses of our whole sample of galaxies
with BC03 and CB07 models. Figure~\ref{fig:compmass1} shows the
results of this comparison. As expected, the masses derived with CB07
models are smaller, since the near-infrared flux of TP-AGB stars was
underestimated in BC03 models. The mean shift is $-0.07$ dex, with
a dispersion of $0.11$ dex, which corresponds to an overestimate
of the stellar masses by a factor $\approx1.2$ with BC03 models.
Their is also a small trend for more massive galaxies to be more affected
by the difference between the two models. This comparison has been
studied in more details by \citet{Eminian:2008MNRAS.384..930E}.

The same comparison have been performed on masses derived only using
photometry: the mean shift is $-0.09$ dex, with a dispersion of $0.19$
dex. The spectral indices are not expected to vary much with the new
treatment of TP-AGB stars introduced by CB07 models. It is thus expected
that masses computed only with photometry vary more from BC03 to CB07
models than those computed also with spectral indices.

We note that the stellar masses derived with BC03 models and a smooth
star formation history are very similar to the ones derived with CB07
models and secondary bursts. Stronger effects have to be expected
on derived star formation rates or mean stellar ages.

\begin{figure*}
\begin{centering}
\subfigure[]{\includegraphics[width=1\columnwidth]{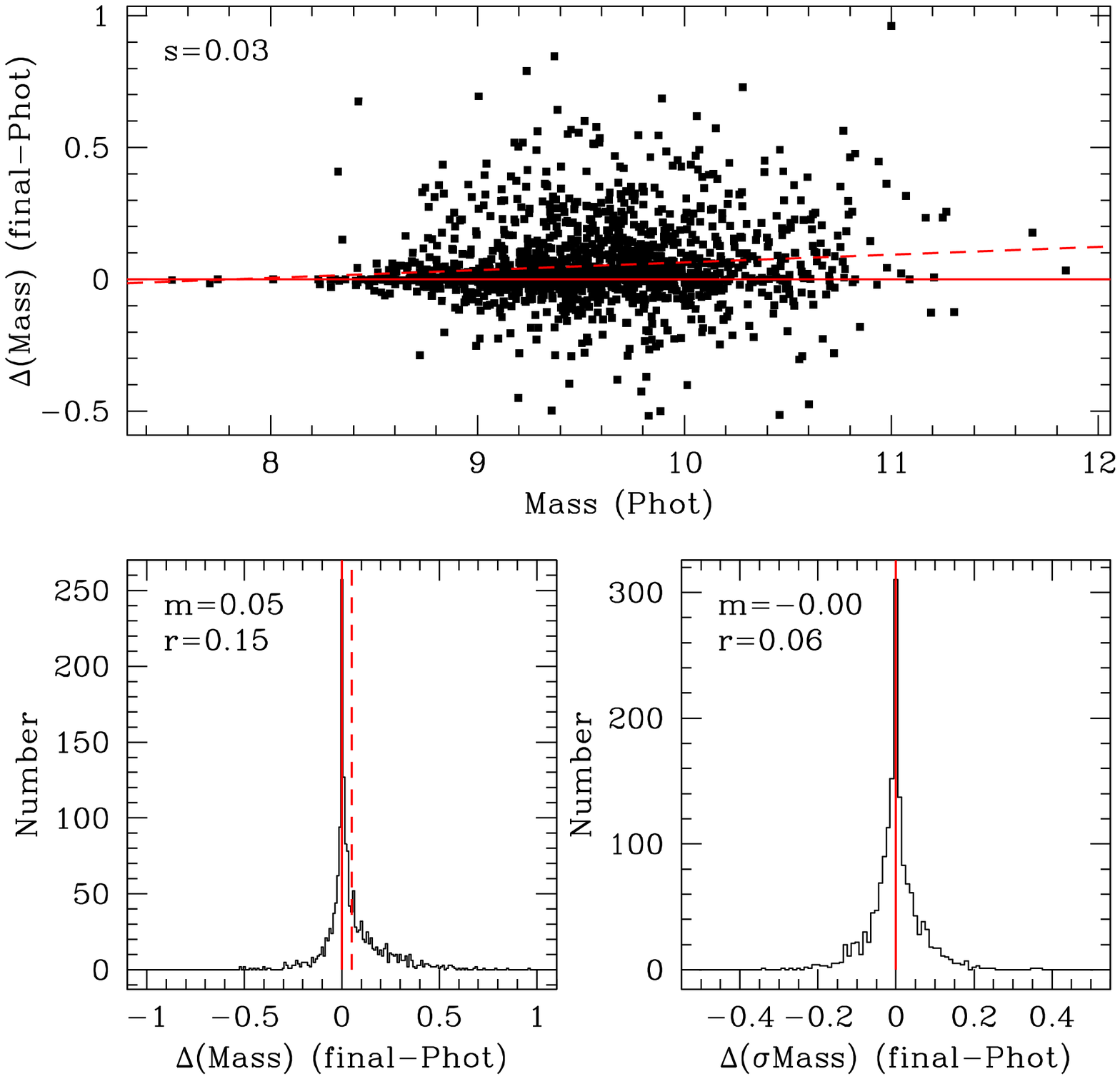}} \subfigure[]{\includegraphics[width=1\columnwidth]{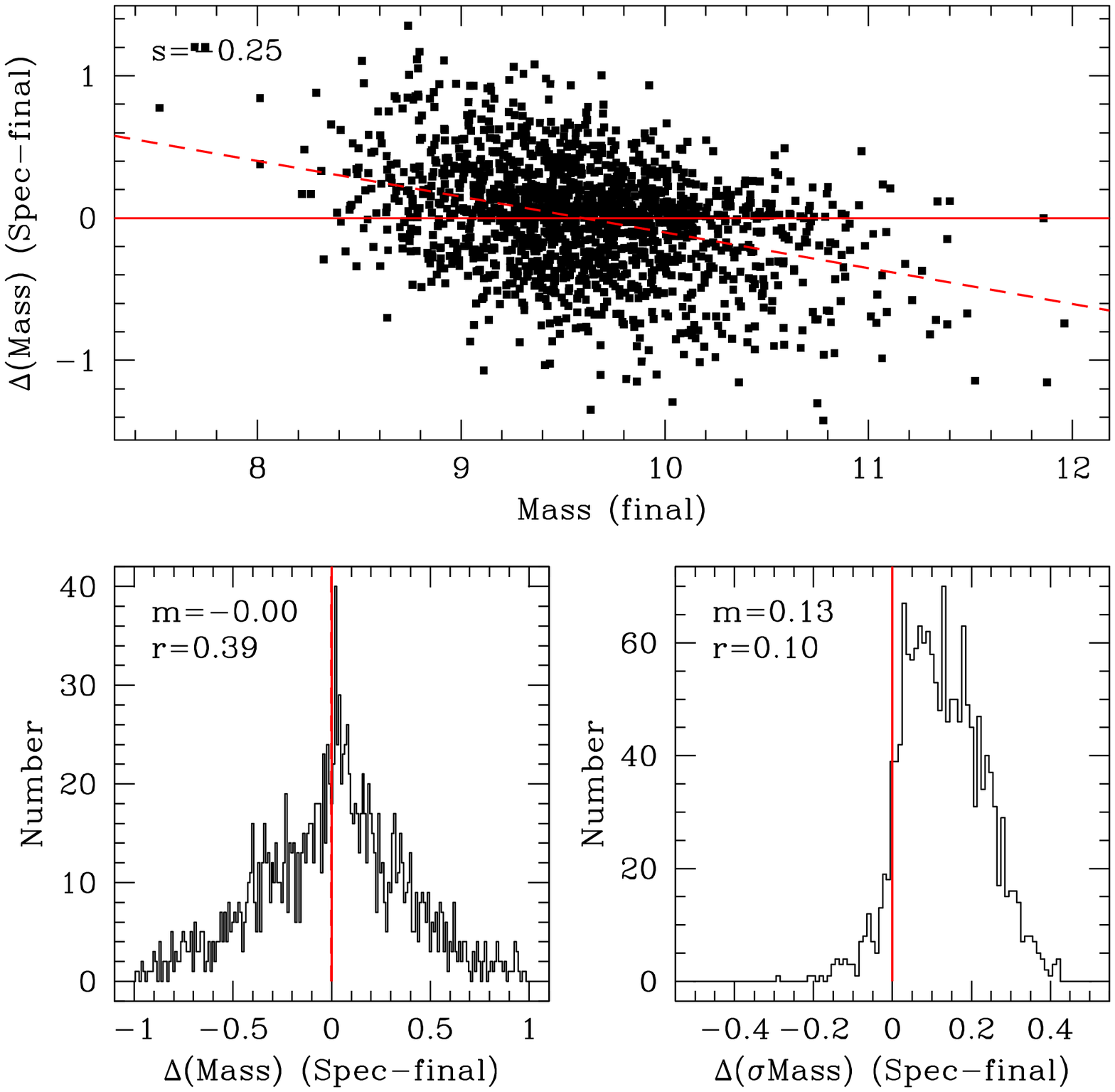}}
\par\end{centering}

\caption{Same legend as in Fig.~\ref{fig:compmass1}, except that we now compare
the stellar masses that we obtain with CB07 models with photometry
and spectroscopy (final) to the ones obtain: (a) using only photometry,
(b) using only spectroscopy. Only objects with a sufficient number
of photometric bands \emph{and} spectral indices (i.e. more than two
of each) are used in these plots.}

\label{fig:compmass2}
\end{figure*}

Figure~\ref{fig:compmass2} now compares the stellar masses obtained
with CB07 models when using only photometry, only spectroscopy, and
both photometry and spectroscopy. Figure~\ref{fig:compmass2}(a)
shows the comparison between the mass obtained using only photometry,
and the one obtained by adding the two spectral indices mentioned
above. The derived mass and associated error are not significantly
biased towards the two different methods (photometric masses are $-0.05$
dex lower). Nevertheless, we have compared the $\chi^{2}$ values
in both cases and found that it never increases when we add the spectral
indices in the analysis. Moreover, the $\chi^{2}$ obtained in this
latter case is lowered by a factor whose mean value is $2.3$. The
dispersion of $0.15$ dex in the comparison between the two masses
has thus to be taken into account as the minimum uncertainty of our
stellar masses.

Figure~\ref{fig:compmass2}(b) shows what would happen if only spectral
indices were used. We see that the derived mass does not suffer from
any bias, but is subject to a large dispersion of $0.39$ dex (a factor
$2.45$). A slope of $-0.25$ dex/decade is found, but is not significant
given the dispersion. The comparison of the derived errors shows that
this additional uncertainty is mostly taken into account: both measurements
have to account for a minimum uncertainty of $0.15$ dex and the measurement
obtained using only spectroscopy show a $0.13$ dex additional uncertainty.

The reason of this comparison is that \citet{Tremonti:2004astro.ph..5537T}
have not used full SEDs to derive stellar mass-to-light but only two
spectral indices: $D_{n}(4000)$ and H$\delta_{\mathrm{A}}$, the
mass being then scaled from $z$-band luminosity. We note that their
spectral indices have been measured at a much better signal-to-noise
and resolution on SDSS spectra as compared to VVDS spectra. We conclude
from Fig.~\ref{fig:compmass2}(b) that \citet{Tremonti:2004astro.ph..5537T}
masses computed only with spectroscopy are directly comparable to
our final masses computed also with photometry, provided that we take
scale them by $-0.07$ dex and $-0.056$ dex to account respectively
for the difference between BC03 and CB07 models, and between \citet{Kroupa:2001MNRAS.322..231K}
and \citet{Chabrier:2003PASP..115..763C} IMF\textbf{.}

\subsection{Defining volume-limited samples}

\begin{table*}
\caption{Limiting masses (in logarithm of solar masses) of our various samples
in order to define volume-limited samples, as a function of the redshift
ranges and for various mass-to-light completeness levels.}

\label{tab:limmass}

\begin{centering}
\begin{tabular}{cccccccccccccccc}
\hline 
\hline & \multicolumn{3}{c}{$z<0.5$} & \multicolumn{3}{c}{$z<0.6$} & \multicolumn{3}{c}{$z<0.7$} & \multicolumn{3}{c}{$z<0.8$} & \multicolumn{3}{c}{$z<0.9$}\tabularnewline
\hline
 & 50\% & 80\% & 95\% & 50\% & 80\% & 95\% & 50\% & 80\% & 95\% & 50\% & 80\% & 95\% & 50\% & 80\% & 95\%\tabularnewline
wide (all) & $9.6$ & $10.0$ & $10.2$ & $9.7$ & $10.2$ & $10.4$ & $9.9$ & $10.3$ & $10.6$ & $10.0$ & $10.5$ & $10.7$ & $10.2$ & $10.6$ & $10.9$\tabularnewline
wide (star-forming) & $9.3$ & $9.6$ & $9.9$ & $9.4$ & $9.7$ & $10.1$ & $9.6$ & $9.9$ & $10.2$ & $9.7$ & $10.0$ & $10.3$ & $9.8$ & $10.1$ & $10.4$\tabularnewline
deep (all) & $8.7$ & $9.2$ & $9.5$ & $8.9$ & $9.3$ & $9.7$ & $9.0$ & $9.5$ & $9.9$ & $9.1$ & $9.6$ & $10.0$ & $9.2$ & $9.7$ & $10.1$\tabularnewline
deep (star-forming) & $8.6$ & $8.9$ & $9.2$ & $8.7$ & $9.0$ & $9.3$ & $8.8$ & $9.0$ & $9.3$ & $8.8$ & $9.1$ & $9.4$ & $8.9$ & $9.2$ & $9.5$\tabularnewline
\end{tabular}
\par\end{centering}
\end{table*}

\begin{figure*}
\begin{centering}
\includegraphics[width=1\columnwidth,keepaspectratio]{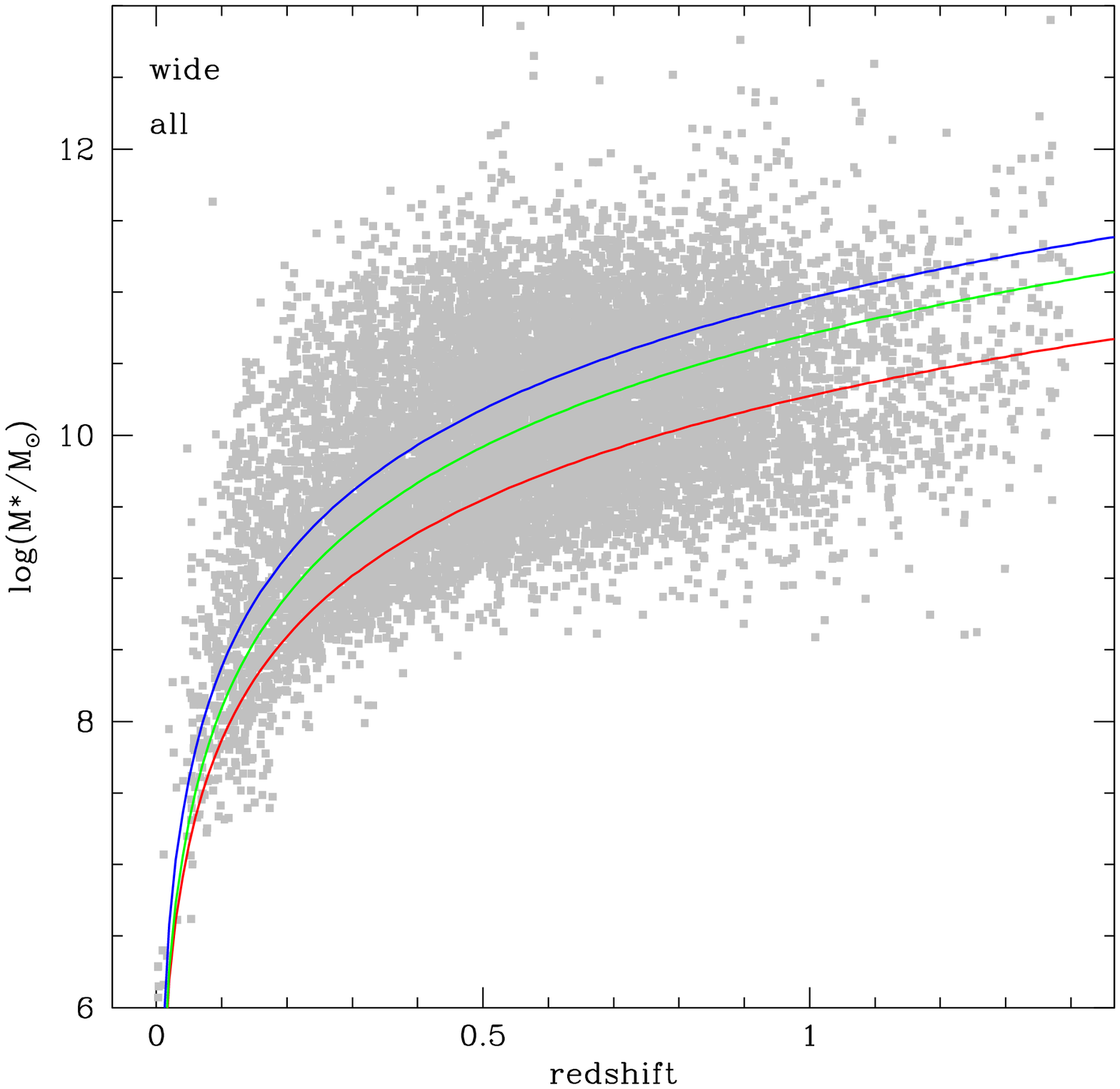}
\includegraphics[width=1\columnwidth,keepaspectratio]{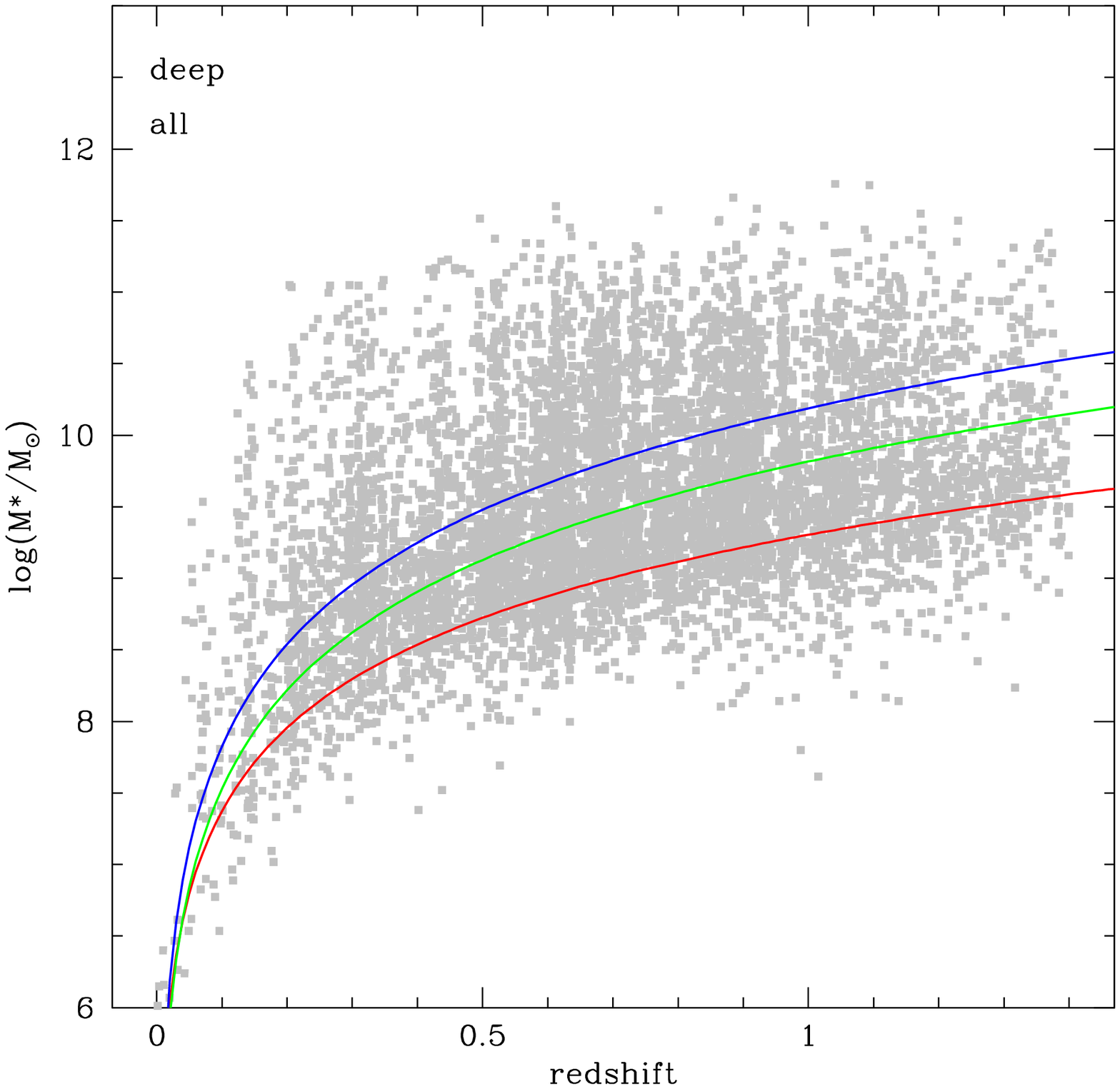}
\includegraphics[width=1\columnwidth,keepaspectratio]{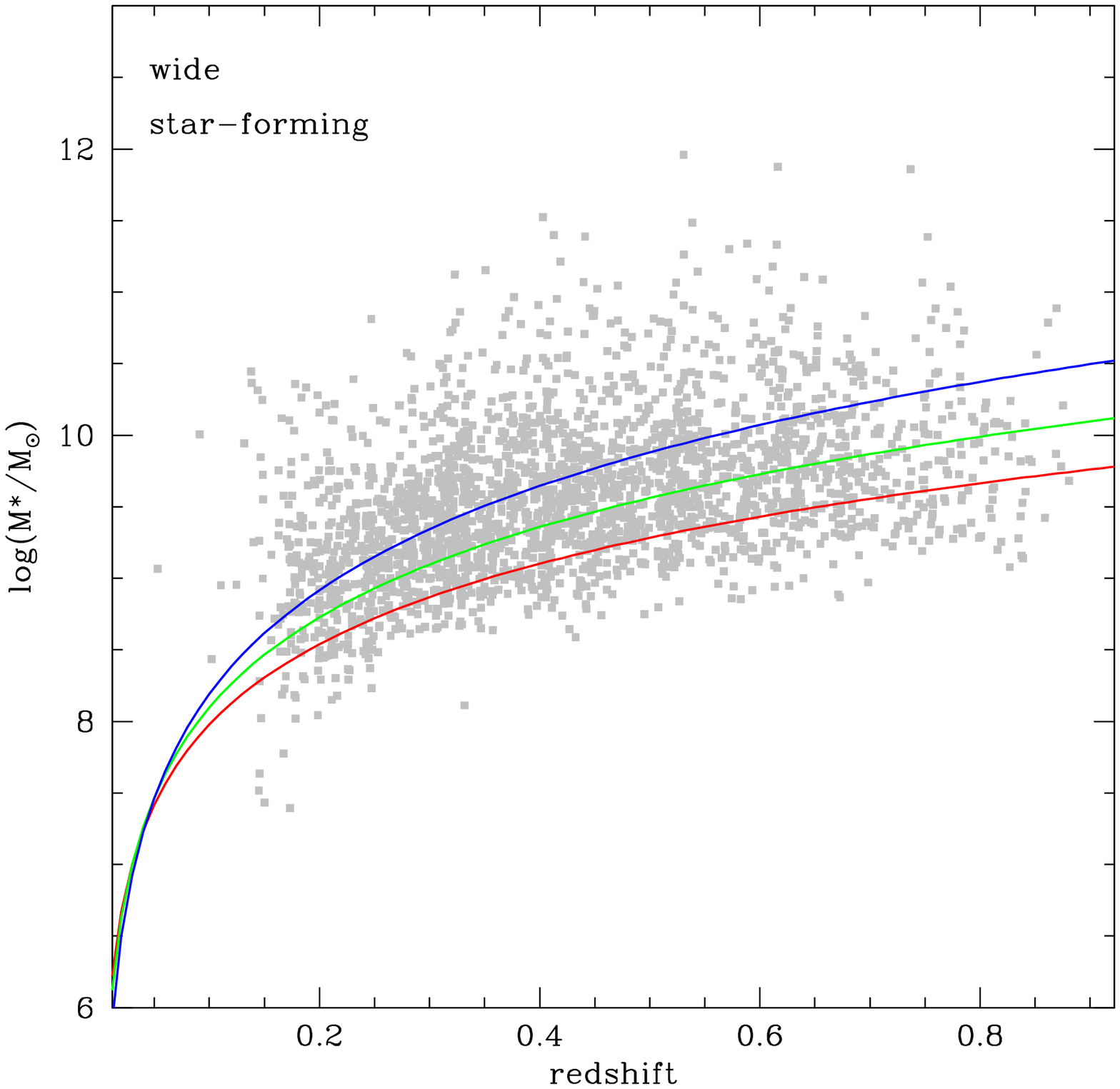}
\includegraphics[width=1\columnwidth,keepaspectratio]{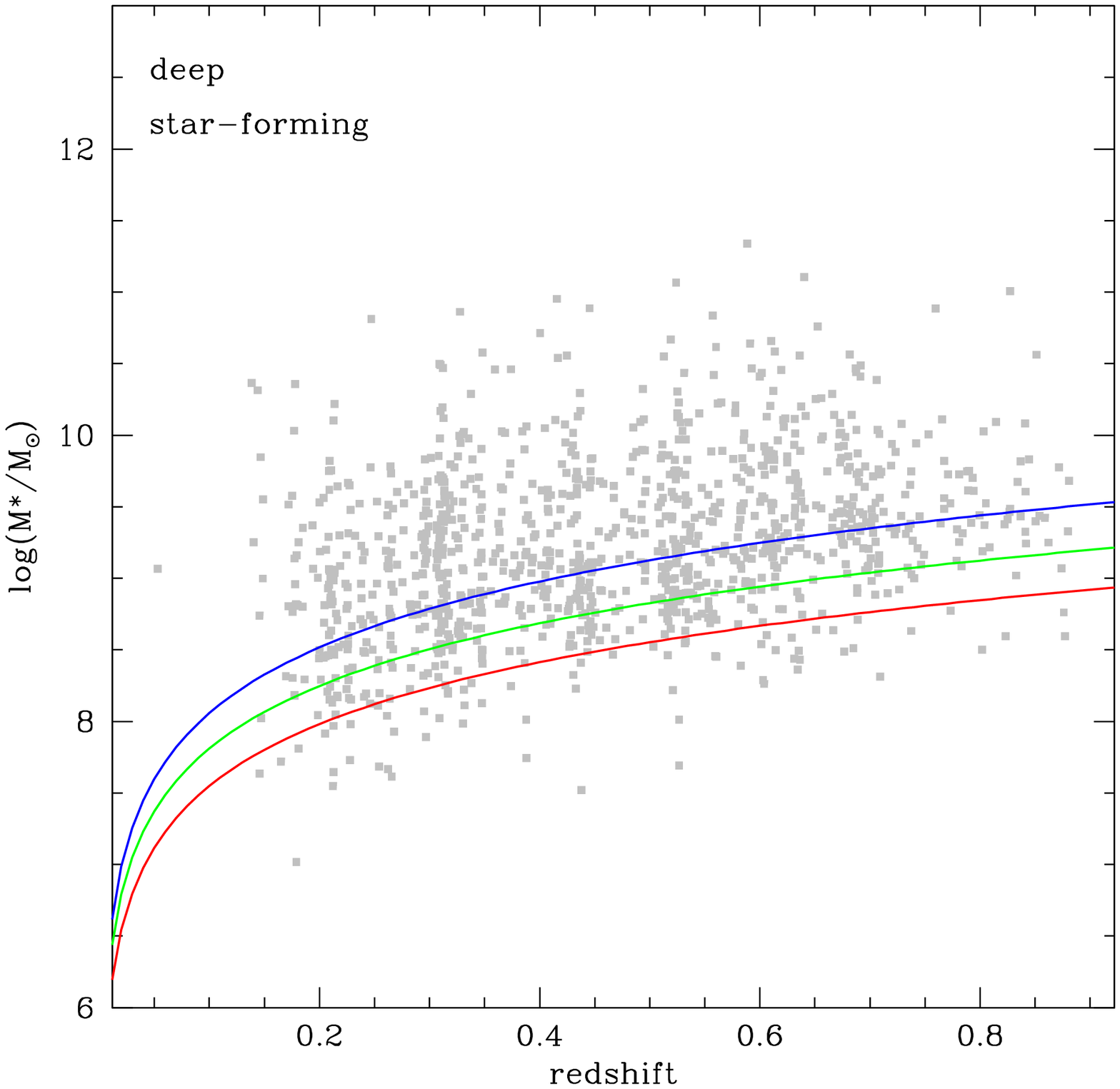}
\par\end{centering}

\caption{Stellar mass estimates (in logarithm of solar masses), as a function
of redshift, of the galaxies in our samples: left: the wide sample,
right: the deep sample, top: all galaxies (except broad-line AGNs),
bottom: only star-forming galaxies (see Sect.~\ref{sub:starforming}).
The red, green and blue curves illustrate respectively the 50\%, 80\%
and 95\% mass-to-light completeness levels (see text for details).}

\label{figselectmass}
\end{figure*}

Figure~\ref{figselectmass} shows the stellar masses obtained, as
a function of redshift, for the galaxies of our various samples (wide
or deep, all galaxies or star-forming galaxies only). As expected
in a magnitude-selected sample, the minimum stellar mass which is
detected depends on the redshift, because of the decrease of apparent
magnitude when the distance increases. This effect is well-known as
the Malmquist bias. The upper-limit of the mass distribution is in
part due to the bright cut-off in the magnitude selection, while in
addition to that a larger volume is observed at higher redshift which
allows a better sampling of the upper mass region. The VVDS is a constant
solid-angle survey: the observed volume increases with distance and
rare objects, like very massive galaxies, get more chances to be detected.

There are two ways to account for these two biases. The first one
is to use the Vmax technique: each object is weighted by the inverse
of the maximum volume in which it could be observed, given the selection
function of the survey. The result is a density which is statistically
corrected for the selection effects, but may suffer from evolution
effects. Another, simpler, technique is to build volume-limited samples:
a redshift range is defined and only objects which may be observed
in the entire volume between these two redshifts (given the selection
function) are kept. We use this last technique in our study.

In practice, for magnitude-selected surveys, one has to calculate
the minimum stellar mass which is detected at the upper limit of the
redshift range, and then to throw out the objects below this limit
since they are not observed in the entire volume. Nevertheless the
minimum mass which may be detected at a given redshift, for a given
limiting apparent magnitude, depends on the stellar mass-to-light
ratio. One way to solve this issue is to calculate, for each galaxy,
the mass $M_{\mathrm{lim}}^{\star}$ it would have if its apparent
magnitude $I$ was equal to the selection magnitude $I_{\mathrm{sel}}$
(i.e. $24$ for the deep sample, and $22.5$ for the wide sample):\begin{equation}
M_{\mathrm{lim}}^{\star}=M^{\star}+0.4\cdot\left(I-I_{\mathrm{sel}}\right)\label{eq:limmass}\end{equation}

The result is a distribution of minimum stellar masses which reflects
the distribution of stellar mass-to-light ratios in the sample. The
red, green, and blue curves in Fig.~\ref{figselectmass} show respectively
the 50\%, 80\% and 95\% levels for the cumulative sum of this distribution.
The limiting masses derived for the 4 different samples mentioned
above are given in Table~\ref{tab:limmass}. As expected, we see
that the limiting mass increases with redshift and is lower for the
deep sample. The derived completeness levels on the deep sample are
similar but higher to the ones found by \citet{Meneux:2008A&A...478..299M}
on the same data. The difference is due to the different cosmology
used in this paper, i.e. $h=1$.

We see also that the limiting mass is lower when we consider only
star-forming galaxies, compared to the whole samples. This is due
to star-forming galaxies having lower mass-to-light ratios. This shows
that if a sub-sample has a smaller range in mass-to-light ratios,
then a less conservative mass limit can be adopted than if a single
fixed maximum mass-to-light ratio is taken.

Finally, we note that the wide sample contains more massive objects
since it covers a larger volume of universe.

\subsection{Absolute magnitudes\label{sub:absmag}}

\begin{figure*}
\begin{centering}
\includegraphics[width=1\columnwidth]{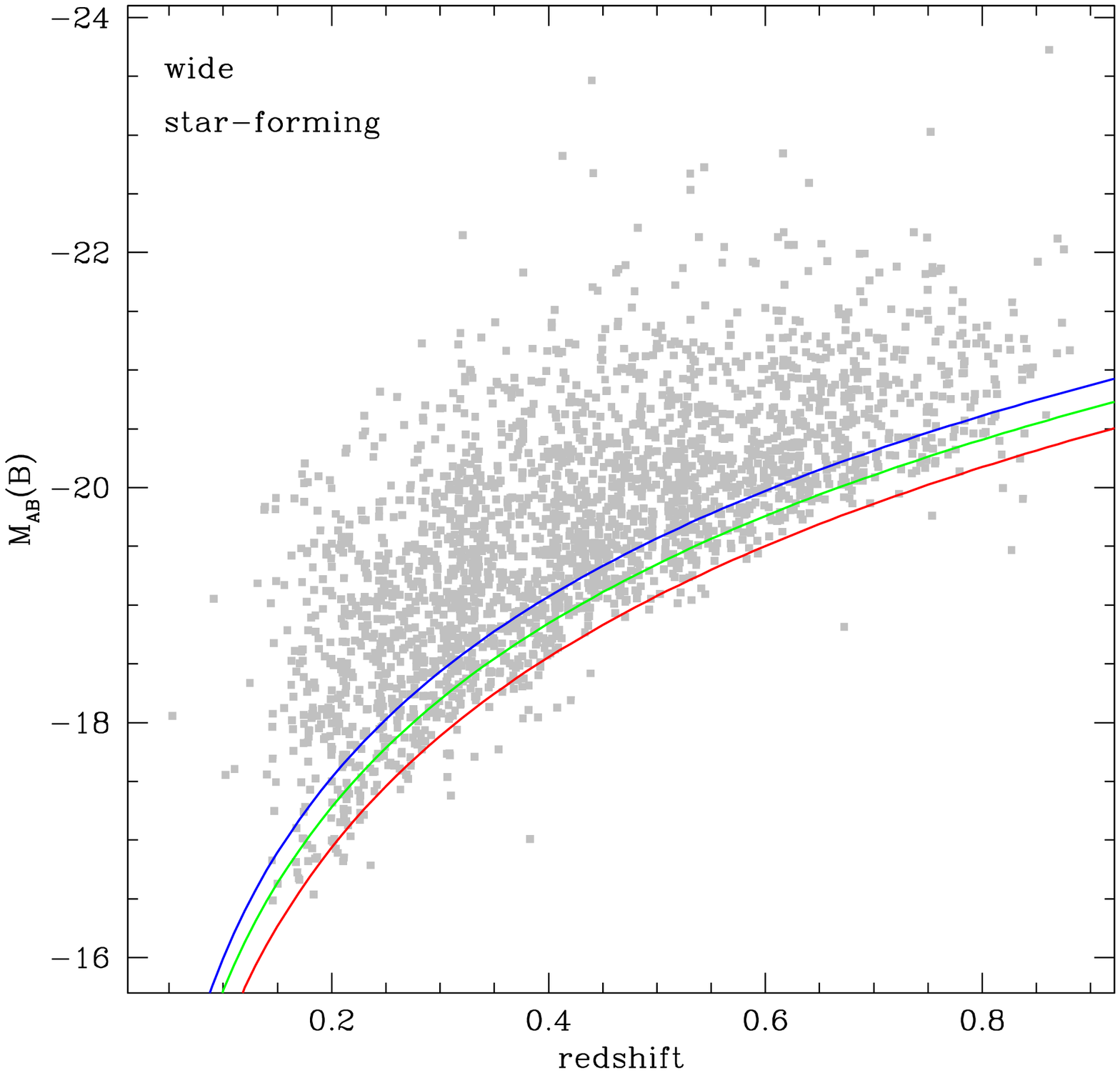} \includegraphics[width=1\columnwidth]{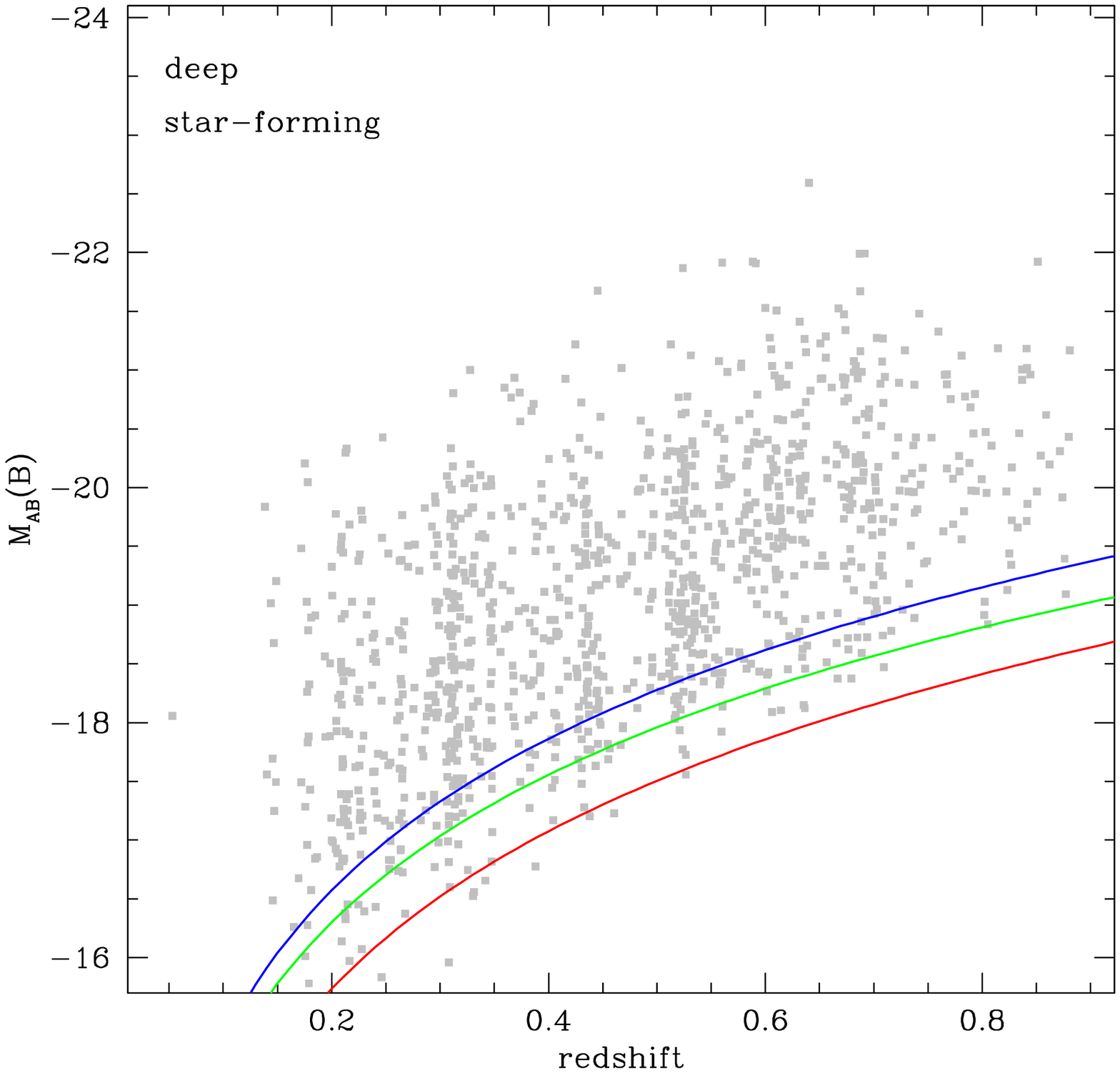}
\par\end{centering}

\caption{Absolute magnitudes estimates in the rest-frame $B$-band (AB system),
as a function of redshift, of the star-forming galaxies of the wide
(left) and deep (right) samples. The red, green and blue lines illustrate
respectively the 50\%, 80\% and 95\% $k$-correction completeness
levels (see text for details).}

\label{figselectmb}
\end{figure*}

\begin{table*}
\caption{Limiting $B$-band absolute magnitudes (AB system) of our wide and
deep samples of star-forming galaxies, in order to define volume-limited
samples, as a function of the redshift ranges and for various $k$-correction
completeness levels.}

\label{tab:limmb}

\begin{centering}
\begin{tabular}{ccccccc}
\hline 
\hline & \multicolumn{2}{c}{$z<0.5$} & \multicolumn{2}{c}{$z<0.7$} & \multicolumn{2}{c}{$z<0.9$}\tabularnewline
\hline
 & 50\% & 95\% & 50\% & 95\% & 50\% & 95\%\tabularnewline
wide & $-19.1$ & $-19.6$ & $-19.9$ & $-20.3$ & $-20.5$ & $-20.9$\tabularnewline
deep & $-17.5$ & $-18.3$ & $-18.2$ & $-18.9$ & $-18.6$ & $-19.3$\tabularnewline
\end{tabular}
\par\end{centering}
\end{table*}

As a by-product of the estimation of stellar masses, we have also
calculated the $k$-corrected absolute magnitudes, in the rest-frame
$B$-band, using the Bayesian approach and the CB07 models. The results
for star-forming galaxies are shown in Fig.~\ref{figselectmb} as
a function of redshift. Because of the $k$-correction, the maximum
absolute magnitude which may be detected, as a function of redshift,
is also affected by the stellar mass-to-light ratio. The absolute
magnitude in rest-frame $B$-band is affected by the ratio between
the flux in this band, and the flux in the observed $I$-band, which
ratio depends on the type of the galaxy and on redshift We also note
that as observed $I$-band is fairly close to rest-frame $B$-band,
the variation in $k$-correction is small compared to the variation
in mass-to-light ratio as can be seen by the reduced spread between
the different completeness level curves in Fig.~\ref{figselectmass}
as compared to Fig.~\ref{figselectmb}.

The limiting $B$-band absolute magnitudes, associated to various
redshift ranges and $k$-correction completeness levels, are given
in Table~\ref{tab:limmb}. We note that the absolute magnitudes used
in this study may differ from previous works performed on the same
dataset \citep[e.g. ][]{Ilbert:2005A&A...439..863I}. For self-consistency,
we have chosen to calculate the absolute magnitudes with the same
models than the ones used for the stellar masses, namely CB07 models
with secondary bursts.

\section{Estimation of the metallicities\label{sec:metals}}

In this study, the metallicities are estimated as the gas-phase oxygen
abundances $12+\log\left(\mathrm{O/H}\right)$.

\subsection{The empirical approach\label{sub:empimeta}}

Strong-line empirical calibrators may be used to compute the gas-phase
metallicities from the available measured emission lines, e.g. H$\alpha$,
H$\beta$, {[}O\noun{iii}]$\lambda$5007, {[}N\noun{ii}]$\lambda$6584,
{[}O\noun{ii}]$\lambda$3727. As shown by \citet{Lamareille:2006A&A...448..893L},
the {[}N\noun{ii}]$\lambda$6584 and H$\alpha$ emission lines are
reliably deblended for sufficiently high signal-to-noise ratio even
at the resolution of the VVDS.

We use the $N2$ calibrator up to $z\sim0.2$ \citep{Denicol:2002MNRAS.330...69D,VanZee:1998AJ....116.2805V},
the $O3N2$ calibrator for the redshift range $0.2<z<0.5$ \citep{Pettini:2004MNRAS.348L..59P},
and the $R_{\mathrm{23}}$ calibrator for the redshift range $0.5<z<0.9$
\citep{McGaugh:1991ApJ...380..140M}. 

At high redshifts, where nitrogen or sulfur lines are not observed,
the metallicity is degenerate because of radiative cooling at high
metallicities which lowers the intensities of oxygen lines. Thus,
low oxygen-to-hydrogen line ratio can be interpreted as either low
oxygen abundance or high oxygen cooling, i.e. high oxygen abundance.
The degeneracy of the $R_{\mathrm{23}}$ calibrator is broken using
the $B$-band absolute magnitude and the reference luminosity-metallicity
relation \citep[see][for a detailed discussion of this method]{Lamareille:2006A&A...448..907L}.

As clearly shown by \citet{Kewley:2008arXiv0801.1849K}, the difficulty
of using different metallicity calibrators at different redshifts
comes from the huge differences, up to $\pm0.7$ dex, which exist
between them. Therefore, we have decided to renormalize the different
calibrators to the \citet[hereafter CL01]{Charlot:2001MNRAS.323..887C}
calibrator, which has been used to estimate the mass-metallicity relation
from SDSS data \citep{Tremonti:2004astro.ph..5537T}. The correction
formulas are found in Table~B3 of \citet{Kewley:2008arXiv0801.1849K}.

Finally, a consequence of the small spectral coverage is also that
H$\alpha$ and H$\beta$ lines are not observed together for most
redshifts, making difficult the estimation of the dust attenuation
from the standard Balmer decrement method. As already shown by \citet{Kobulnicky:2003ApJ...599.1031K},
and \citet{Lamareille:2006A&A...448..907L}, metallicities can be
reliably derived using equivalent width measurements instead of dust-corrected
fluxes.

Nevertheless, \citet{Liang:2007A&A...474..807L} have shown that the
{[}O\noun{iii}]$\lambda$5007/{[}O\noun{ii}]$\lambda$3727 emission
line ratio, which is used in $R_{\mathrm{23}}$ calibrator, does not
give the same value if it is calculated using equivalent widths or
dust-corrected fluxes: the ratio between the two results depends on
the differential dust attenuation between stars and gas, and on the
slope of the stellar continuum. They have found that the derived metallicities
are affected by a factor ranging from $-0.2$ to $0.1$ dex, but they
also showed that this factor has a mean value of only $-0.041$ dex.
It is thus not significant for the estimation of the mass-metallicity
relation on statistical samples. Nevertheless, in order to avoid any
possible bias on a different sample, we decided to use their corrective
formula based on the $D_{n}(4000)$ index.

\subsection{The Bayesian approach}

The metallicities may also be estimated using again the Bayesian approach.
The relative fluxes of all measured emission lines are compared to
a set of photoionization models, which predicts the theoretical flux
ratios given four parameters: the gas-phase metallicity, the ionization
level, the dust-to-metal ratio and the reddening \citep{Charlot:2001MNRAS.323..887C}.
The CL01 models are based on population synthesis for the ionizing
flux \citep{Bruzual:2003MNRAS.344.1000B}, emission line modeling
\citep[Cloudy, ][]{Ferland:2001PASP..113...41F} and a two-component
dust attenuation law \citep{Charlot:2000ApJ...539..718C}.

We calculate the $\chi^{2}$ of each model and summarize them in the
PDF of the metallicity using a similar method than described above
in Eq.~\ref{eq:chi2} and~\ref{eq:pdf}, but now applied to emission-line
fluxes instead of photometric points. Only the emission lines with
enough signal-to-noise (S/N>4) are used in the fit. This method is
applied directly on observed line fluxes: the correction for dust
attenuation is included self-consistently in the models.

The O/H degeneracy at high redshift ends up with double-peaked PDFs:
one peak for the low metallicity solution and another one at high
metallicity. However, other information such as dust extinction, ionization
level, or star formation rate help affecting to the two peaks two
different probabilities. Thus, we solve the degeneracy by fitting
two peaks in the PDFs, and by keeping the one with the highest probability.
As already discussed in \citet{Lamareille:2006A&A...448..907L}, this
method cannot be used to chose the metallicity of one single galaxy,
but it can be used statistically to derive a mean metallicity, e.g.
as a function of mass.

In this study, given the relatively low spectral resolution and spectral
coverage of our spectra, we use the CL01 method only as a check of
the quality of empirically-calibrated metallicities. The three main
advantages of checking our results with the CL01 method are: \emph{(i)}
the use of one unique calibrator for all redshift ranges; \emph{(ii)}
the different method to deal with dust extinction, i.e. the self-consistent
correction instead of the use of equivalent widths; \emph{(iii)} the
different method to break the O/H degeneracy, i.e. the fit of the
double-peaked PDF instead of the use of the luminosity.

\begin{figure*}
\begin{centering}
\subfigure[$z<0.4$]{\includegraphics[width=1\columnwidth]{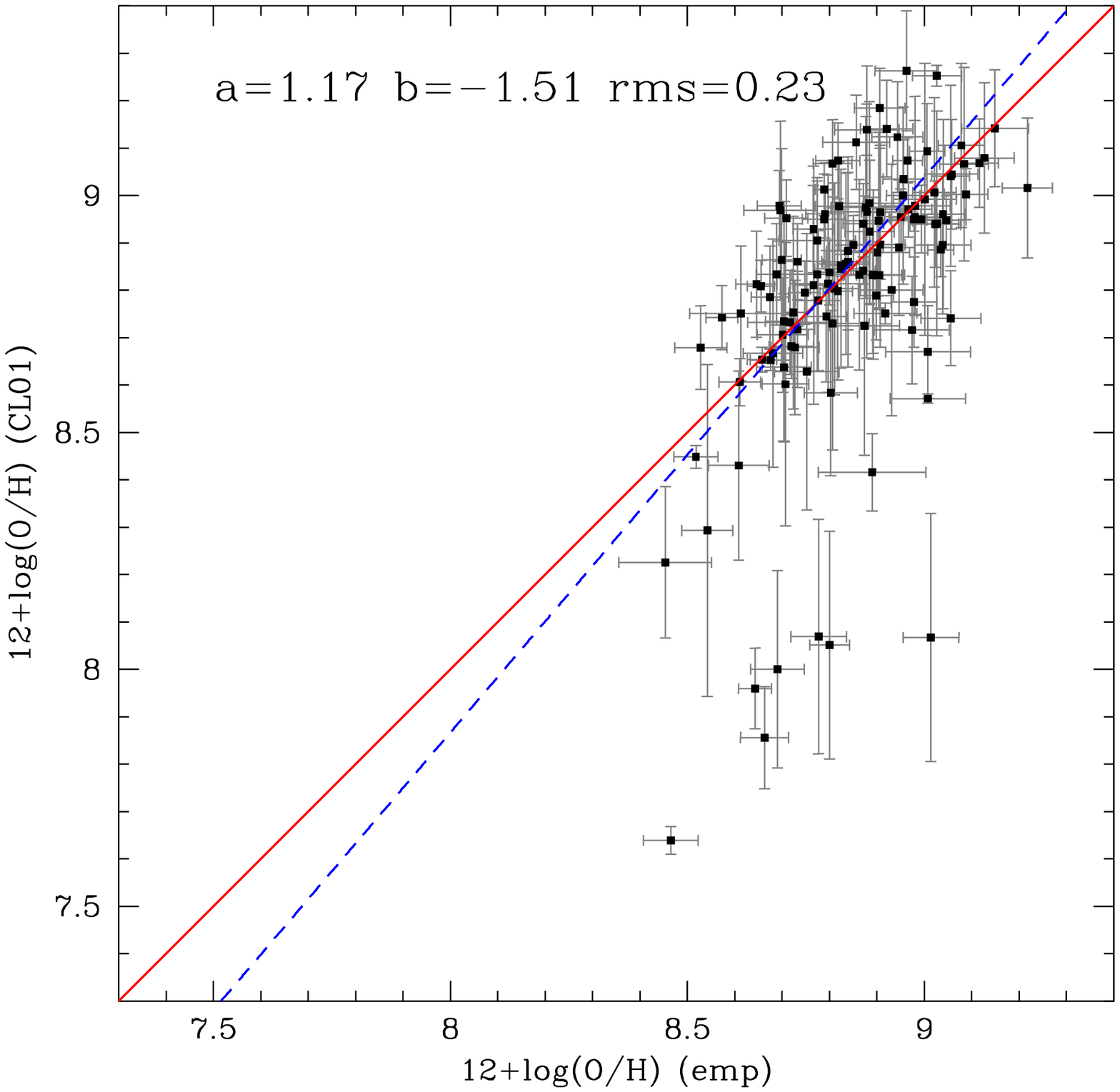}}
\subfigure[$z>0.4$]{\includegraphics[width=1\columnwidth]{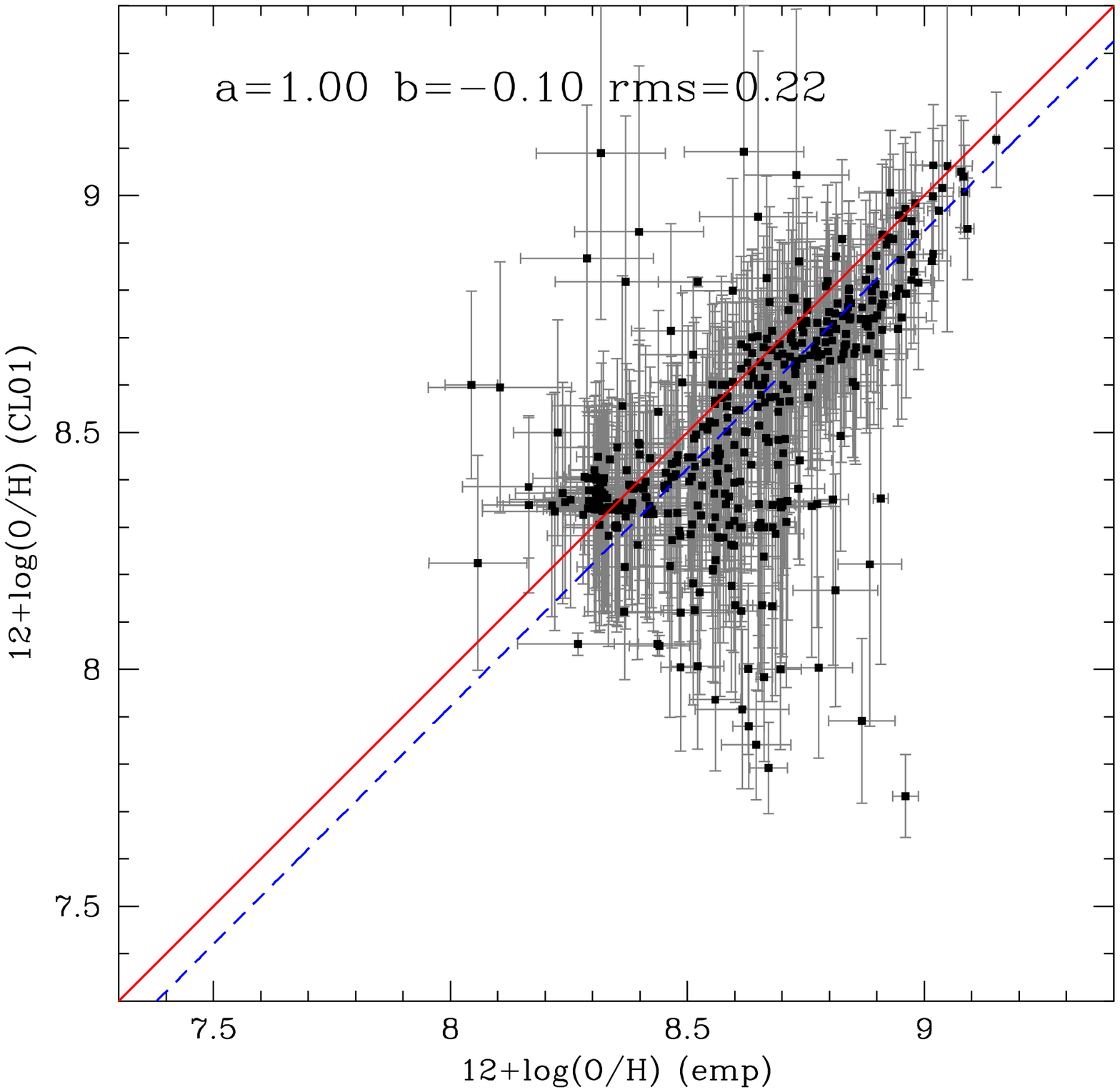}}
\par\end{centering}

\caption{Comparison between empirically-calibrated ($x$-axis), and CL01 metallicities
($y$-axis) for our whole sample of star-forming galaxies. The red
solid line shows the $y=x$ relation, while the blue dashed line is
a least-square fit to the data points. The parameters of the fit are
given in the two plots (\emph{a}: slope, \emph{b}: zero-point).}

\label{figcompz}
\end{figure*}

Fig.~\ref{figcompz} shows the comparison between the empirically-calibrated
and the CL01 metallicities. Apart from a small number of outliers
(7\%), the two results are in good agreement within a dispersion of
approximately $0.22$ dex. This dispersion can be compared with the
intrinsic dispersion of the mass-metallicity relation, which is also
of $\sim0.22$ dex. We have also calculated the mean of the error
bars for the two methods: $0.06$ dex for the empirical method, and
$0.18$ dex for the CL01 method. 

Let us summarize all the contributions to the global dispersion. \emph{(i)}
The empirically-calibrated errors have been estimated from basic propagation
of the line measurement errors. We know from tests performed on duplicated
observations that \emph{platefit\_vimos} errors are not underestimated.
They reflect the negligible contribution of noise to the global dispersion.
\emph{(ii)} Conversely, the CL01 errors have been estimated from the
width of the PDF. They also reflect the additional dispersion due
to the degeneracies in the models, which are taken into account thanks
to the Bayesian approach. \emph{(iii)} Finally the difference between
the mean CL01 error and the global dispersion reflects real variations
in the physical parameters of galaxies, which are not taken into account
in both models.

The comparison between the two methods have shown that: \emph{(i)}
the various empirical calibrators used at different redshift ranges
give consistent results with the CL01 metallicities, thanks to the
correction formulas provided by \citet{Kewley:2008arXiv0801.1849K};
\emph{(ii)} the empirical correction for dust attenuation using equivalent
widths, and the correction formula provided by \citet{Liang:2007A&A...474..807L},
is consistent with CL01's results, obtained directly from observed
line fluxes and a self-consistent reddening correction; \emph{(iii)}
apart from a relatively small number of outliers, the O/H degeneracy
is correctly broken using the luminosity diagnostic with the empirical
method.

\subsection{Possible biases}

\begin{figure}
\begin{centering}
\includegraphics[width=0.9\columnwidth]{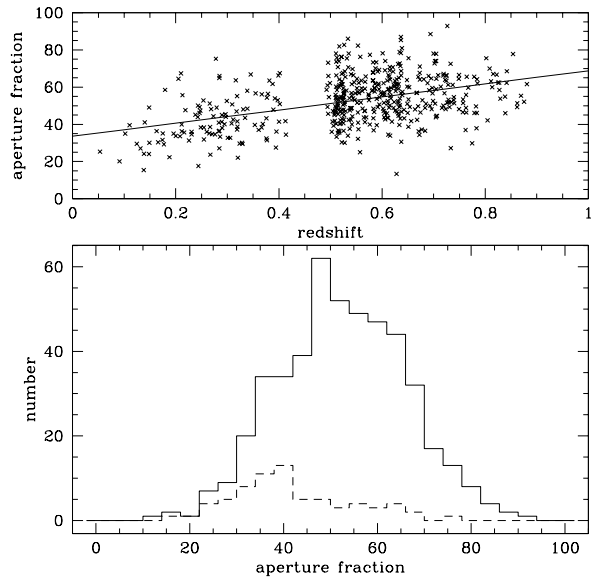}
\par\end{centering}

\caption{Bottom: distribution of the aperture fractions (in percent) for the
star-forming galaxies in our sample. The dashed histogram shows the
distribution for the galaxies at low redshift ($z<0.5$) with a high
metallicity ($12+\log(\mathrm{O/H})>8.8$). Top: evolution of the
apertures fraction as a function of redshift. The solid line is the
fit to the data points.}

\label{fig:histoape}
\end{figure}

Fig.~\ref{figcompz} also indirectly shows that the metallicities
obtained with $N2$ and $R_{23}$calibrators at different redshifts
are directly comparable, even if there is no galaxy in our sample
where these two calibrators can be applied at the same time. Indeed
we see that both $N2$ and $R_{23}$metallicities are in agreement
with CL01 metallicities, apart from the small number of outliers.
Moreover we have investigated the possible biases introduced by the
use of different sets of lines, at different redshifts, with the CL01
method: we have confirmed that this method gives, as expected, comparable
results at different redshifts.

Another possible bias comes from the fact that the slit used for the
spectroscopy not necessarily covers all the light of the observed
galaxy. We have taken the spatial extension of the galaxies calculated
from photometry, and compared it to the width of the slit, i.e. one
arcsecond. This gives the aperture fraction of the galaxy, expressed
in percent. Fig.~\ref{fig:histoape}(bottom) shows the distribution
of the aperture fractions for the star-forming galaxies in our sample
(not that this information was not available for the CDFS sample).
The mean aperture fraction is 52\%, and we see that the large majority
of objects has an aperture fraction greater than 20\%, which is the
minimum aperture fraction given by \citet*{Kewley:2005PASP..117..227K}
in order to minimize aperture effects on derived parameters like the
metallicity or the star-formation rate. 

In Fig.~\ref{fig:histoape}, we also observe the expected correlation
between redshift and the aperture fraction, both increasing at the
same time. The aperture fractions show an increase of 35\% per unit-redshift.
Since the aperture fractions are already high enough at low redshift
not to affect the derived metallicities, and since they increase with
redshift, this slope is likely to have a marginal effect on the derived
evolution of metallicity as a function of redshift.

\section{The luminosity-metallicity relation\label{sec:The-luminosity-metallicity-relation}}

\begin{figure*}
\begin{centering}
\includegraphics[width=0.25\paperwidth]{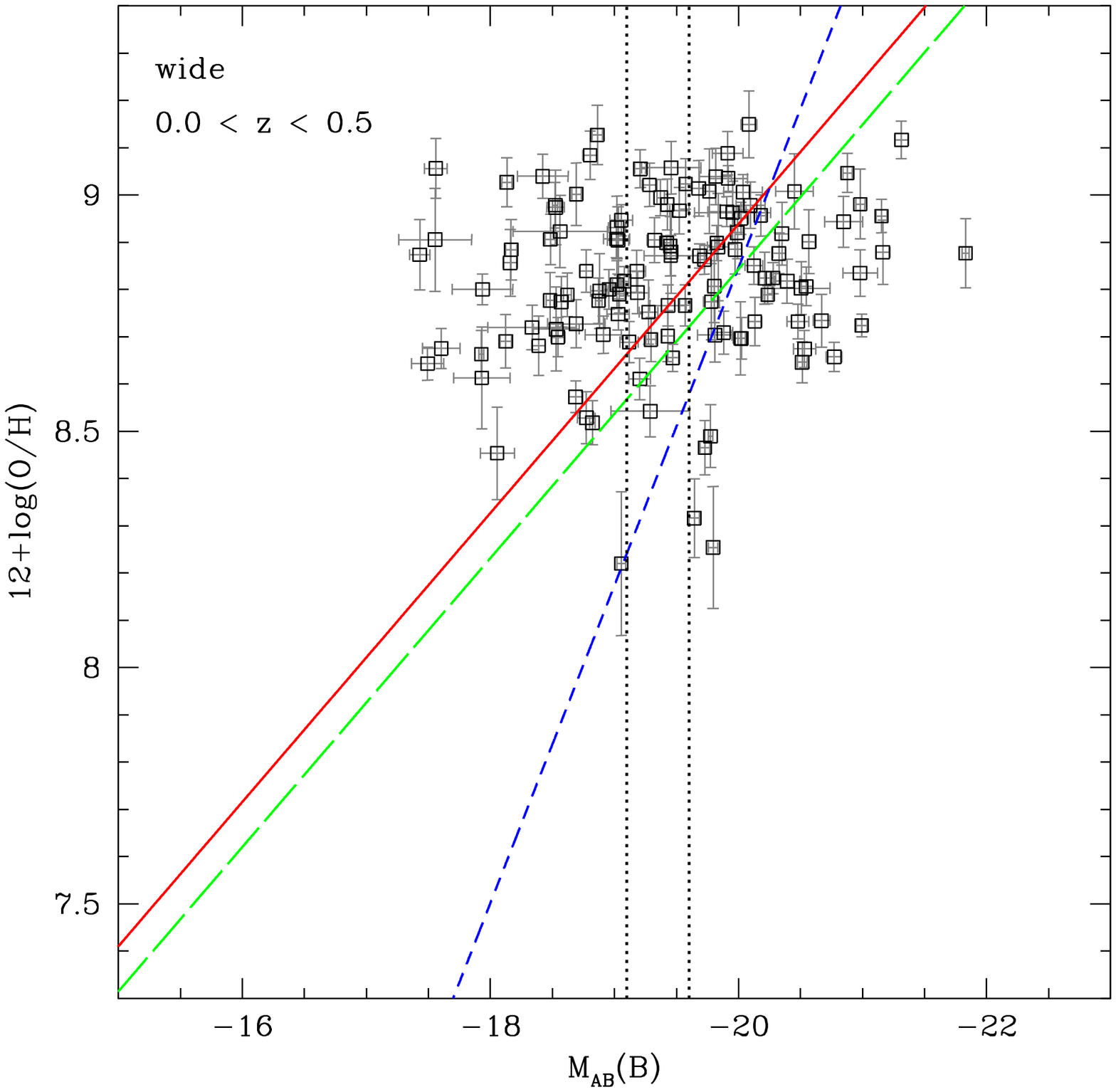} \includegraphics[width=0.25\paperwidth]{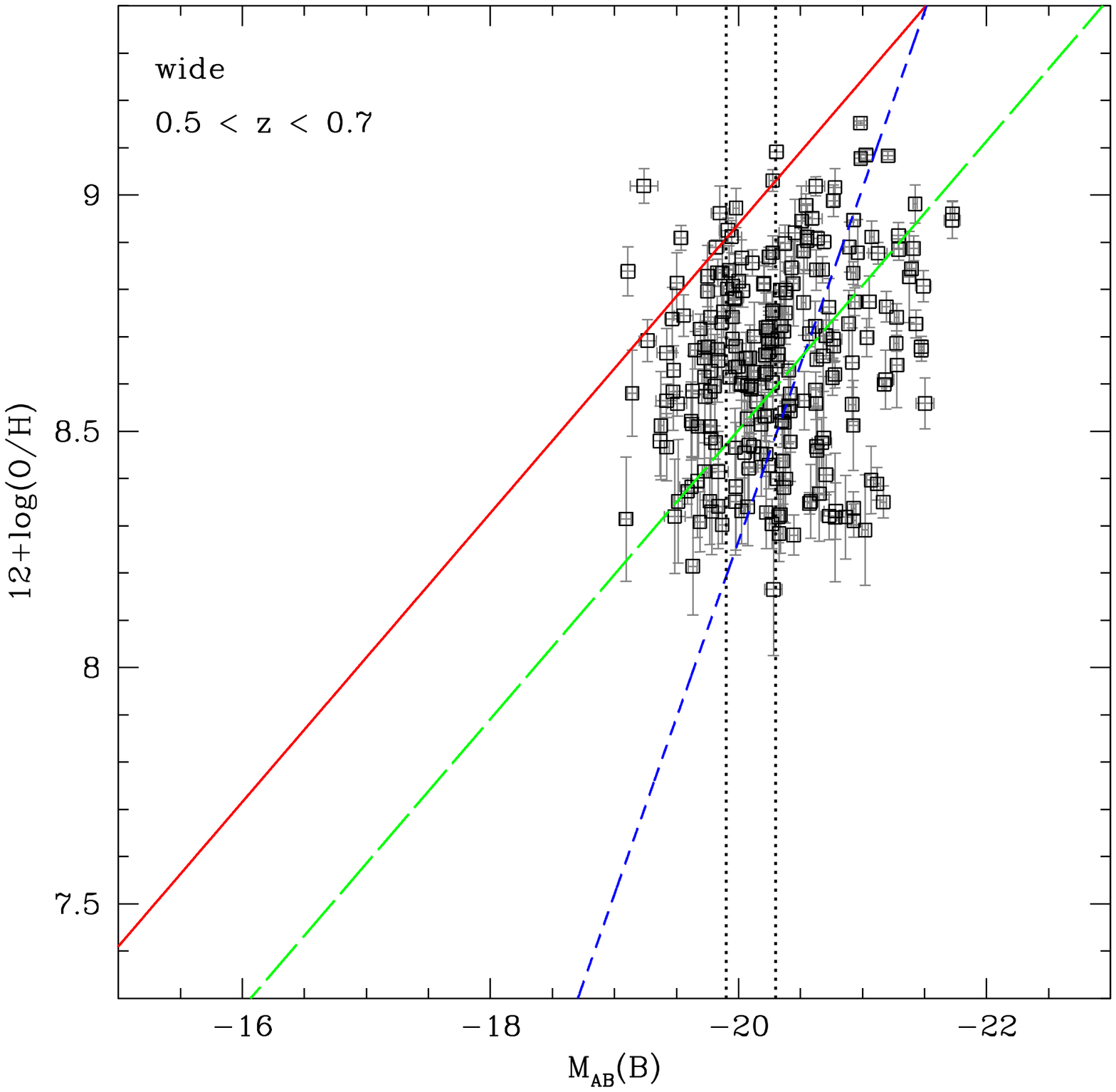}
\includegraphics[width=0.25\paperwidth]{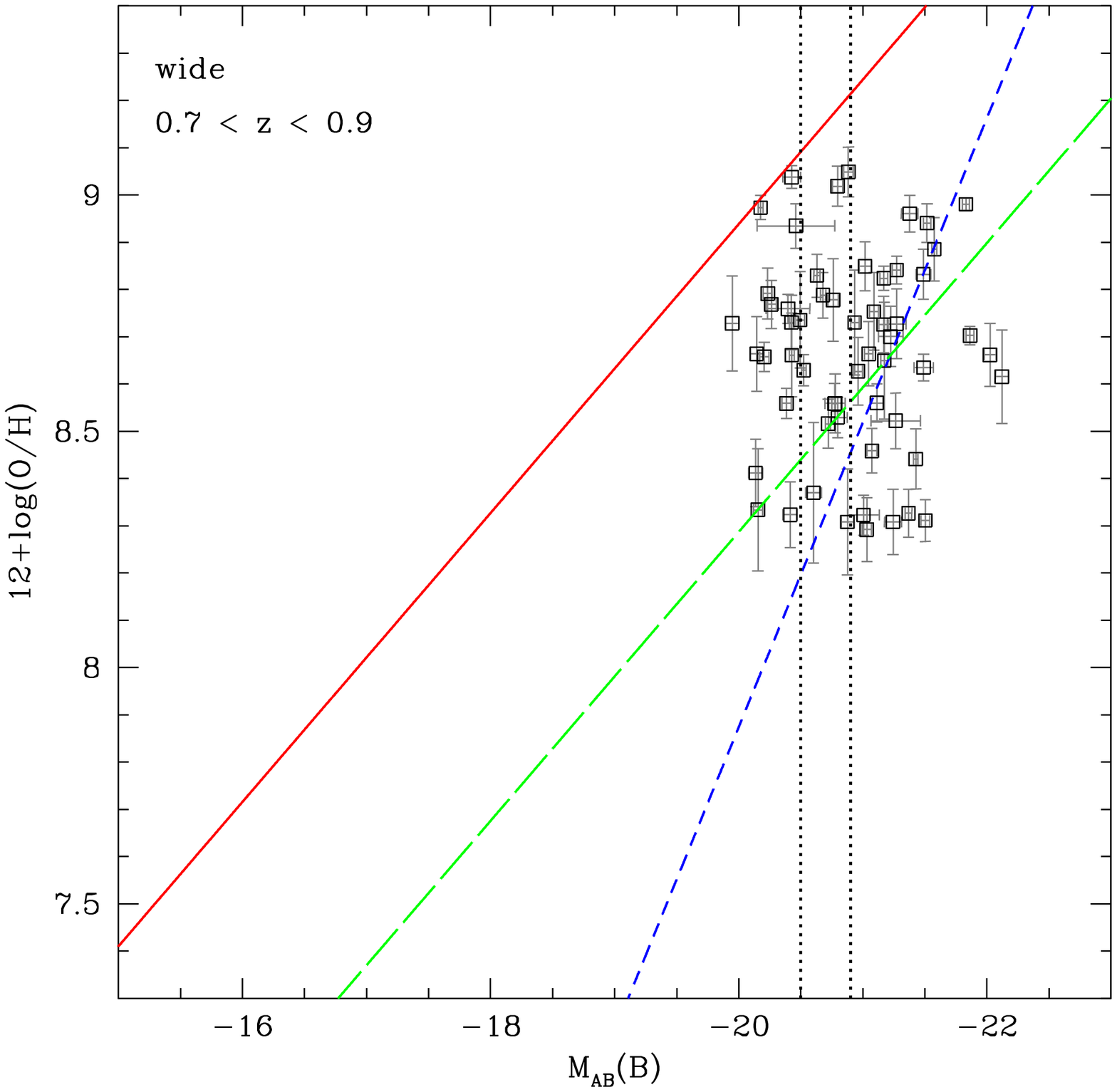} \includegraphics[width=0.25\paperwidth]{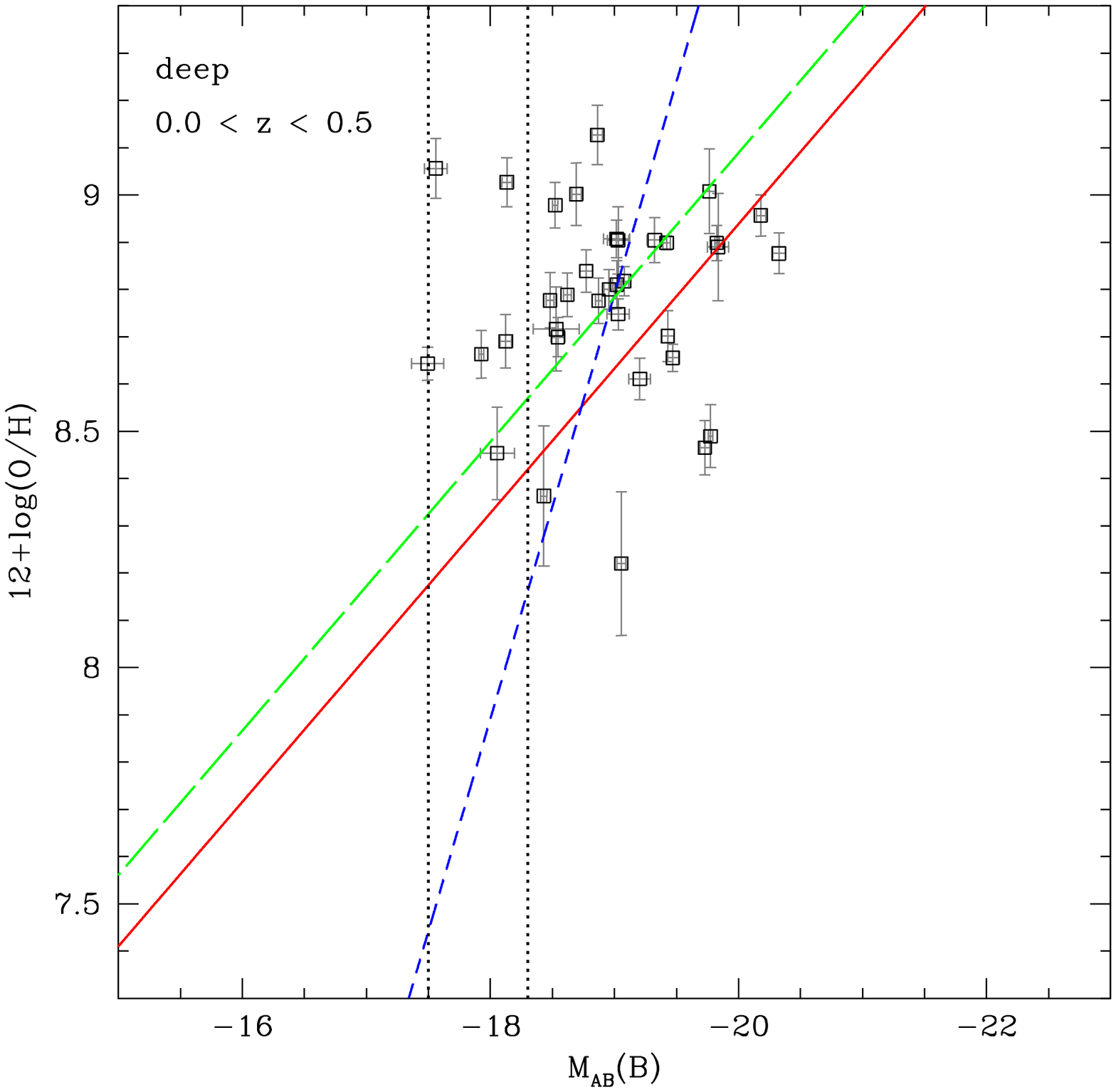}
\includegraphics[width=0.25\paperwidth]{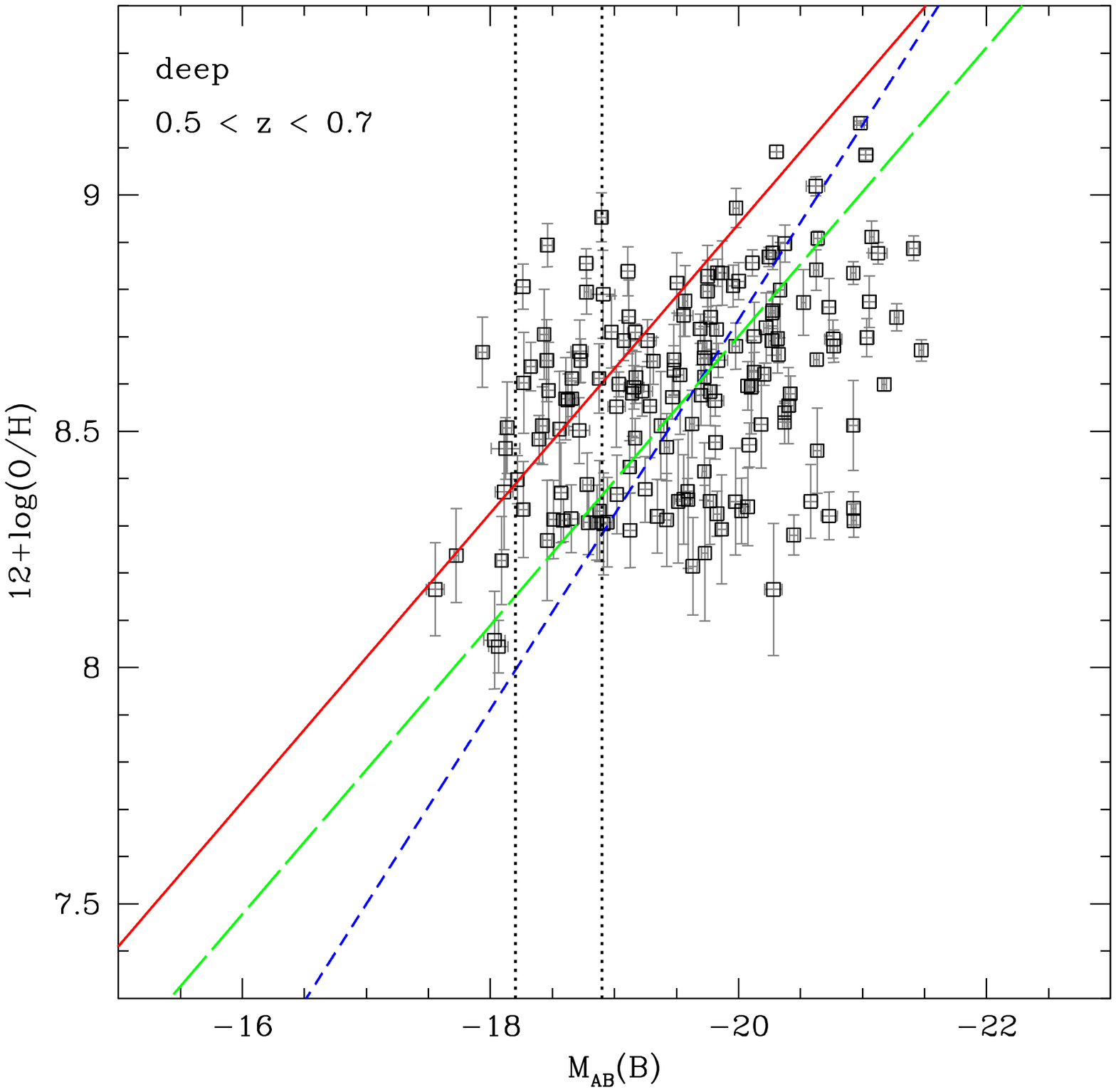} \includegraphics[width=0.25\paperwidth]{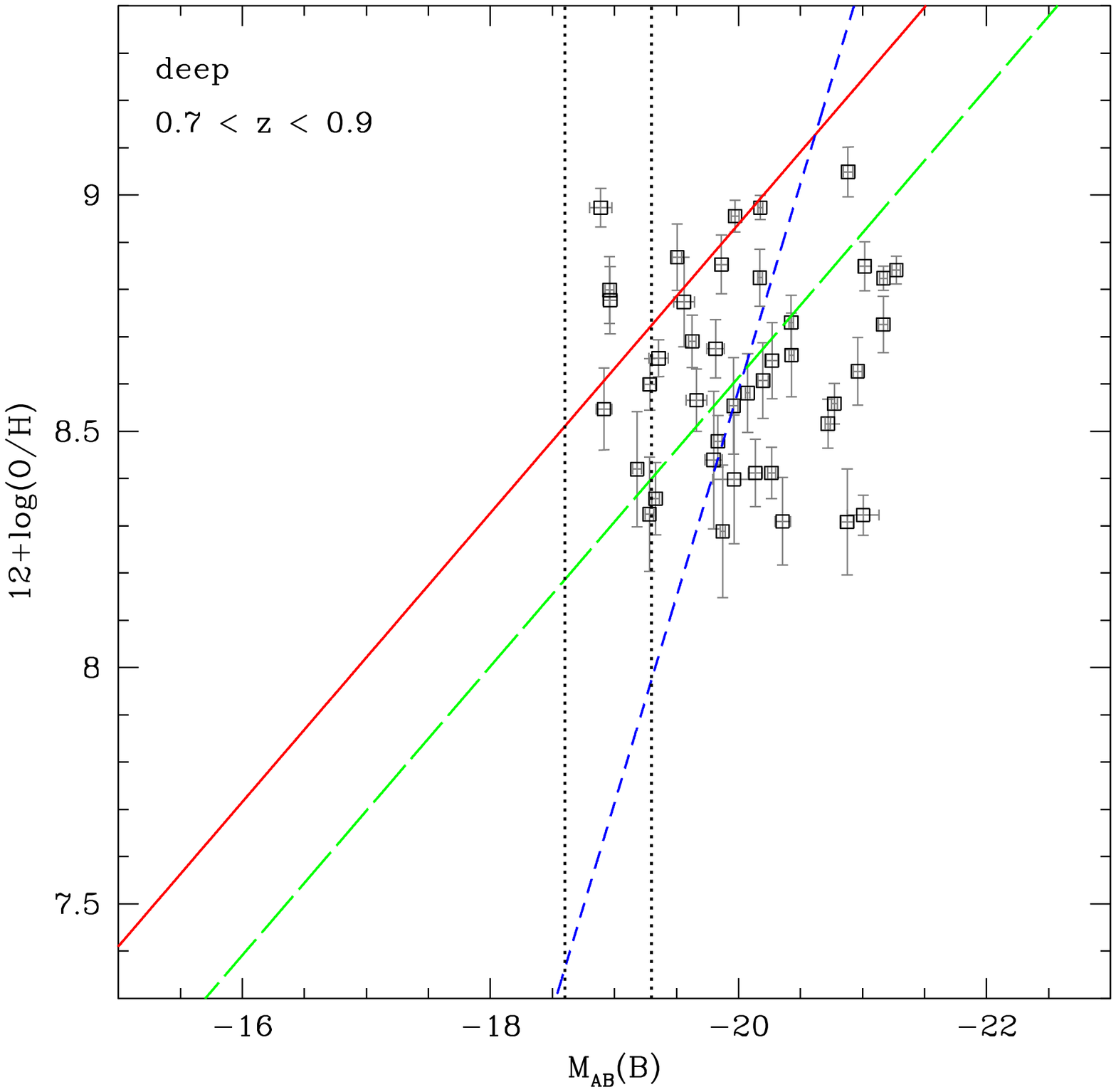}
\par\end{centering}

\caption{Rest-frame $B$-band luminosity-metallicity relation for the wide
(top) and deep (bottom) samples, for three redshift ranges: from left
to right $0.0<z<0.5$, $0.5<z<0.7$ and $0.7<z<0.9$. The metallicities
have been estimated using the empirical approach (see Sect.~\ref{sub:empimeta}).
The solid line shows the luminosity-metallicity relation at low redshift
derived by \citet{Lamareille:2004MNRAS.350..396L}, and renormalized
to the CL01 method. The short-dashed line shows the fit to the data
points using the ols bisector method (see text). The long-dashed lines
shows the fit to the data points assuming a constant slope. }

\label{fig:lz}
\end{figure*}

\begin{table*}
\caption{Evolution of the luminosity-metallicity relation for the wide and
deep samples. The reference relation is the one obtained by \citet{Lamareille:2004MNRAS.350..396L}
with 2dFGRS data and renormalized to the CL01 method. In each redshift
ranges, we give the results of the ols bisector fit (slope and zero-point)
and of the constant-slope fit (mean shift, see Fig.~\ref{fig:lz}),
together with the mean redshift and magnitude, and the dispersion
of the relation. The metallicity shift is given in three cases: \emph{a)}
using only star-forming galaxies, \emph{b)} adding candidate star-forming
galaxies, and \emph{c)} adding also candidate AGNs.}

\begin{centering}
\begin{tabular}{lrrrrrrrr}
\hline 
\hline
sample & slope & zero-point & $z$ & $M_{\mathrm{AB}}(B)$ & \multicolumn{3}{r}{$\Delta\log(\mathrm{O/H})^{L}$} & rms\tabularnewline
 &  &  &  &  & \emph{a} & \emph{b} & \emph{c} & \tabularnewline
\hline
\emph{reference} & $-0.31$ & $2.83$ &  &  &  &  &  & \tabularnewline
\emph{wide} &  &  &  &  &  &  &  & \tabularnewline
~$0.0<z<0.5$ & $-0.67\pm0.17$ & $-4.60\pm3.4$ & $0.31$ & $-19.99$ & $\mathit{-0.08}$ & $-0.10$ & $\mathit{-0.10}$ & $0.23$\tabularnewline
~$0.5<z<0.7$ & $-0.75\pm0.08$ & $-6.67\pm1.6$ & $0.59$ & $-20.53$ & $\mathit{-0.39}$ & $-0.44$ & $\mathit{-0.45}$ & $0.23$\tabularnewline
~$0.7<z<0.9$ & $-0.64\pm0.21$ & $-5.01\pm4.4$ & $0.77$ & $-21.22$ & $\mathit{-0.58}$ & $-0.65$ & $\mathit{-0.70}$ & $0.24$\tabularnewline
\emph{deep} &  &  &  &  &  &  &  & \tabularnewline
~$0.0<z<0.5$ & $-0.90\pm0.45$ & $-8.28\pm8.61$ & $0.30$ & $-18.99$ & $\mathit{0.17}$ & $0.15$ & $\mathit{0.15}$ & $0.28$\tabularnewline
~$0.5<z<0.7$ & $-0.41\pm0.06$ & $0.50\pm1.29$ & $0.59$ & $-19.67$ & $\mathit{-0.20}$ & $-0.24$ & $\mathit{-0.25}$ & $0.27$\tabularnewline
~$0.7<z<0.9$ & $-0.87\pm0.41$ & $-8.89\pm8.20$ & $0.76$ & $-20.05$ & $\mathit{-0.24}$ & $-0.32$ & $\mathit{-0.39}$ & $0.29$\tabularnewline
\end{tabular}
\par\end{centering}

\label{tab:lz}
\end{table*}

\subsection{Study of the derived fits}

We study in this section the relation between the rest-frame $B$-band
luminosity and gas-phase metallicity, for the star-forming galaxies
of our wide and deep samples. Here we use the metallicities estimated
with the empirical approach, and renormalized to the CL01 method (see
Sect.~\ref{sub:empimeta}). Figure~\ref{fig:lz} shows the luminosity-metallicity
relation of the wide and deep samples in three redshift ranges: $0.0<z<0.5$,
$0.5<z<0.7$ and $0.7<z<0.9$. The results are compared to the luminosity-metallicity
relation in the local universe derived by \citet{Lamareille:2004MNRAS.350..396L}
with 2dFGRS data, and renormalized to the CL01 method. In order to
do this comparison, we have done a linear fit to the data points.

As already shown before, the luminosity-metallicity relation is characterized
by an non-negligible dispersion of the order of $\approx0.25$ dex
(higher than the one of the mass-metallicity relation). Therefore,
the method used to perform the fit has a huge impact on the results,
and it is mandatory to use the same method before doing comparisons
between two studies. Thus, we have used the same method than \citet{Lamareille:2004MNRAS.350..396L},
i.e. the ols bisector fit \citep{Isobe:1990ApJ...364..104I}, starting
from the 50\% $k$-correction completeness level (see Sect.~\ref{sub:absmag}).
The results are shown in Fig.~\ref{fig:lz} and in Table~\ref{tab:lz}. 

The slope is by $\sim2\sigma$ steeper than the one of the reference
relation in the local universe, this slope being similar in all redshift
ranges ($\approx-0.75$), except $0.5<z<0.7$ in the deep sample.
As already shown by \citet{Lamareille:2004MNRAS.350..396L} who have
compared the slopes of the luminosity-metallicity relation for low
and high metallicity objects, this slope is steeper for high metallicity
objects. This effect has been confirmed by \citet{Lee:2006ApJ...647..970L}
who have extended the mass-metallicity relation of \citet{Tremonti:2004astro.ph..5537T}
to lower masses and metallicities and found a flatter slope. We emphasize
that we are not discussing here the saturation at $12+\log(\mathrm{O/H})\approx9.2$
as observed by \citet{Tremonti:2004astro.ph..5537T}, and which might
be confused with a flatter slope at very high metallicity.

Thus, given that we expect a steeper slope at high metallicity, we
analyze our results as a lack of data points in the low metallicity
region: the wide sample does not go deep enough to detect low-metallicity
and low-luminosity objects above the completeness limit. Moreover,
in the $0.0<z<0.5$ redshift range of wide and deep samples, the S/N>4
cut on emission lines introduces a bias towards high metallicity objects:
the faint and blended {[}N\noun{ii}]$\lambda$6584 line, used to compute
metallicities in this redshift range, becomes rapidly undetectable
at low metallicities. 

The apparently high number of galaxies with low redshift, low luminosity
and high metallicity seen in Fig.~\ref{fig:lz}(left) may be explained
by two effects. First, as said in previous paragraph, the {[}N\noun{ii}]$\lambda$6584
line is more likely to be observed for high metallicity objects. There
is consequently a lack of objects with low luminosity and low metallicity
which makes the distribution apparently wrong (as compared to previous
studies). Second, these objects at low redshift may have lower aperture
fractions, which may introduce a bias towards higher metallicities
given that we preferentially observe their central part. Indeed, we
have plot the distribution of the aperture fractions for objects with
a low redshift and a high metallicity in Fig.~\ref{fig:histoape}(bottom,
dashed histogram): they are clearly lower, with a mean of 40\%.

We recognize that the results obtained by doing a direct linear fit
to data points suffer from a drawback: the correlations between metallicity
and luminosity shown in Fig.~\ref{fig:lz} are weak. This weakness
is characterized by the non-negligible error bars for the slopes and
the zero-points of the relations provided in Table~\ref{tab:lz},
and also by low Spearman correlation rank coefficients: of the order
of $-0.3$ for the bottom-center panel, and$-0.1$ for the other panels.

\subsection{Global metallicity evolution}

Despite the weakness of the correlations found in our data, we know
from previous studies that the luminosity-metallicity exists. Thus
we can use the existence of this relation as an assumption and find
new results.

In order to quantify the evolution of the luminosity-relation, we
can also derive the mean evolution of metallicity. This is done with
the additional assumption that the slope of the luminosity-relation
remains constant slope at zero-order. The results are shown in Fig.~\ref{fig:lz}
and in Table~\ref{tab:lz}. As expected, the evolution is barely
significant in the $0.0<z<0.5$ redshift range. In the $0.5<z<0.7$
and $0.7<z<0.9$ redshift ranges, the evolution is stronger. It is
similar in the wide sample to what has been found by \citet{Lamareille:2006A&A...448..907L}.
In the $0.5<z<0.7$ redshift range of the deep sample, the results
are similar with both methods, which confirms the hypothesis that
a steeper slope is found only when low-metallicity points are not
included in the fit.

Finally, the comparison of the wide and deep samples show a stronger
evolution of the metallicity with redshift in the wide sample. This
result tells us that the slope \emph{does not actually remain constant},
and that the evolution of the metallicity is stronger in more luminous
objects. At $z\sim0.76$, galaxies with an absolute $B$-band magnitude
of $\sim-20.1$ have $-0.32$ dex lower metallicities than galaxies
of similar luminosities in the local universe, while galaxies with
an absolute $B$-band magnitude of $\sim-21.2$ have $-0.65$ dex
lower metallicities.

We have checked that the results are stable when breaking the O/H
degeneracy at high redshift with a different reference relation (e.g.
lowered in metallicity). As stated before by \citet{Lamareille:2006A&A...448..907L},
the use of a lower reference relation to break the degeneracy only
changes the metallicities of the galaxies in the intermediate region
($12+\log(\mathrm{O/H})\approx8.3$), thus non-significantly changing
the whole luminosity-metallicity relation. 

We have also checked the effect of not including the candidate star-forming
galaxies in the fit. As stated in Sect.~\ref{sub:starforming}, the
contamination of candidate star-forming galaxies by AGNs is less than
1\%, and is thus not expected to affect significantly the luminosity-metallicity
relation. Therefore, the results of not including the candidate star-forming
galaxies are only given as information in Table~\ref{tab:lz}. We
have also checked the effect of including candidate AGNs. In both
cases, the effect on the luminosity-metallicity relation is very small.

We need to disentangle between the two effects which make these objects
more luminous, i.e. a higher mass or a lower mass-to-light ratio,
which one is responsible of the stronger evolution of the metallicity.
Thus, we now study the mass-metallicity relation in the next section.

\section{The mass-metallicity relation\label{sec:The-mass-metallicity-relation}}

\subsection{Global metallicity evolution\label{sub:Globalmzevol}}

\begin{figure*}
\begin{centering}
\includegraphics[width=0.25\paperwidth]{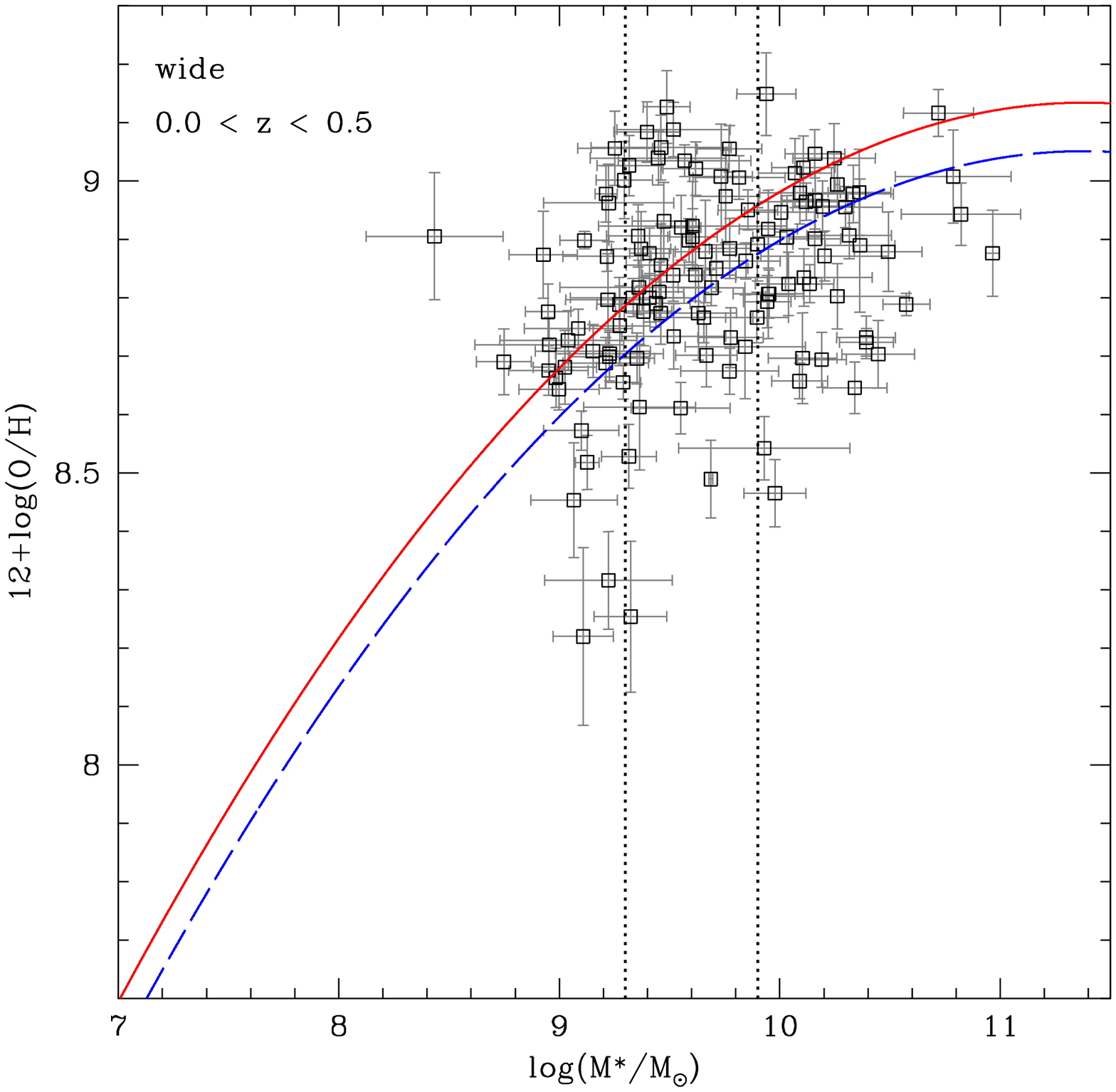} \includegraphics[width=0.25\paperwidth]{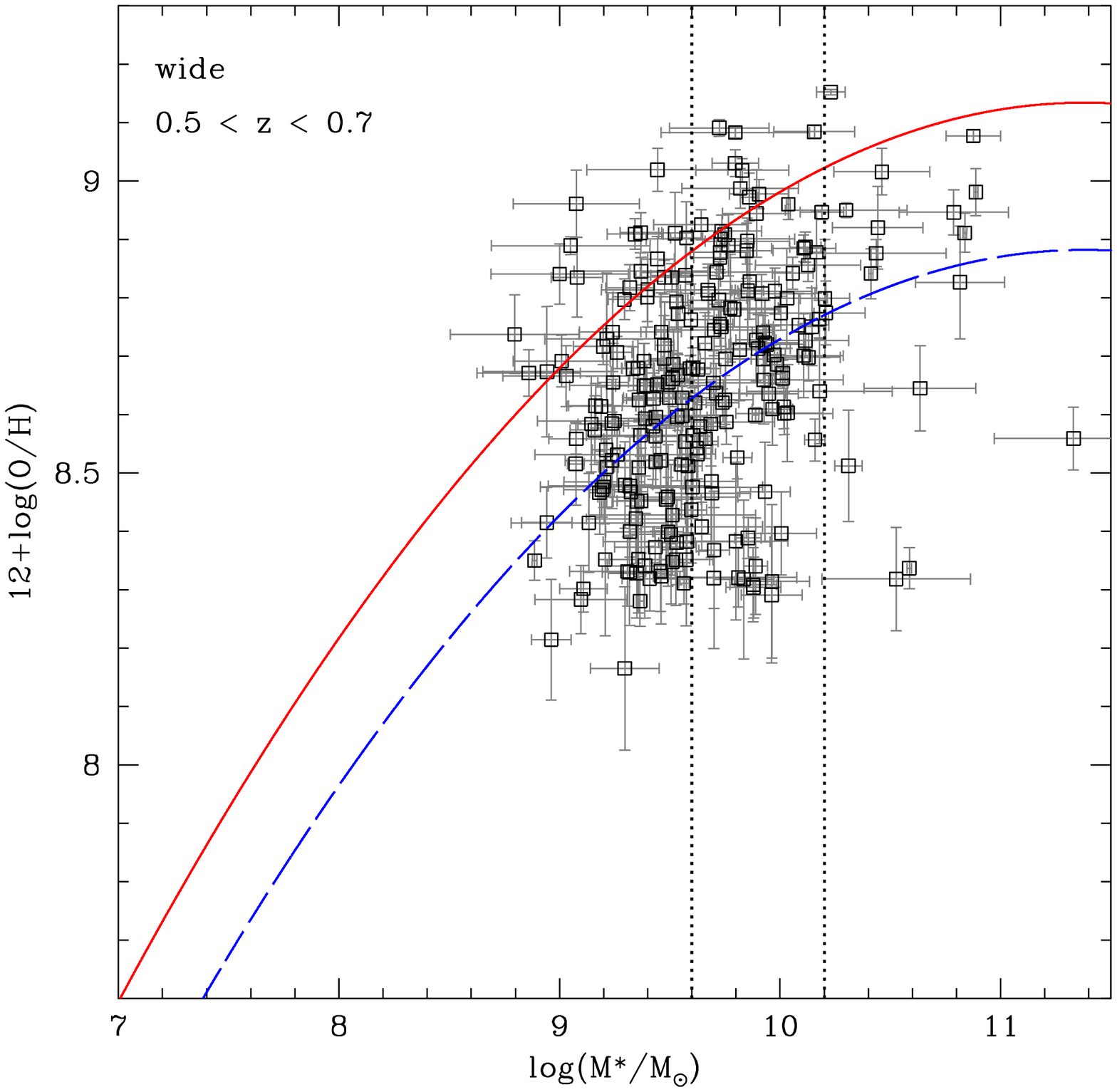}
\includegraphics[width=0.25\paperwidth]{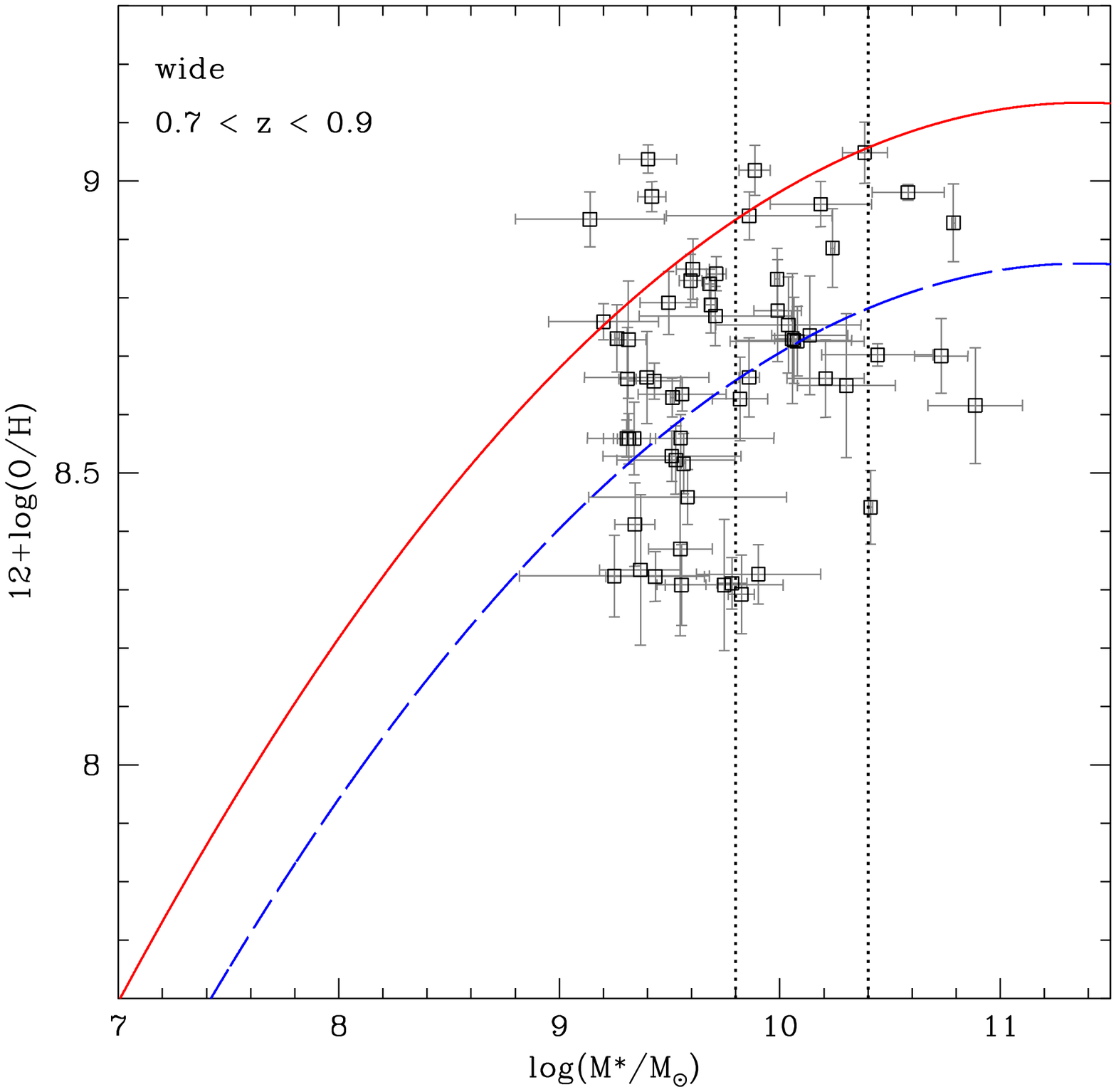} \includegraphics[width=0.25\paperwidth]{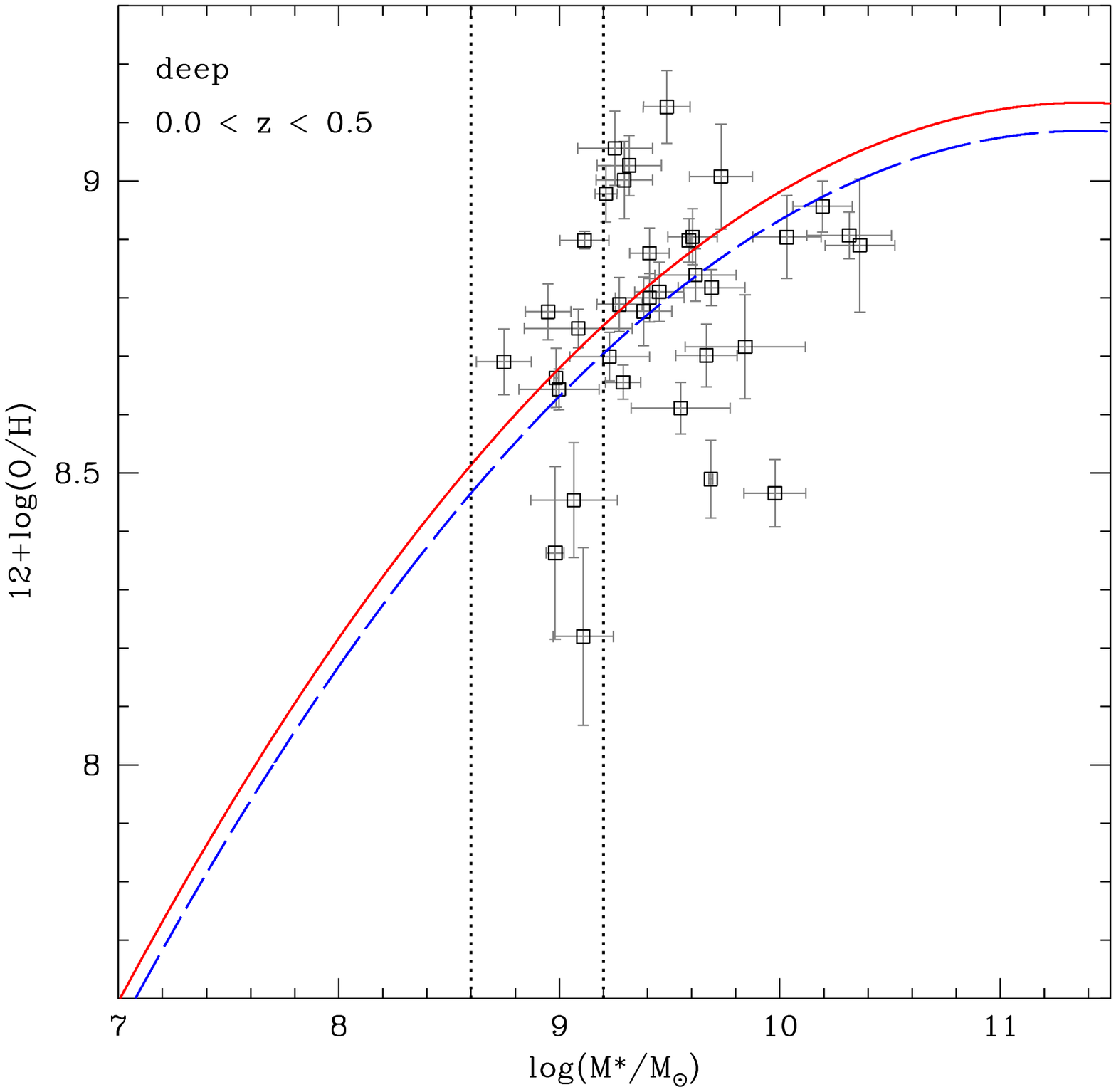}
\includegraphics[width=0.25\paperwidth]{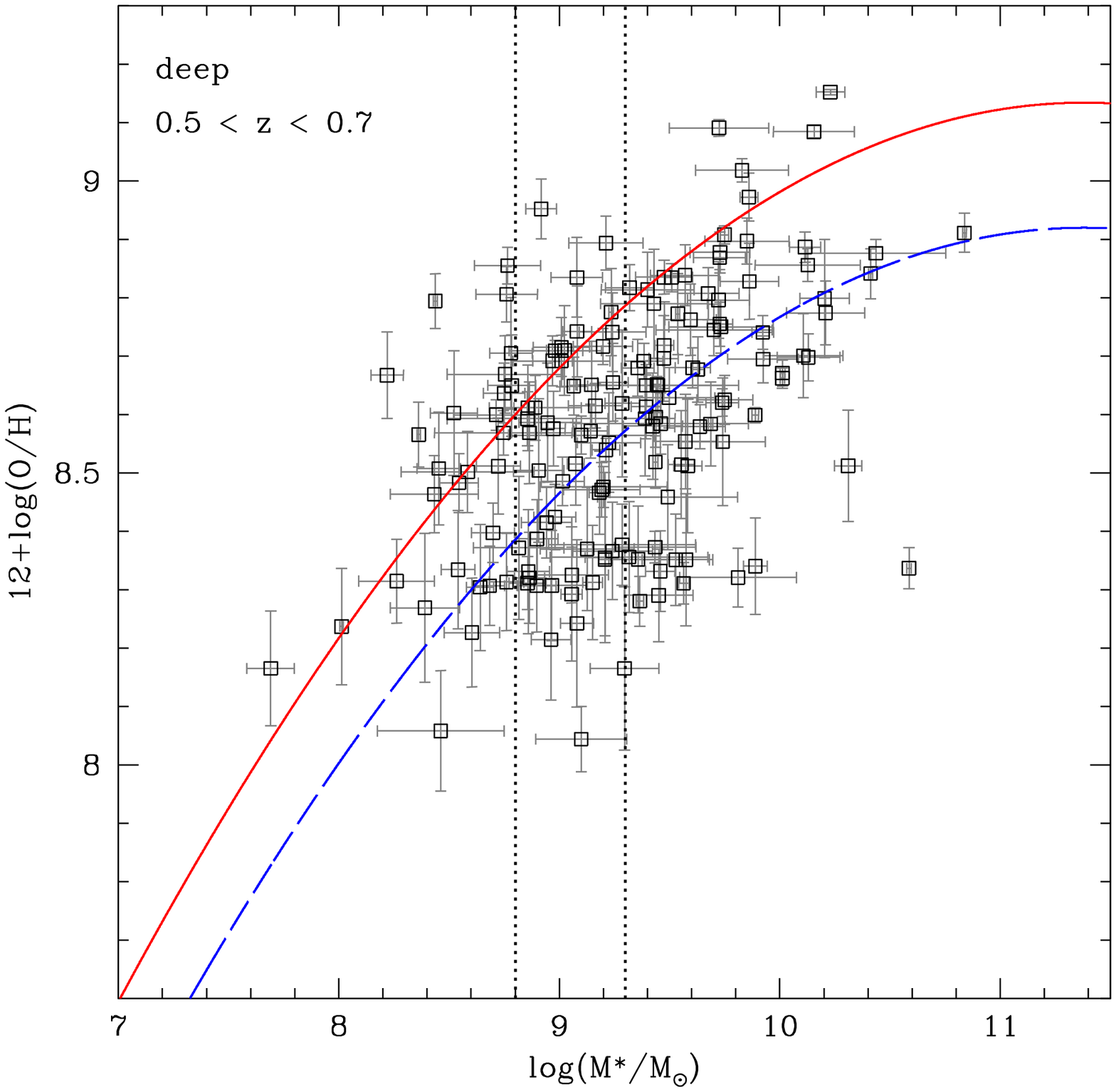} \includegraphics[width=0.25\paperwidth]{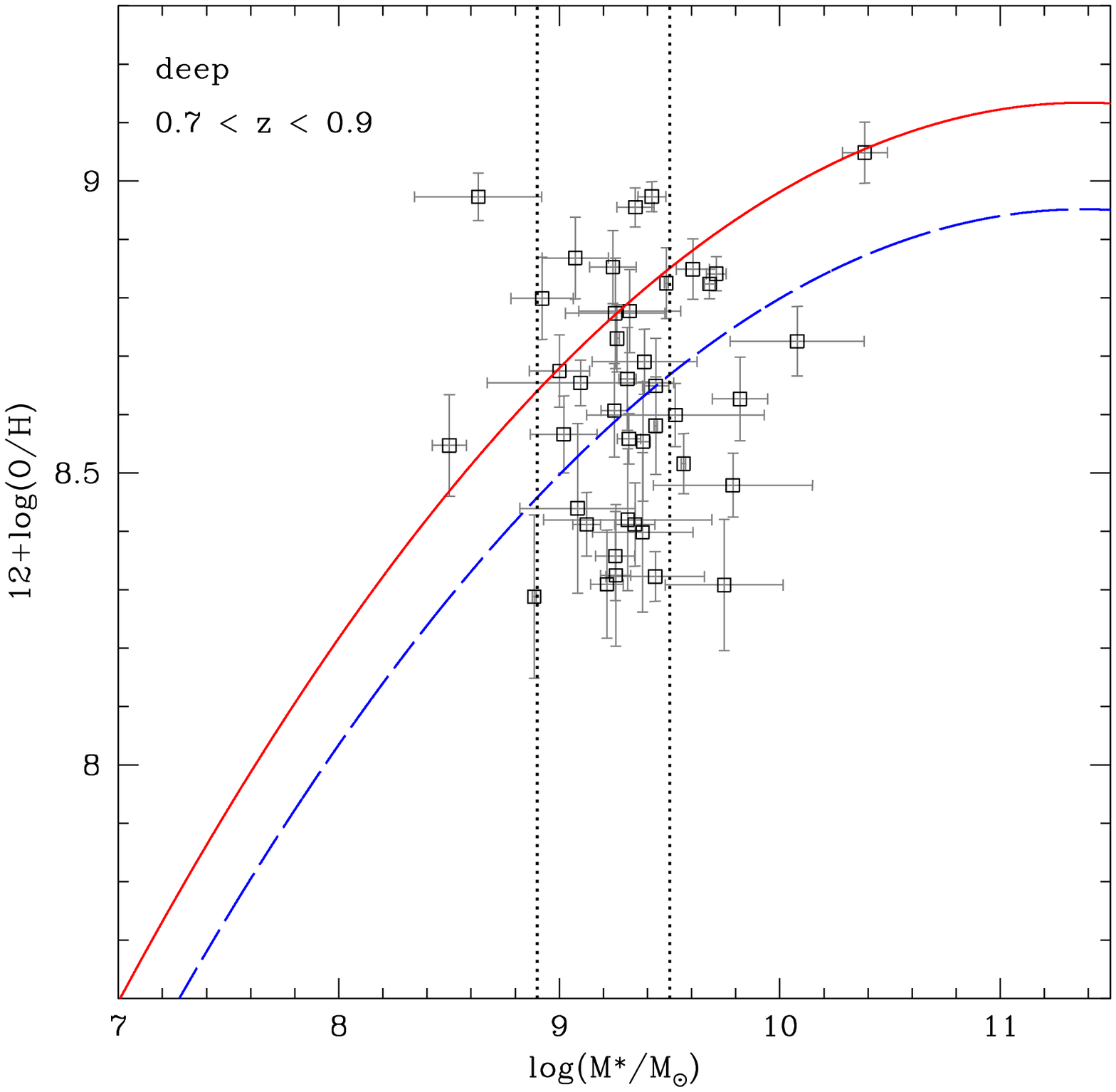}
\par\end{centering}

\caption{The mass-metallicity relation of star-forming galaxies for the wide
(top) and deep (bottom) samples, for three redshift ranges: from left
to right $0.0<z<0.5$, $0.5<z<0.7$ and $0.7<z<0.9$. The metallicities
have been estimated using the empirical approach (see Sect.~\ref{sub:empimeta}).
The solid curve shows the mass-metallicity relation at low redshift
derived by \citet{Tremonti:2004astro.ph..5537T}. The long-dashed
curves shows the fit to the data points assuming that the SDSS curve
is only shifted down in metallicity. }

\label{fig:mz}
\end{figure*}

\begin{table}
\caption{Evolution of the mass-metallicity relation for the wide and deep samples.
The reference relation is the one obtained by \citet{Tremonti:2004astro.ph..5537T}
with SDSS data and the CL01 method, renormalized in mass. In each
redshift ranges, we give the mean redshift, stellar mass, and metallicity
shift (assuming that the shape of the relation remains constant, see
Fig.~\ref{fig:mz}), and the dispersion of the relation. The metallicity
shift is given in three cases: \emph{a)} using only star-forming galaxies,
\emph{b)} adding candidate star-forming galaxies, and \emph{c)} adding
also candidate AGNs.}

\begin{centering}
\begin{tabular}{lrrrrrr}
\hline 
\hline
sample & $z$ & $\lg(M_{\star})$ & \multicolumn{3}{r}{$\Delta\log(\mathrm{O/H})^{M}$} & rms\tabularnewline
 &  &  & \emph{a} & \emph{b} & \emph{c} & \tabularnewline
\hline
\emph{wide} &  &  &  &  &  & \tabularnewline
~$0.0<z<0.5$ & $0.30$ & $9.87$ & $\mathit{-0.08}$ & $-0.08$ & $\mathit{-0.09}$ & $0.17$\tabularnewline
~$0.5<z<0.7$ & $0.59$ & $9.97$ & $\mathit{-0.22}$ & $-0.25$ & $\mathit{-0.26}$ & $0.20$\tabularnewline
~$0.7<z<0.9$ & $0.78$ & $10.19$ & $\mathit{-0.23}$ & $-0.28$ & $\mathit{-0.36}$ & $0.19$\tabularnewline
\emph{deep} &  &  &  &  &  & \tabularnewline
~$0.0<z<0.5$ & $0.29$ & $9.45$ & $\mathit{-0.04}$ & $-0.05$ & $\mathit{-0.05}$ & $0.20$\tabularnewline
~$0.5<z<0.7$ & $0.59$ & $9.45$ & $\mathit{-0.17}$ & $-0.21$ & $\mathit{-0.23}$ & $0.19$\tabularnewline
~$0.7<z<0.9$ & $0.76$ & $9.40$ & $\mathit{-0.12}$ & $-0.18$ & $\mathit{-0.23}$ & $0.20$\tabularnewline
\end{tabular}
\par\end{centering}

\label{tab:mz}
\end{table}

We now derive the mass-metallicity relation for the star-forming galaxies
of the wide and deep samples, using the empirical approach for computing
the metallicities, and in three redshift ranges: $0.0<z<0.5$, $0.5<z<0.7$
and $0.7<z<0.9$. The results are shown in Fig.~\ref{fig:mz} and
Table~\ref{tab:mz}, and are compared to the reference mass-metallicity
relation in the local universe, derived by \citet{Tremonti:2004astro.ph..5537T}
with SDSS data. The reference relation has been shifted in mass, in
order to take into account the effect of using different models (see
Sect.~\ref{sub:Description-of-the}). 

As in Fig.~\ref{fig:lz}, the data points shown in Fig.~\ref{fig:mz}
do not show strong Spearman rank correlation coefficients. We thus
skip the step of doing a fit to these data points. We directly probe
the global evolution in metallicity of star-forming galaxies compared
to the reference relation, doing the likely assumption that the mass-metallicity
relation exists also at high redshift. To do so, we calculate the
mean shift in metallicity by fitting to the data points the same curve
than Eq.~3 of \citet{Tremonti:2004astro.ph..5537T}, allowing only
a different zero-point. The zero-order assumption is indeed that the
shape of the mass-metallicity relation does not vary with redshift.
The results are shown in Fig.~\ref{fig:mz} and in Table~\ref{tab:mz}.
The fit is performed above the 50\% mass-to-light completeness level.

There are some caveats in the interpretation of the mass and metallicity
evolution of the deep sample, because of selection and statistical
effects. The mean observed stellar seams to decrease with redshift,
which is in contradiction of what one would expect from the Malmquist
bias.\textbf{ }First, we find higher stellar masses than expected
in the lowest redshift bin. The sample is actually not complete down
to the limiting mass: lower mass galaxies would have lower metallicities,
and low metallicities are indeed difficult to measure because of the
blending of {[}N\noun{ii}]$\lambda$6584 and H$\alpha$ lines. Conversely,
the highest redshift bin has a lower mean mass than expected, which
is more probably due to a statistical effect because of the small
solid angle of the deep sample. This later effect also explains why
the metallicity evolution seems smaller in the highest redshift bin.

As observed on the luminosity-metallicity relation in previous section,
we clearly see a stronger metallicity evolution in the wide sample
than in the deep sample. The wide and deep samples span interestingly
different ranges in masses, but are otherwise identical. This effect
thus shows that the most massive galaxies have experienced the most
significant evolution in metallicity. At $z\sim0.77$, galaxies at
$10^{9.4}$ solar masses have $-0.18$ dex lower metallicities than
galaxies of similar masses in the local universe, while galaxies at
$10^{10.2}$ solar masses have $-0.28$ dex lower metallicities. We
therefore conclude that the shape of the mass-metallicity relation
varies with redshift, so that it was flatter in an earlier universe.

We also remark that like for the luminosity-metallicity relation,
the potential influence of candidate AGNs or candidate star-forming
galaxies is negligible (see Table~\ref{tab:mz}).

\subsection{Evolution of the shape of the mass-metallicity relation}

\begin{figure*}
\begin{centering}
\includegraphics[width=0.25\paperwidth]{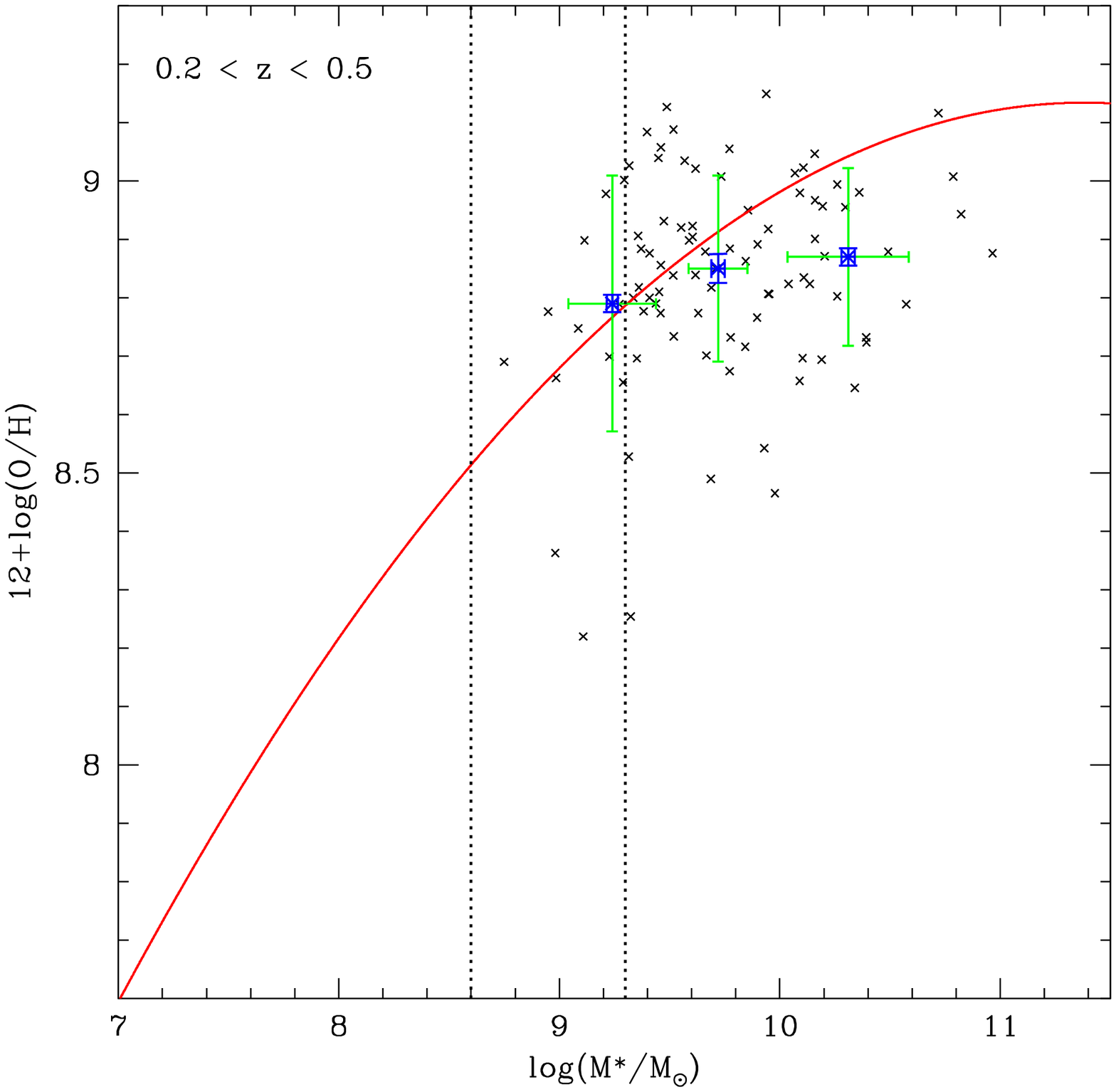}
\includegraphics[width=0.25\paperwidth]{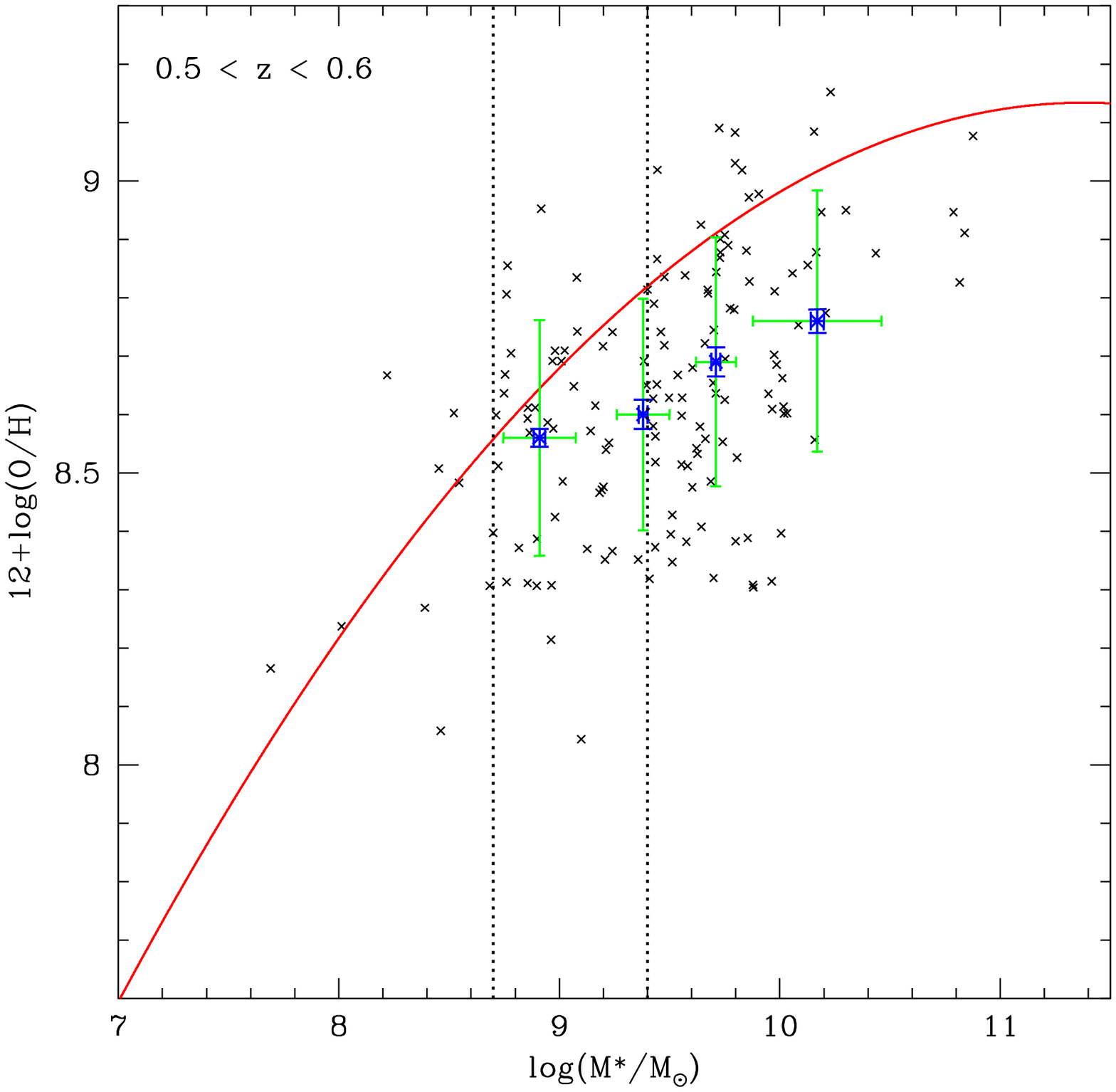}
\includegraphics[width=0.25\paperwidth]{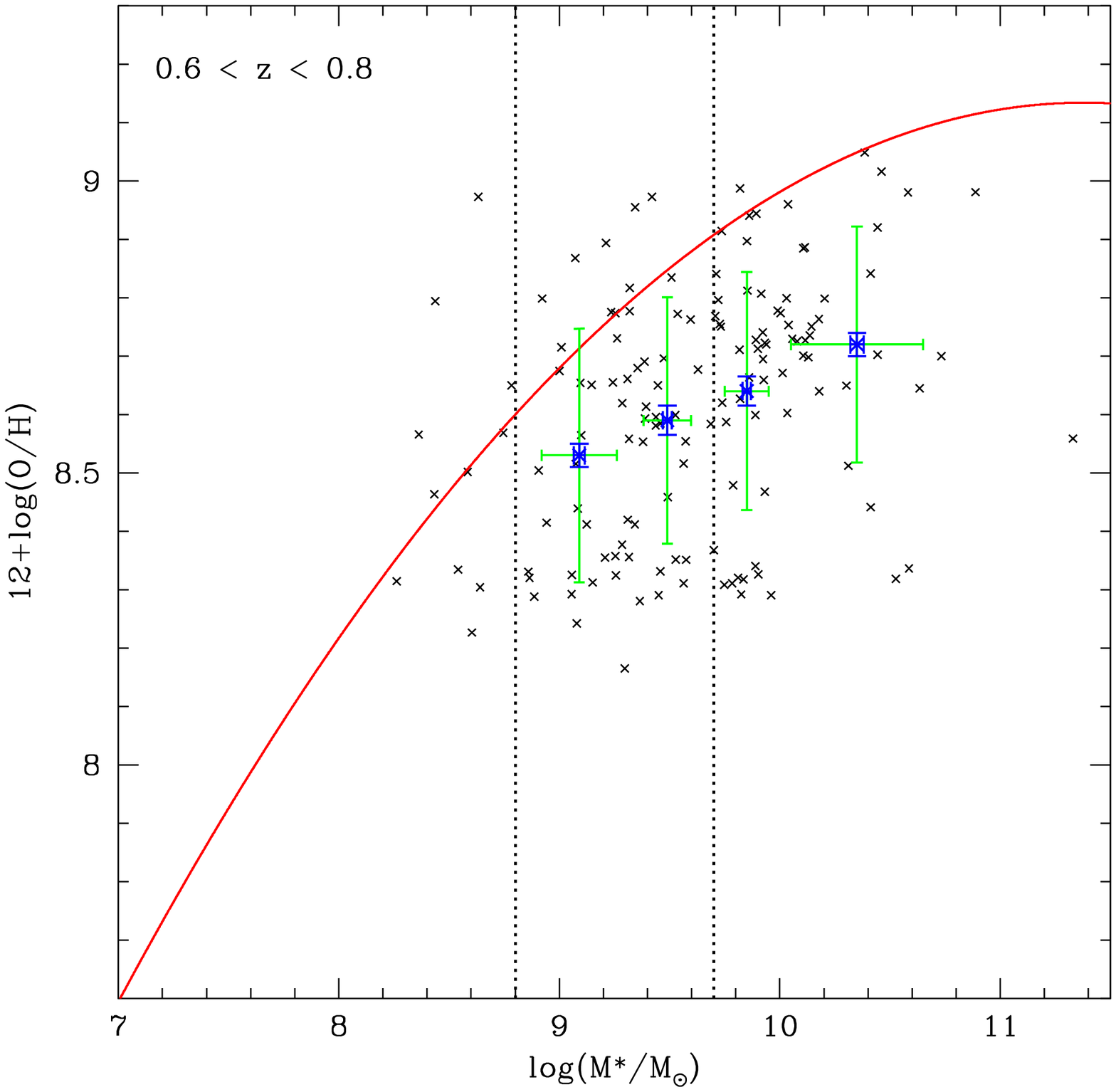}
\par\end{centering}

\caption{The mass-metallicity relation of star-forming galaxies for three redshift
ranges: from left to right $0.2<z<0.5$, $0.5<z<0.6$ and $0.6<z<0.8$.
The metallicities have been estimated using the empirical approach
(see Sect.~\ref{sub:empimeta}). The solid curve shows the mass-metallicity
relation at low redshift derived by \citet{Tremonti:2004astro.ph..5537T}.
We show the mean metallicities by bins of stellar masses. The blue
error bars represent the uncertainty on the mean, while the green
error bars represent the dispersion of the data points. The 50\% mass-to-light
completeness levels for the deep and wide samples (see Table~\ref{tab:limmass})
are shown as vertical dotted lines.}

\label{fig:mzevol}
\end{figure*}

\begin{figure}
\begin{centering}
\includegraphics[width=1\columnwidth]{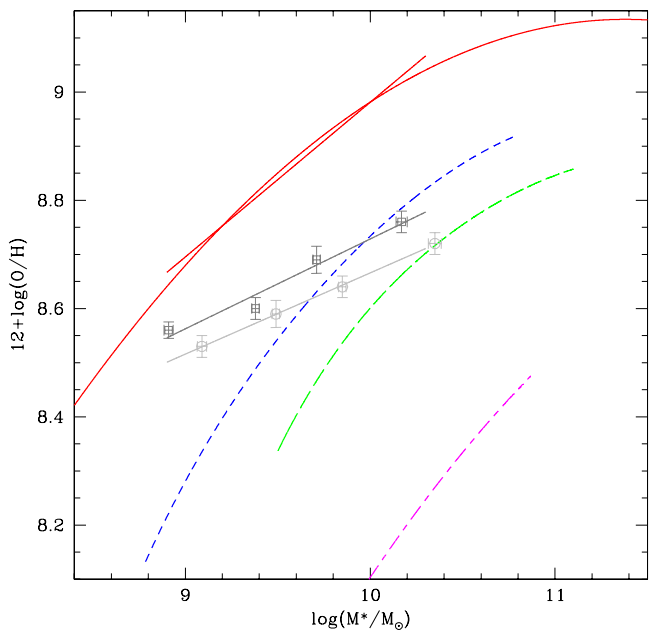}
\par\end{centering}

\caption{The mass-metallicity relation of star-forming galaxies for two redshift
ranges: $0.5<z<0.6$ (dark grey) and $0.6<z<0.8$ (light grey). The
metallicities have been estimated using the empirical approach (see
Sect.~\ref{sub:empimeta}). The solid curve shows the mass-metallicity
relation at low redshift derived by \citet{Tremonti:2004astro.ph..5537T}.
This relation has also been linearized in the range $8.9<\log(M_{\star}/M_{\odot})<10.3$.
The short-dashed curve is the relation derived by \citet{Savaglio:2005ApJ...635..260S}
at $z\sim0.7$, the long-dashed line is the relation derived by \citet{erb:2006ApJ...644..813E}
at $z\sim2.0$, and the long-dashed--short-dashed line is the relation
derived by \citet{Maiolino:2008A&A...488..463M} at $z\sim3.5$.}

\label{fig:mzevol2}
\end{figure}

We now evaluate the evolution of the shape of the mass-metallicity
relation as a function of redshift, by co-adding a number of data
points together in bins of mass, thus increasing the signal-to-noise
ratio. Figure~\ref{fig:mzevol} shows the results in three redshift
ranges: $0.2<z<0.5$, $0.5<z<0.6$ and $0.6<z<0.8$. The mean redshifts
are respectively $z\sim0.33$, $z\sim0.54$ and $z\sim0.67$. The
mean masses are respectively $10^{9.75}$, $10^{9.54}$ and $10^{9.70}$
solar masses. We have joined for this study the wide and deep samples:
only the deep sample is used from its 50\% completeness limit to the
50\% completeness of the wide sample, above which both samples are
used.

The mean masses and metallicities have been estimated in respectively
3, 4, and 4 stellar mass bins. This approach is similar to the one
used at low redshift by \citet{Tremonti:2004astro.ph..5537T}. The
bins are not equally spaced in mass,\textbf{\textcolor{red}{ }}but
always contain a similar number of data points, i.e. $\sim45$ in
our case. We have set two consecutive mass bins to have 25\% of their
data points in common. We evaluate the errors on the means and the
dispersions of the data points, which are both shown in Fig.~\ref{fig:mzevol}.
The errors on the means are very low thanks to the number of available
data points. The dispersion in metallicity stays almost constant and
equal to $\approx0.22$ dex. All results are calculated as the median
of 1000 bootstrap estimates.

Fig.~\ref{fig:mzevol2} shows the evolution of the slope in the two
last redshift bins. We confirm that the mass-metallicity relation
tends to become flatter at higher redshifts: the slope is $0.166\pm0.02$
dex/decade at $z\sim0.54$ and $0.150\pm0.01$ dex/decade at $z\sim0.67$,
which has to be compared to the slope $0.29$ dex/decade of the \citet{Tremonti:2004astro.ph..5537T}
curve linearized in the same mass range. The slope evolves by $-0.25$
dex/decade by unit of redshift in the range $0<z<0.7$. These results
also confirm the assumption, used in previous section, that the mass-metallicity
relation exists at high redshift.

\subsection{Comparison with previous works}

\begin{table*}
\caption{Different setups for different mass-metallicity relations found in
the literature, and associated shifts to be applied to their stellar
masses in order for them to be comparable with our results. This table
shows the type of data, the models, the presence or not of secondary
bursts, and the IMF used to compute the stellar masses. The last column
gives the global shift that has to be applied to the logarithm of
the stellar mass.}

\begin{centering}
\begin{tabular}{llllll}
\hline 
\hline
References$^{\dagger}$ & data$^{\star}$ & model & bursts & IMF$^{\dagger}$ & sum\tabularnewline
\hline
\emph{our study} & P+S & CB07 & yes & C03 & \tabularnewline
T04 & S & BC03 & yes & K01 & \tabularnewline
 & $+0.00$ & $-0.07$ & - & $-0.056$ & $-0.126$\tabularnewline
S05 & P & Pégase & yes & BG03 & \tabularnewline
 & $+0.05$ & $-0.09$ & - & $+0.024$ & $-0.016$\tabularnewline
E06 & P & BC03 & no & C03 & \tabularnewline
 & $+0.05$ & $-0.09$ & $+0.14$ & - & $+0.1$\tabularnewline
M08 & P & BC03 & no & S55 & \tabularnewline
 & $+0.05$ & $-0.09$ & $+0.14$ & $-0.232$ & $-0.132$\tabularnewline
\hline
\end{tabular}
\par\end{centering}

\medskip{}

{\footnotesize $^{\star}$S stands for spectroscopy, P stands for
photometry.}{\footnotesize \par}

{\footnotesize $^{\dagger}$T04: }\citet{Tremonti:2004astro.ph..5537T}{\footnotesize ;
S05: }\citet{Savaglio:2005ApJ...635..260S}{\footnotesize ; E06: }\citet{erb:2006ApJ...644..813E}{\footnotesize ;
M08: }\citet{Maiolino:2008A&A...488..463M}{\footnotesize ; C03: \citet{Chabrier:2003PASP..115..763C};
K01: \citet{Kroupa:2001MNRAS.322..231K}; BG03: \citet{Baldry:2003ApJ...593..258B};
S55: \citet{Salpeter:1955ApJ...121..161S}.}{\footnotesize \par}

\label{tab:shiftsmass}
\end{table*}

Fig.~\ref{fig:mzevol2} shows also the comparison between our results,
and other studies performed at high redshifts \citep{Savaglio:2005ApJ...635..260S,erb:2006ApJ...644..813E,Maiolino:2008A&A...488..463M}.
All curves have been renormalized in stellar masses using the shifts
summarized in Table.~\ref{tab:shiftsmass}. Metallicities derived
by \citet{Savaglio:2005ApJ...635..260S}, \citet{erb:2006ApJ...644..813E}
and \citet{Maiolino:2008A&A...488..463M} have been converted respectively
from \citet{Kobulnicky:2004ApJ...617..240K}, \citet{Pettini:2004MNRAS.348L..59P}
$N2$ , and \citet{Kewley:2002ApJS..142...35K} methods to the CL01
method.

The mass-metallicity relation at redshift $z\sim0.7$ is given by
Eq.~8 of \citet{Savaglio:2005ApJ...635..260S}, shifted by $0.47$
dex in stellar mass. The mass-metallicity relation at redshift at
redshift $z\sim2.0$ is given by Eq.~3 of \citet{Tremonti:2004astro.ph..5537T},
shifted by $-0.56$ dex in metallicity, as found by \citet{erb:2006ApJ...644..813E}.
The mass-metallicity relation at redshift $z\sim3.5$ is given by
Eq.~2 of \citet{Maiolino:2008A&A...488..463M}, with the parameters
given in their Table~5.

Contrary to \citet{Savaglio:2005ApJ...635..260S}, we find a flatter
slope of the mass-metallicity relation at $z\sim1$. Nevertheless,
the comparison of data points shows that their results and ours are
in good agreement for the highest mass bins. The larger difference
comes from the lowest mass bins, in which they are probably not complete.
We know indeed that lower mass-to-light ratio galaxies are preferentially
observed in the lowest incomplete mass bins, and that such galaxies
show smaller mean metallicities \citep{Ellison:2008ApJ...672L.107E}. 

The comparison with the data of \citet{erb:2006ApJ...644..813E},
which are taken at $z\sim2$, is less straightforward: there is a
fairly good agreement in metallicity with our high-mass end data,
but actually at a rather different redshift. This could mean that
there have been very little metallicity evolution from $z\sim2$ to
$z\sim1$, but this would be hard to understand when looking also
at the other results. The metallicity evolves indeed strongly from
$z\sim3.5$ \citep{Maiolino:2008A&A...488..463M}, and between $z\sim1$
and the local universe \citep[our data and][]{Savaglio:2005ApJ...635..260S}.
Nevertheless we note that this later discrepancy with \citet{erb:2006ApJ...644..813E}
results may be understood: according to the downsizing scenario, the
evolution of the most massive galaxies plotted here should be actually
smaller between $z=2$ and $z=1$ than between $z=1$ and $z=0$.
Among others, \citet{PerezGonzalez:2008ApJ...687...50P} have quantified
that galaxies below $\log(M_{\star}/M_{\odot})=11.5$ have formed
half of their stars at $z<1$.\textbf{ }The main reason of a possible
overestimate of \citet{erb:2006ApJ...644..813E} metallicities is
probably statistical variation effects. They have indeed based their
results on stacked spectra of very few galaxies. The effect of the
selection function is therefore difficult to analyze.

The type of galaxies observed by \citet{erb:2006ApJ...644..813E}
at $z\sim2$, which are active galaxies, may also have later evolved
to {}``red and dead'' passive galaxies, and got higher metallicities
than the ones actually observed by us or by \citet{Tremonti:2004astro.ph..5537T}
at lower redshifts. Such dead galaxies would unfortunately not satisfy
any more the selection function of any work based on emission-line
measurements, and would not be observed.

\subsection{Evolution of the mass-to-light ratio\label{sub:Evolutionml}}

\begin{figure}
\begin{centering}
\includegraphics[width=0.9\columnwidth]{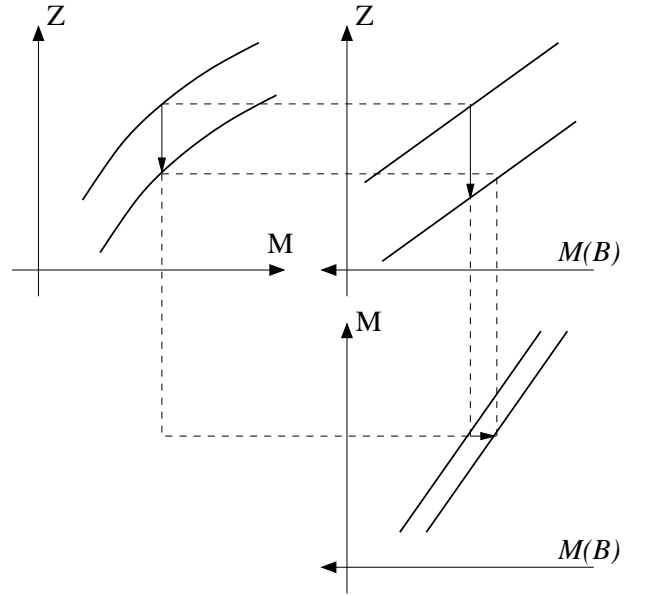}
\par\end{centering}

\caption{This plot shows, for galaxies with similar stellar masses, their evolution
when redshift increases: for metallicity in the mass-metallicity plane
(top-left), for metallicity in the luminosity-metallicity plane (top-right),
and for luminosity in the mass-luminosity plane (bottom-right). }

\label{fig:explainevol}
\end{figure}

\begin{table}
\caption{Evolution of the mass-to-light ratio for the wide and deep samples.
This evolution is computed as an absolute rest-frame $B$-band magnitude
evolution at constant stellar mass, from the comparison of the luminosity-metallicity
(see Fig.~\ref{fig:lz} and Table~\ref{tab:lz}) and mass-metallicity
(see Fig.~\ref{fig:mz} and Table~\ref{tab:mz}) relations. }

\begin{centering}
\begin{tabular}{lrr}
\hline 
\hline
 & \multicolumn{2}{r}{$\Delta M_{\mathrm{AB}}(B)$}\tabularnewline
redshift range & \emph{wide} & \emph{deep}\tabularnewline
\hline
~$0.0<z<0.5$ & $-0.06$ & $0.65$\tabularnewline
~$0.5<z<0.7$ & $-0.61$ & $-0.10$\tabularnewline
~$0.7<z<0.9$ & $-1.19$ & $-0.45$\tabularnewline
\end{tabular}
\par\end{centering}

\label{tab:evolmtl}
\end{table}

The metallicity evolution on the wide sample is very similar to what
has been found by \citet{Lamareille:2006A&A...448..907L} on the luminosity-metallicity
relation, after having applied a correction for luminosity evolution
\citep{Ilbert:2005A&A...439..863I}. This tells us that the stronger
evolution of the luminosity-metallicity relation, compared to the
mass-metallicity relation, is effectively due to an additional luminosity
evolution. This luminosity evolution has to be understood for galaxies
with similar masses, which means we can measure the evolution of the
mass-to-light ratio in galaxies using metallicity as a pivot. 

Fig.~\ref{fig:explainevol} shows schematically the global evolution
of metallicity in the luminosity-metallicity plane (top-right). When
there is both an evolution of the metallicity in the mass-metallicity
plane (top-left), and an evolution of the luminosity in the mass-luminosity
plane (bottom-right), galaxies being more luminous at a given mass,
the result is a stronger evolution of the metallicity in the luminosity-metallicity
plane.

The luminosity evolution at constant mass of our galaxies can be calculated
using the following formula:\begin{equation}
\Delta M_{\mathrm{AB}}(B)=a^{-1}\times\left(\Delta\log(\mathrm{O/H})^{M}-\Delta\log(\mathrm{O/H})^{L}\right)\label{eq:diffl}\end{equation}
where $a$ is the slope of the luminosity-metallicity relation, i.e.
$-0.31$ dex/mag. The results are given in Table~\ref{tab:evolmtl}.
The positive evolution for the lowest redshift bin in the deep sample
is due to the incompleteness of the $N2$ calibration: at a given
metallicity, our sample is biased towards higher luminosity. In the
other redshift ranges and the whole wide sample, we see that galaxies
at higher redshifts have lower mass-to-light ratios than today, and
that this evolution is more significant for massive galaxies (i.e.
the wide sample compared to the deep). These two results are in good
agreement with the general scenario of downsizing \citep{Cowie:1996AJ....112..839C}
for the evolution of the star formation rates in galaxies. Less massive
galaxies show a smaller evolution in their star formation rate at
$z<1$ since they are still actively forming stars. 

It will be further analyzed in a subsequent paper of this series.

\subsection{Derived star formation rates}

\begin{table*}
\caption{Derived star formation activities needed to explain the observed values
quoted in Table~\ref{tab:mz}, assuming the closed-box model, and
$h=0.7$, $\Omega_{\mathrm{m}}=0.3$ and $\Omega_{\Lambda}=0.7$ cosmology.
The values are given in the wide and deep samples, and for two different
assumptions: \emph{a)} constant star formation rate, \emph{b)} exponentially
declining star formation rate (see text). In case \emph{b}, each value
is the instantaneous star formation rate at the specified redshift
(see Eq.~\ref{eq:massevol2}).}

\begin{centering}
\begin{tabular}{lrrrr}
\hline 
\hline
mean redshift & \multicolumn{4}{r}{$\log(\mathrm{SFR}/M_{\star})$}\tabularnewline
 & \multicolumn{2}{r}{\emph{wide}} & \multicolumn{2}{r}{\emph{deep}}\tabularnewline
 & \emph{a} & \emph{b} & \emph{a} & \emph{b}\tabularnewline
\hline
~$0.3$ & $-11.06\pm0.7$ & $-10.93\pm0.7$ & $-11.05\pm0.9$ & $-10.99\pm0.9$\tabularnewline
~$0.6$ & $-10.64\pm0.5$ & $-10.36\pm0.5$ & $-10.51\pm0.4$ & $-10.40\pm0.5$\tabularnewline
~$0.77$ & $-10.73\pm0.4$ & $-10.27\pm0.5$ & $-10.65\pm0.5$ & $-10.53\pm0.5$\tabularnewline
\end{tabular}
\par\end{centering}

\label{tab:evolSFR}
\end{table*}

We now discuss the evolution of metallicity at constant stellar mass,
in terms of star formation rates.

Assuming constant star formation rates (SFR) and the closed-box model,
one can calculate a stellar mass evolution and relates it to a metallicity
evolution using the following equations:\begin{equation}
M_{\star}(t)=M_{\star}(0)+SFR\times t\label{eq:massevol}\end{equation}
\begin{equation}
M_{\star}(t)+M_{g}(t)=M_{\mathrm{tot}}\label{eq:mtot}\end{equation}

\begin{equation}
Z(t)=y\times\ln\left(1+\frac{M_{\star}(t)}{M_{g}(t)}\right)\label{eq:zevol}\end{equation}
where $M_{\star}$, $M_{g}$ and $M_{\mathrm{tot}}$ are respectively
the stellar mass, the gas mass, and the total baryonic mass of the
galaxy (which remain constant in the closed-box model); and where
$Z$ and $y$ are respectively the metallicity and the true yield
(which depends only on the stellar initial mass function).

We know the mass-metallicity relation at redshift $z=0$ \citep{Tremonti:2004astro.ph..5537T},
and the Table~\ref{tab:mz} gives the metallicity evolution for various
stellar masses and redshifts. Thus, we can revert Eq.~\ref{eq:massevol},
\ref{eq:mtot} and~\ref{eq:zevol} to derive the star formation rate
which explains the observed values. The results are quoted in Table~\ref{tab:evolSFR}
(case \emph{a}). They are calculated for a given cosmology, and for
a true yield $y=0.0104$ \citep{Tremonti:2004astro.ph..5537T}. 

A better modeling can be performed by taking into account the time
evolution of the star formation rates in galaxies. We may for example
assume an exponentially decreasing star formation rate, i.e. $SFR(t)=SFR(0)\times\exp(-t/\tau)$,
where $\tau$ gives the characteristic \emph{e}-folding time of the
galaxy. The exponentially decreasing law is a good choice when analyzing
a population of galaxies, with respect to the global cosmic evolution
of star formation rate. Eq.~\ref{eq:massevol} is then replaced by
the following formula:\begin{equation}
M_{\star}(t)=M_{\star}(0)+SFR(0)\times\tau\times\left(1-e^{-t/\tau}\right)\label{eq:massevol2}\end{equation}

Table~\ref{tab:evolSFR} (case \emph{b}) gives the results obtained
when assuming the relation given by Eq.~12 of \citet{Savaglio:2005ApJ...635..260S}
between the \emph{e}-folding time and the total baryonic mass of the
galaxy. Comparing cases \emph{a} and \emph{b}, we see that the results
are not dramatically affected by the assumption of a decreasing SFR.
The star formations activities in case \emph{b} are systematically
higher than in case \emph{a}, which is expected as with a decreasing
SFR a higher initial value is needed to explain the same metallicity
evolution, as compared to a constant SFR.

We note also that the\textbf{ }derived star formation activities in
the wide and deep samples\textbf{ }are very close, despite their different
ranges in stellar masses. In case \emph{a}, star formation activities
in the deep sample are nevertheless higher than in the wide sample,
which is expected as the wide sample spans higher stellar masses.
Conversely, in case \emph{b}, we find higher star formation activities
for more massive objects. This comes from the assumption, made by
\citet{Savaglio:2005ApJ...635..260S} and used in our equations, that
less massive galaxies have a longer \emph{e}-folding times. Consequently
the less massive galaxies of the deep sample, which are assigned long
\emph{e}-folding times ans thus can be approximated as constant SFR,
do not vary much from case \emph{a} to case \emph{b} while the more
massive galaxies of the wide sample, which are assigned short \emph{e}-folding
times, are more affected by the decreasing SFR hypothesis.

Preliminary results on the evolution of star formation activities
of VVDS galaxies, which will be presented in a subsequent paper \textbf{\citep[see also][]{Walcher:2008arXiv0807.4636W}},
give at redshift $z\approx0.6$ (our best sampled redshift bin) the
following numbers: $\approx10^{-9.8}$ yr$^{-1}$ for galaxies at
$\approx10^{9.5}$ solar masses (mean mass of the deep sample), and
$\approx10^{-10.3}$ yr$^{-1}$ for galaxies at $\approx10^{10}$
solar masses (mean mass the wide sample).\textbf{ }

Assuming the closed-box hypothesis, our results in the wide sample
are in fairly good agreement with the observed star formation activities
at similar masses, this agreement being better in case \emph{b}. But\textbf{
}in the deep sample, the inferred star formation activities which
explain our observed metallicity evolution seem significantly underestimated
by a factor $\approx5$. Although the varying \emph{e}-folding times
hypothesis is in agreement with the slower evolution of the mass-to-light
ratio found in the deep sample in previous section, it fails in providing
a good agreement between our inferred star formation activities in
the deep sample and observed values at similar masses. Moreover this
hypothesis make more massive galaxies forming stars more actively,
which is in contradiction with observations.

We therefore conclude that the closed-box hypothesis is not valid
to explain our observed metallicity evolutions (assuming that the
preliminary results mentioned above are confirmed).

\section{Conclusion}

We have calculated rest-frame luminosities, stellar masses and gas-phase
oxygen abundances of a statistically significant sample of star-forming
galaxies, selected from the VIMOS VLT Deep Survey. This has allowed
us to derive luminosity-metallicity and mass-metallicity relations
in various redshift ranges up to $z\sim0.9$. These relations have
also been derived in two sub-samples: the deep sample which extends
to lower observed luminosities thanks to a deeper magnitude selection,
and the wide sample which extends to higher luminosities thanks to
a larger solid angle. The selection function have been taken into
account in order to define volume- and mass-limited samples.

For both the wide and the deep samples, we have studied as a function
of redshift: the evolution in metallicity at constant luminosity,
the evolution in metallicity at constant stellar mass, the associated
evolution in luminosity at constant stellar mass, and the evolution
of the slope of the mass-metallicity relation.

Additionally, we have also:

\begin{itemize}
\item Measured the fraction of star-forming galaxies and AGNs as a function
of redshift.
\item Calculated the difference in terms of stellar masses between the old
BC03 and new CB07 stellar population models.
\item Studied the effect on the computation of stellar mass of the use of
spectral indices rather than only photometric points.
\item Checked that gas-phase oxygen abundance measurements in low and high
redshift ranges are comparable even if calculated with different methods.
\item Studied the marginal effect of candidate star-forming galaxies and
candidate AGNs on the derived luminosity-metallicity or mass-metallicity
relations.
\end{itemize}
In the wide sample, the mean evolution of metallicity at constant
luminosity, and of luminosity at constant stellar mass, are in good
agreement with previous studies by e.g. \citet{Lamareille:2006A&A...448..907L,Ilbert:2005A&A...439..863I,Hammer:2005A&A...430..115H}
among others. 

By doing the assumption that the slope of the luminosity-metallicity
relation, or the shape of the mass-metallicity relation, remain constant
with redshift, we finally found different metallicity evolutions for
the wide and deep samples. The wide and deep samples span different
ranges in masses, which shows that the assumption is wrong and that
this slope or shape actually evolves with redshift. We have found
that the most massive galaxies show the strongest metallicity evolution.
At $z\sim0.77$, galaxies at $10^{9.4}$ solar masses have $-0.18$
dex lower metallicities than galaxies of similar masses in the local
universe, while galaxies at $10^{10.2}$ solar masses have $-0.28$
dex lower metallicities.

We have then studied the mass-metallicity relation on co-added data
points by bins of stellar masses. The inferred slopes show an evolution
to a flatter mass-metallicity relation at $z\sim1$ as compared to
the local universe, which confirms that higher mass galaxies show
a stronger metallicity evolution during this period. The slope evolves
by $-0.25$ dex/decade by unit of redshift in the range $0<z<0.7$.

Moreover, we have inferred from our observations the star formation
activities which explain this evolution in metallicity. To do so,
we have assumed the closed-box hypothesis. We have found that the
observed metallicity evolution would be explained by similar star
formation activities at any mass, which is in contradiction with the
well-known correlation between stellar mass and star formation activity.
Assuming preliminary observations of the star formation activities
in VVDS galaxies, we conclude that our results are underestimated
by a factor $\approx5$ in the deep sample, while their are in fair
agreement in the wide sample.

These results make a strong evidence against the closed-box model.
Indeed, we know from previous studies that the star formation activity
of galaxies decreases while stellar mass increases. The only way to
explain the smaller metallicity evolution of the less massive galaxies
is thus to replace in Eq.~\ref{eq:zevol} the true yield by a smaller
effective yield. Assuming a smaller effective yield allows galaxies
to show smaller metallicity evolution, even with a stronger star formation
activity.

The smaller effective yield of less massive galaxies can be understood
in the {}``open-closed'' model. In this model, galaxies with small
stellar masses evolve like open-boxes: most of the metals produced
during star formation are ejected in the intergalactic medium by stellar
winds and supernovae feedback, the effective yield is thus very small.
Conversely, galaxies with high stellar masses evolve in the open-closed
model like closed-boxes: the metals are retained in the galaxy thanks
to a high gravitational potential, the effective yield is thus close
to the true yield. 

The dependence of the effective yield with the gravitational potential
of the galaxies, itself related to their total baryonic mass, has
been already shown by \citet{Tremonti:2004astro.ph..5537T}. It is
naturally explained in the hierarchical galaxy formation scenario:
in contrary to the closed-box model, the total baryonic mass does
not remain constant and increases with time, together with the stellar
mass, thanks to galaxy merging and accretion. This relation between
the stellar mass and the total baryonic mass is at the origin of the
mass-metallicity relation in the open-closed model, and is clearly
supported by our data.

\emph{In the open-closed model, the smaller metallicity evolution
of less massive is naturally explained despite their stronger star
formation activities.}\textbf{\emph{ }}

We emphasize that Eq.~\ref{eq:massevol}, \ref{eq:mtot} and~\ref{eq:massevol2}
are not valid any more in the open-closed model. It is not enough
to replace, in Eq.~\ref{eq:zevol}, the true yield by the effective
yield in order to infer the right star formation rates which explain
the observed metallicity evolution. These equations must be modified
to take also the difference between the true and the effective yield
into account. The way to modify these equations depends on the model
used to explain smaller effective yields (e.g. gas loss). Replacement
equations can be found in other studies \citep[e.g.][]{erb:2006ApJ...644..813E,erb:2008ApJ...674..151E,Dalcanton:2007ApJ...658..941D,Finlator:2008MNRAS.385.2181F}.
We also refer the reader to subsequent papers of this series, for
a more detailed discussion of the relation between the present results
and the evolution of the star formation rates of galaxies, and for
comparisons with simulations \citep[e.g.][]{DeLucia:2004MNRAS.349.1101D}.

\begin{acknowledgements}
We thank C. Tremonti for having made the original \emph{platefit}
code available to the VVDS collaboration. F. Lamareille thanks the
Osservatorio di Bologna for the receipt of a post-doctoral fellowship.\\
 This work has been partially supported by the CNRS-INSU and its Programme
Nationaux de Galaxies et de Cosmologie (France), and by Italian Ministry
(MIUR) grants COFIN2000 (MM02037133) and COFIN2003 (num.2003020150)
and by INAF grants (PRIN-INAF 2005).\\
 The VLT-VIMOS observations have been carried out on guaranteed time
(GTO) allocated by the European Southern Observatory (ESO) to the
VIRMOS consortium, under a contractual agreement between the Centre
National de la Recherche Scientifique of France, heading a consortium
of French and Italian institutes, and ESO, to design, manufacture
and test the VIMOS instrument.\\
This work is based in part on data products produced at TERAPIX and
the Canadian Astronomy Data Centre as part of the Canada-France-Hawaii
Telescope Legacy Survey, a collaborative project of NRC and CNRS.
\end{acknowledgements}
\bibliographystyle{aa}
\bibliography{my}

\end{document}